\documentclass{jfm}
\usepackage{graphicx}
\usepackage{epstopdf, epsfig}

\usepackage{url}
\usepackage{amsmath}
\usepackage{natbib}
\usepackage{float}
\usepackage{indentfirst}
\usepackage{hyperref}

\usepackage{booktabs}
\usepackage{tabularx}
\usepackage{color}
\usepackage{caption}
\usepackage[position=top, singlelinecheck=false]{subfig}
\usepackage{setspace}
\usepackage{verbatim}
\usepackage{xcolor}
\usepackage{bm}

\shorttitle{Fragmentation of Colliding Liquid Rims}
\shortauthor{K. Tang, T.A.A. Adcock and W. Mostert}

\title{Fragmentation of Colliding Liquid Rims}

\author{K. Tang\aff{1}
  \corresp{\email{kaitao.tang@eng.ox.ac.uk}},
  T.A.A. Adcock\aff{1}
 \and W. Mostert\aff{1}}

\affiliation{\aff{1}Department of Engineering Science, University of Oxford, Oxford OX1 3PJ, UK}

\begin{document}

\maketitle

\begin{abstract}
We present direct numerical simulations of the splashing process between two cylindrical liquid rims. This belongs to a class of impact and collision problems with a wide range of applications in science and engineering, and motivated here by splashing of breaking ocean waves. Interfacial perturbations with a truncated white noise frequency profile are introduced to the rims before their collision, whose subsequent morphological development is simulated by solving the two-phase incompressible Navier-Stokes equation with the adaptive mesh refinement (AMR) technique, within the Basilisk software environment. We first derive analytical solutions predicting the unsteady interfacial and velocity profiles of the expanding sheet forming between the two rims, and develop scaling laws for the evolution of the lamella rim under capillary deceleration. We then analyse the formation and growth of transverse ligaments ejected from the lamella rims, which we find to originate from the initial corrugated geometry of the perturbed rim surface. Novel scaling models are proposed for predicting the decay of the ligament number density due to the ongoing ligament merging phenomenon, and found to agree well with the numerical results presented here. The role of the mechanism in breaking waves is discussed further and necessary next steps in the problem are identified.
\end{abstract}

\section{Introduction}
\label{sec:introduction}

Liquid atomisation is a class of challenging multiphase problems \citep{Obenauf2021} featuring a large separation of scales and various interacting physical mechanisms, which is of significance to numerous fields of application including meteorology \citep{Villermaux2009}, ink-jet printing \citep{castrejon2015plethora, castrejon2021formulation, lohse2022fundamental}, internal combustion engines \citep{yarin2006drop}, and pharmaceutical manufacturing \citep{Mehta2017}. Within the context of air-sea interactions, liquid fragmentation is primarily associated with wave breaking events, and gives rise to ocean sprays. These spray drops are then transported within the atmospheric boundary layer while exchanging with the latter moisture, momentum and heat during their lifetime; thus leaving their impact on both global and regional climates \citep{lhuissier2012bursting, Veron2015, deike2022mass}. Atomisation involves topological changes of the liquid bulk driven by external forces, typically followed by formation of corrugated ligaments subject to capillary breakup, and ends with a number of fragments featuring a broad size distribution, the knowledge of which is crucial for various areas of applications listed above \citep{Villermaux2007}. 

Among various types of liquid atomisation problems, the impact of liquid droplets has received much scholarly attention for its ubiquitous presence, rich dynamics and vast range of applications \citep{yarin2006drop, Villermaux2007, cheng2022drop} since the pioneering experimental study of \cite{worthington1877xxviii}. The impact process features a competition between inertial and capillary forces, which together with the characteristics of the impacting object and the surrounding gas phase shapes the final outcome of the original droplet: bouncing, spreading or splashing. Empowered by the rapid pace of sensor developments and increasing computational capacities, past works have elucidated considerable details about the ephemeral kinematic and morphological development of drops impacting with various types of surfaces \citep{cheng2022drop}, including liquid films \citep{thoraval2013drop}, deep pools \citep{agbaglah2015drop, wang2023analysis}, smooth solid surfaces \citep{wildeman2016spreading, cimpeanu2018three}, rough solid surfaces with friction \citep{garcia2021spreading}, or an identical droplet \citep{he2019non, he2022spin}. Some recent works have also probed the dynamic properties of drop impact including the distribution of impact force and stresses, providing an alternative approach to investigate impact dynamics at early times \citep{cheng2022drop}. Specifically, there have been a series of recent experimental works studying the high-speed impact of droplets with a surface of comparable size as a canonical unsteady fragmentation problem \citep{wang2018universal, wang2017drop, wang2018unsteady, wang2021growth, wang2022mass}, providing valuable insights into various aspects of the impact process in the limit of large $We$, including the self-similar evolution of the liquid-phase thickness and velocity profile \citep{wang2017drop}, the dynamics of the rim bordering the expanding liquid sheet \citep{wang2018universal}, the growth of liquid ligaments and the detachment of liquid drops from their tips \citep{wang2018unsteady, wang2021growth}, and the partition of mass, momentum and energy during the entire collision process \citep{wang2022mass}.

However, the impact-induced fragmentation of liquid bulks featuring non-spherical initial shapes is also attested, which has received considerably less attention and remains less understood compared to drop impact problems \citep{liu2021role}. Among these is the collision of liquid rims, which has seen some recent investigations experimentally \citep{neel2020fines} and numerically \citep{agbaglah2021breakup} and is also the focus of the present work. In the two works cited above, an initially intact liquid film is perforated to form small holes, which then expand under surface tension and develop bordering rims travelling at the Taylor-Culick velocity \citep{taylor1959dynamics, culick1960comments}. Neighbouring film holes merge with one another when their bordering rims collide, oscillate and break up into small fragments, a process commonly observed during the rupture of films, which may be induced by rapid radial expansion of liquid shells \citep{vledouts2016explosive} or the inflation of a liquid drop interacting with a surrounding airflow \citep{jackiw2022prediction, tang2022bag, ling2023detailed}. \cite{agbaglah2021breakup} placed the two holes immediately adjacent to each other so that the two liquid rims collide at low impact Weber number $We$; and the fused liquid bridge is found to pinch off under oscillation and form only a few small droplets. \cite{neel2020fines} were able to investigate the rim collision phenomena by varying the initial distance between the two perforation sites, thus varying the impact $We$ value within the range of $50 \leq We \leq 200$. Apart from the primary capillary breakup mechanism of fused liquid bridges as discussed by \cite{agbaglah2021breakup}, \cite{neel2020fines} also identified a critical $We$ value of 66 beyond which the decelerating transverse lamella rims develop elongating ligaments under the Rayleigh-Taylor (RT) instability, which then produce many secondary fine drops under capillary instabilities featuring a skewed size distribution. Additionally, there have been some early theoretical analyses on the capillary-driven coalescence of two liquid cylinders \citep{hopper1993acoalescence, hopper1993bcoalescence, eggers1999coalescence}; although these are conducted at the creeping flow limit with negligible liquid bulk velocity, and thus may not be directly applicable to the current problem featuring finite collision speeds.

\begin{figure}
	\centering
	\subfloat[]{
		\label{fig:splashing-wang16}
		\includegraphics[height=.27\textwidth]{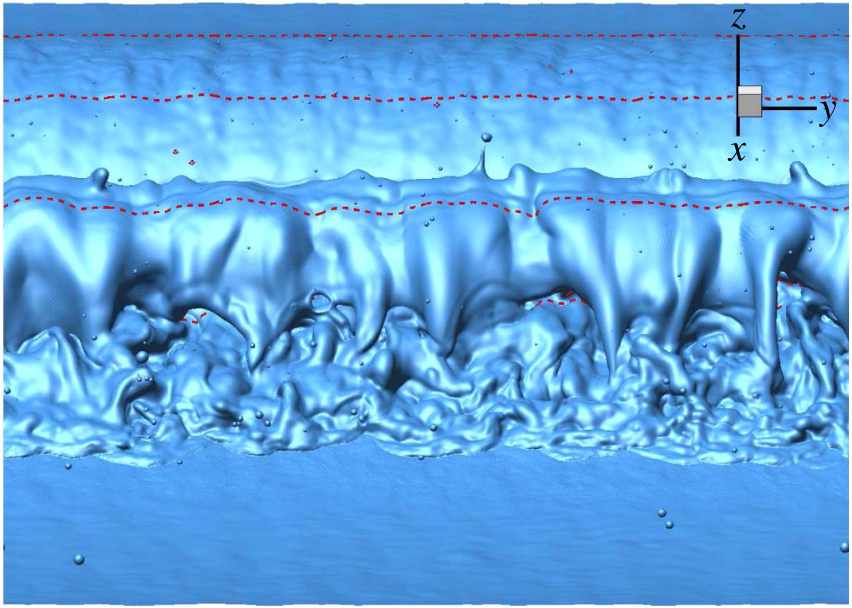}}
    \centering
	\subfloat[]{
		\label{fig:splashing-mostert22}
		\includegraphics[height=.27\textwidth]{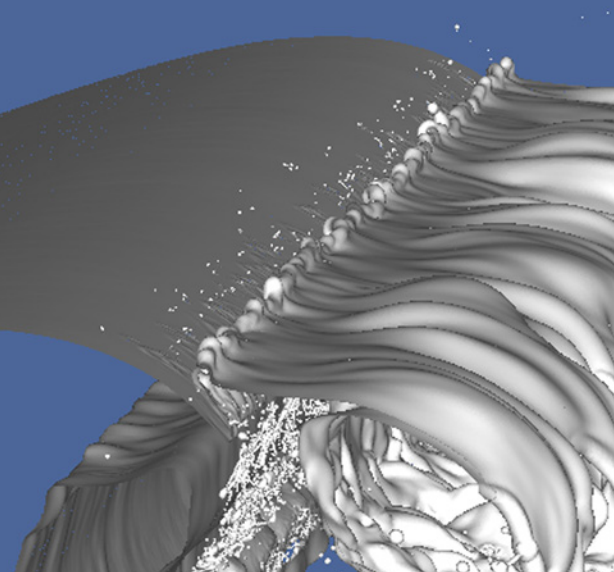}}
    \centering
	\subfloat[]{
		\label{fig:splashing-erinin23}
		\includegraphics[height=.27\textwidth]{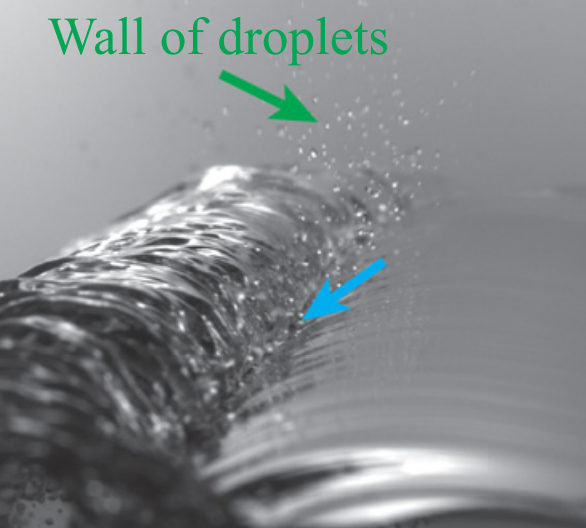}}
  
    \centering
	\subfloat[]{
		\label{fig:splashing-erinin23-sketch}
		\includegraphics[width=.75\textwidth]{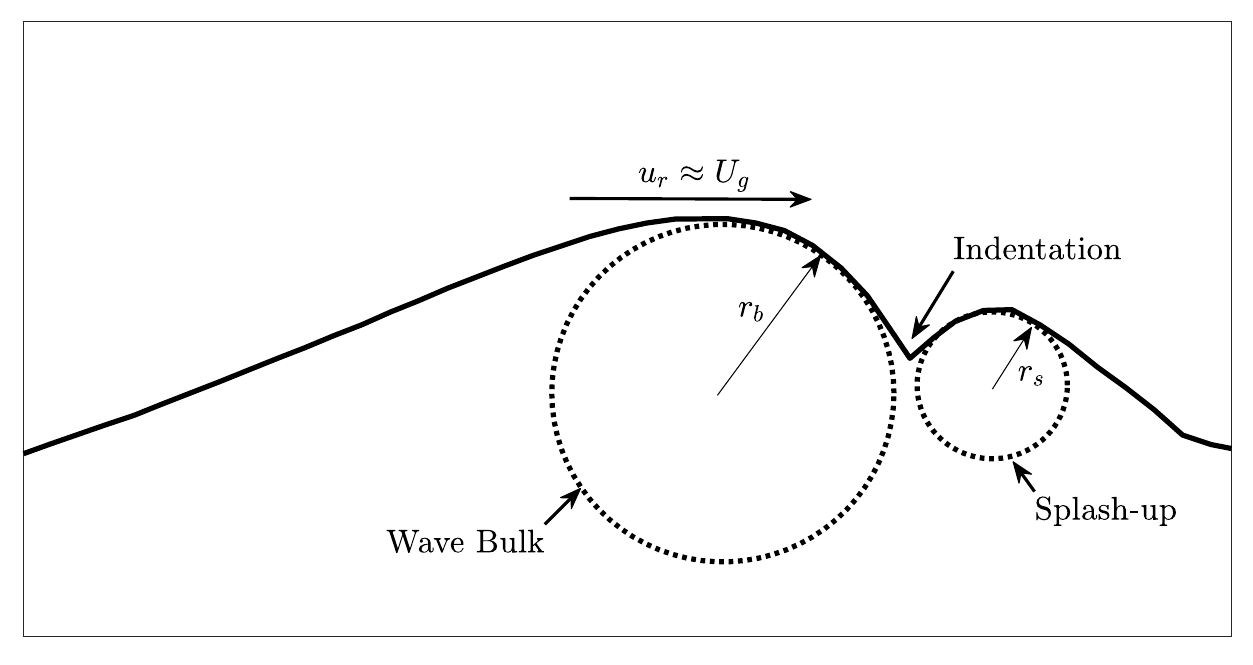}}
	\caption{(a)-(c): Wave splashing observed in previous numerical (a,b) and experimental (c) studies, adapted from \cite{wang2016high} (a), \cite{Mostert2021} (b) and \cite{erinin2023plunging_b} (c), respectively. (d): Sketch showing the ensemble-averaged breaking wave profile taken from \cite{erinin2023plunging_a} after the initial moment of impact in the breaking wave, at the moment of secondary splashing, where dashed lines indicate our simplification of the problem as the collision of two cylindrical rims with radii $r_b$ and $r_s$. In this study we consider the basic case $r_s = r_b$.}
	\label{fig:splash-observs}
\end{figure}

Apart from its presence during the rupture of thin films, the collision of liquid rims is also of specific interest due to its strong resemblance to the secondary wave splashing phenomena observed in ocean wave breaking events \citep{kiger2012air, Mostert2021, erinin2023plunging_b}. Namely, two consecutive well-defined splashing phases have been identified \citep{kiger2012air} during the lifetime of a deep-water plunging breaker. The first is the ‘forward splashing’ mechanism occurring right upon the reconnection of the overturning wave front and the sea surface below, which according to \cite{Mostert2021} only produces small amounts of droplets. After this, as shown in fig.~\ref{fig:splashing-erinin23-sketch}, the wave bulk catches up with the decelerated splash-up generated from the initial impact at a relative speed $u_r$ close to the phase speed of the unbroken wave $U_g$. At the indentation region between these two structures, the shape of the wave bulk and the splash-up can be approximated as two cylinders, whose cross sections feature radii of curvature $r_b$ and $r_s$ which are typically different. The rapid closure of the indentation region leads to the `secondary splashing’ phenomenon, which is characterised by a wall of vertically projected small droplets along the transverse direction as reproduced in figs.~\ref{fig:splashing-wang16}-\ref{fig:splashing-erinin23}, accompanied by air entrainment within the former indentation region \citep{kiger2012air}. Under experimental conditions, fragments generated via this mechanism comprise about one third of the total amount of droplets produced over the entire wave-breaking process \citep{erinin2023plunging_b}; and \cite{Mostert2021} found that this splashing mechanism produces many fragments and can be curbed by strong surface tension (small $Bo$ values). While \cite{wang2016high} noted the connection between the corrugated surface of the splash-up and the vortical structures beneath the wave surface, to our knowledge no existing study has analysed the physical mechanism governing the formation of these fragments, and their contribution to the overall droplet distribution associated with wave breaking remains unknown \citep{andreas1995spray, kiger2012air, Veron2015}. Furthermore, this splashing mechanism is found to produce many fragments close to the minimum grid size of \cite{Mostert2021}; together with the highly transient nature of wave breaking and the presence of other fragmentation mechanisms, this indicates considerable difficulty in investigating the secondary splashing phenomena directly within the context of wave breaking. While we do not yet reproduce the fragment statistics seen in the studies above, the present work serves as a first step towards understanding the more complex wave splashing phenomena by retaining the major generation mechanism of splash fragments while leaving out many complicating factors, including size difference between the wave bulk and the splash-up evolving with time, and internal turbulent flow within the liquid phase; as a similar approach taken by \cite{gao2021bubble} reveals the connection between the bubble size distributions of destabilising air cylinders and air cavities entrained by plunging breakers.

In this work, we conduct a comprehensive investigation of the collision between two liquid cylinders with identical size, covering the entire deformation and fragmentation period. The direct comparison and establishment of connections between the rim collision results and the statistics of the secondary wave splashing phenomenon are left for future work, together with the role of gravity and the difference between the sizes of the wave bulk ($r_b$) and the splash-up ($r_s$) which complicate the early-time rim dynamics. Two-phase numerical simulations are conducted to derive detailed flow field information during this highly transient collision process. Our study is structured as follows. We first present in \S\ref{subsec:description} the problem configuration and the parameter space of the current work, and then introduce the numerical method in \S\ref{subsec:num-method}. After providing an overview of the rim collision phenomena in \S\ref{sec:overview}, we quantitatively analyse the development of each part of the expanding liquid bulk successively following a spatial order, namely the kinematics of the spreading liquid sheet (\S\ref{subsec:sheet-kinematics}) and its bordering rim (\S\ref{subsec:rim-evolution}), the growth and merge of transverse ligaments topping the rim (\S\ref{sec:lig-dynamics}), and the statistics of fragments shed from the ligaments (\S\ref{sec:frag-stats}). We conclude the study in \S\ref{sec:conclusions} with some remarks on future work.

\section{Formulation and methodology}
\label{sec:formulation}

\subsection{Problem description}
\label{subsec:description}

\begin{figure}
	\centering
	\subfloat[]{
		\label{fig:cyl-col-config}
		\includegraphics[height=.35\textwidth]{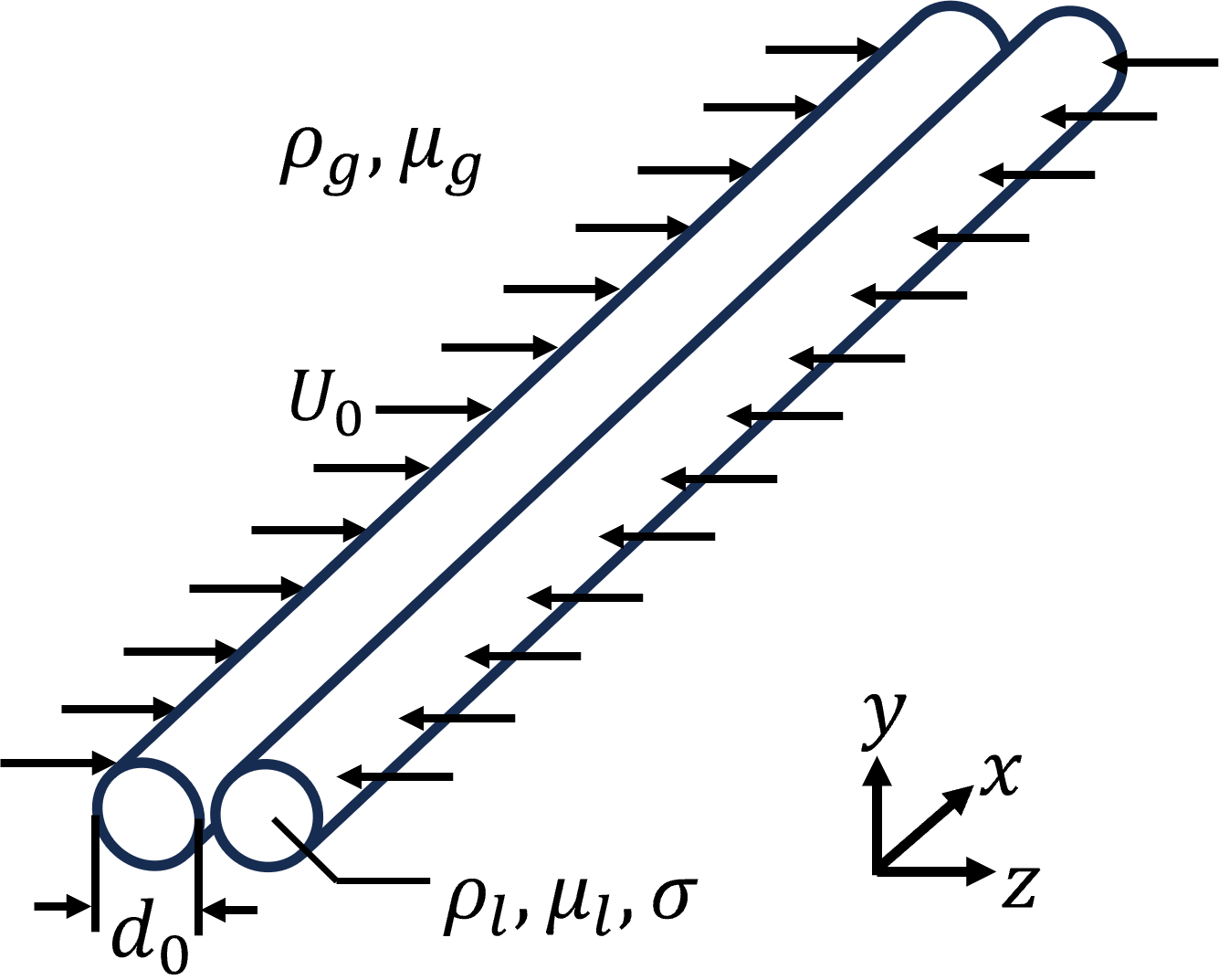}}
    \centering
	\subfloat[]{
		\label{fig:noise-spectrum}
		\includegraphics[height=.35\textwidth]{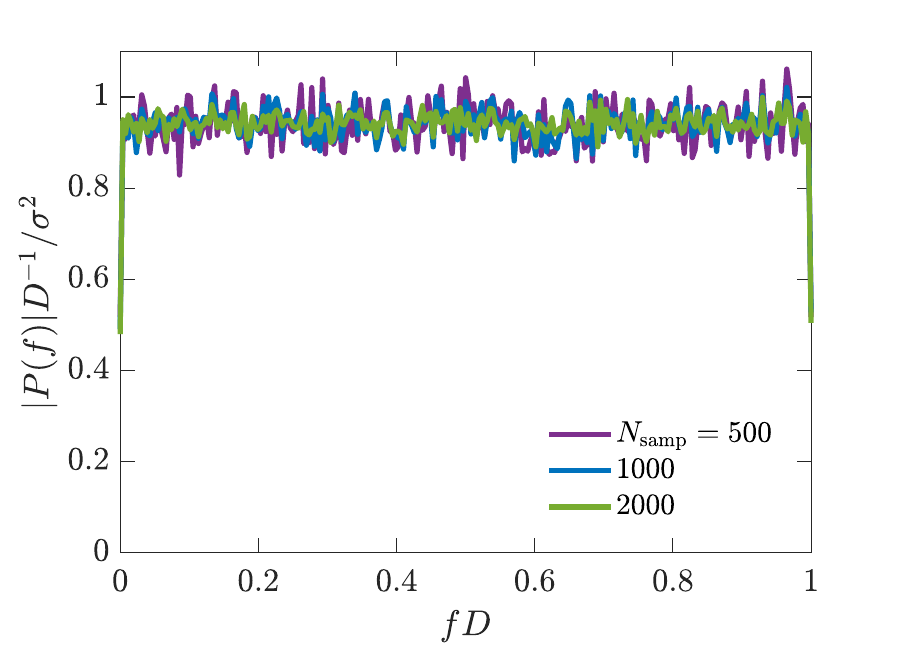}}
	\caption{(a): Sketch showing the configuration of the liquid rim collision problem; (b): ensemble-averaged power density spectrum of the white noise signal for generating initial interface perturbations on the cylindrical rims.}
	\label{fig:configurations}
\end{figure}

The geometrical configuration for the rim collision problem is shown in fig.~\ref{fig:cyl-col-config}, where two infinitely long cylindrical liquid rims with diameter $d_0$, density $\rho_l$ and viscosity $\mu_l$ are aligned along the $x$ axis, and set to travel along the $z$ axis with uniform velocities of opposite signs and the same magnitude $U_0$. The liquid cylinders are surrounded by an inert gas phase with density $\rho_g$ and viscosity $\mu_g$, and the liquid-gas interface is characterised by a surface tension coefficient $\sigma$. It is noted that gravitational effects have been neglected in the current setup; and differing from the configurations of \cite{neel2020fines} and \cite{agbaglah2021breakup}, there is no interstitial film connecting the two approaching cylinders. Consequently, four non-dimensional controlling parameters can be written for this problem:
\begin{equation}
    We \equiv \frac{\rho_l (2U_0)^2 d_0}{\sigma}, \quad Oh \equiv \frac{\mu_l}{\sqrt{\rho_l d_0 \sigma}}, \quad \rho^* \equiv \frac{\rho_l}{\rho_g}, \quad \mu^* = \frac{\mu_l}{\mu_g},
    \label{for:non-dimensional-groups}
\end{equation}
where $We$ and $Oh$ are respectively the Weber and Ohnesorge number comparing inertial and viscous effects to capillary forces, and $\rho^*$ and $\mu^*$ are respectively the density and viscosity ratios of the liquid and gas phase. In this work, $Oh$ is set as 0.01 in most of the simulations, whereas its influence on the fragment statistics is briefly discussed in \S\ref{sec:statistics-converge}. $\rho^*$ and $\mu^*$ are set as 830 and 55, respectively, which are typical values for the air-water system \citep{Pairetti2018}.

We set the width $D$ of the cubic simulation domain as $10d_0$ to allow enough space for the morphological evolution of the coalesced liquid structure. Utilising the symmetry of the splashing phenomena about the $xz$-plane, we only model the merging of the upper halves of the two liquid cylinders to save computational resources. A symmetric boundary condition is therefore applied at the bottom, and an outflow boundary condition is imposed on the top boundary so that fragments produced from the collision can leave the domain from there at late time; while the other boundary conditions are set as periodic. 

To investigate the sensitivity of the fragmentation process to the initial conditions \citep{liu2016numerical, berny2022size}, and also taking into account the surface corrugation of the splash-up in wave breaking events \citep{kiger2012air} which still has not been quantified according to our knowledge; we introduce random transverse perturbation within a certain wavelength range on the cross-sectional radius of the two cylindrical rims \citep{pal2021statistics}, in the form of filtered white noise signals characterised by the following two parameters,
\begin{equation}
    \varepsilon_0 \equiv \frac{2\epsilon_0}{d_0}, \quad N_{\rm max},
\end{equation}
where $\varepsilon_0$ is the non-dimensionalised characteristic amplitude of perturbation, and $N_{\rm max}$ defines the highest wavenumber among the spectrum of the perturbation signal. The filtered white noise signal is the default type of initial interface perturbation we impose on the rims, as \cite{zhang2010wavelength} did for analysing the linear stability of the crown splash. Occasionally we also impose single-wavelength sinusoidal perturbations, or a superposition of sinusoidal perturbations with wavelengths $\lambda = D/8, \, D/16$ and $D/32$ for comparison, which will be explicitly denoted by 'Sing.' and 'Sup.' when reported. In the case of single-wavelength perturbations, $N_{\rm max}$ corresponds to the perturbation wavenumber.

As discussed above, $We$, $\varepsilon_0$ and $N_{\rm max}$ constitute the parameter space for the present study. Among these, $We$ is varied between 60 and 280 where the coalesced liquid bulk expands vertically to form a lamella, and $\varepsilon_0$ is set as 0.02, 0.04 and 0.06, within the limit of small radial perturbations. $N_{\rm max}$ varies between 15 and 80, whose influence will be discussed in detail in \S\ref{subsec:lig-merge}.

\subsection{Numerical method}
\label{subsec:num-method}

The open-source scientific computation toolbox Basilisk \citep{Popinet2019basilisk} is used in this work to solve the two-phase nonlinear, incompressible, variable-density Navier-Stokes equations. A second-order accurate discretisation is applied in both space and time, and a geometric volume-of-fluid (VOF) method in a momentum-conserving formulation is used to maintain a sharp representation of the liquid-gas interface while minimising the parasitic currents induced by surface tension \citep{popinet2018numerical, tang2021effects}. Capillary effects are modelled as source terms in the Navier-Stokes equations using an adaptation of Brackbill's method \citep{brackbill1992continuum, popinet2009accurate}, which calculates the interface curvature by taking the finite-difference discretisation of the derivatives of interface height functions \citep{popinet2009accurate}. The octree-based adaptive mesh refinement (AMR) scheme based on the estimation of local discretisation errors of the VOF function $f$ and flow velocity $\boldsymbol{u}$ is adopted so as to reduce the computational cost at high resolution levels $L$, which is defined using the minimum grid size,
\begin{equation}
    \Delta = \frac{D}{2^L}.
\end{equation}
The results presented in the main body of this work are obtained from three-dimensional simulations at $L = 10$, at which the late-time evolution of interface profile and liquid-phase energetics have reached grid independence. The numerical convergence of fragment statistics is also established for fragments with diameter larger than $4\Delta_{10}$, which is discussed in detail in \S\ref{sec:statistics-converge}. Results from some two-dimensional simulations are also presented for comparison with three-dimensional rim dynamics (\S\ref{subsec:rim-evolution}), and to investigate lamella foot formation at very early time (\S\ref{sec:pos-yneck}).

Since the present study focuses on the liquid-phase morphological development \emph{after} the rims begin to coalesce, the distance between the symmetric axes of the two cylindrical rims are set at initialisation as $\Delta z_c = 0.95d_0$, so that the slightly overlapped rims form a line of contact. The interfacial perturbations on the rims are introduced as follows. Firstly, discrete white noise signals with unit variance $\sigma^2$ are produced using the random number generator provided in Basilisk. Fig.~\ref{fig:noise-spectrum} shows the ensemble-averaged power density spectra of the white noise signals generated in Basilisk at different number of realisations $N_{\rm samp}$. It is observed that the power density $P(f)$ is close to the theoretical value of $\sigma^2$ at all frequencies $f$, matching the requirement of frequency independence for white noise signals. Next, we apply a low-pass filter on these signals so that only the lowest $N_{\rm max}$ wavelengths are preserved, and the filtered signal is normalised so that its standard deviation becomes $\epsilon_0 = 0.5\varepsilon d_0$. The normalised signals $\eta$ are then mapped onto the transverse radius profile $R(x)$ of the liquid rims, in the form of $R(x) = R_0 + \eta(x)$. The two liquid cylinders being positioned close to each other ensures that they coalesce immediately when the simulation starts; and there is not sufficient time for capillary effects to smooth out the perturbations, or amplify them to trigger the Rayleigh-Plateau (RP) instability \citep{pal2021statistics} so that the cylinders break up prematurely.

\section{Overview of rim splashing}
\label{sec:overview}

\begin{figure}
    \centering
	\subfloat[]{
		\label{fig:snapshot-We-200-0.06-iso-1}
		\includegraphics[width=.32\textwidth]{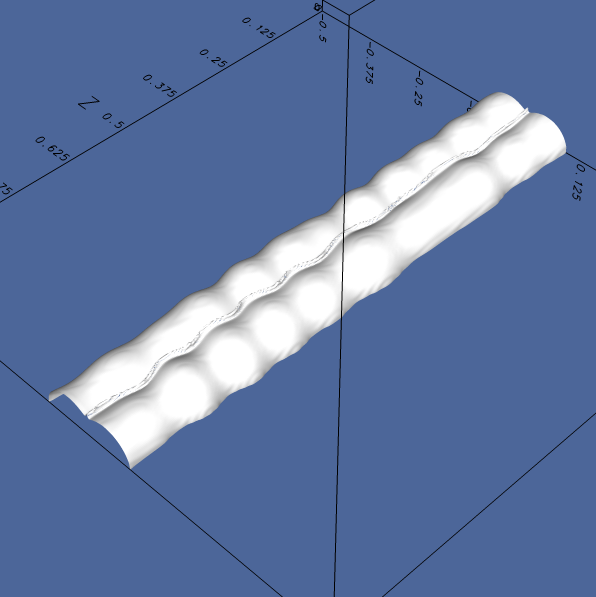}}
	\centering
	\subfloat[]{
		\label{fig:snapshot-We-200-0.06-iso-2}
		\includegraphics[width=.32\textwidth]{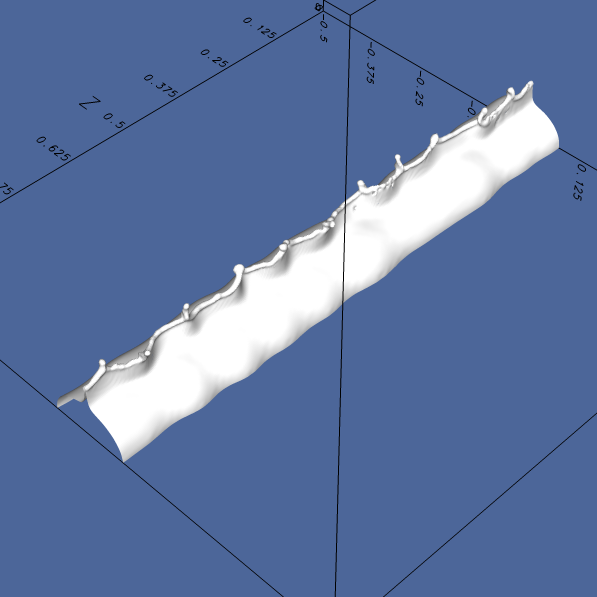}}
    \centering
	\subfloat[]{
		\label{fig:snapshot-We-200-0.06-iso-3}
		\includegraphics[width=.32\textwidth]{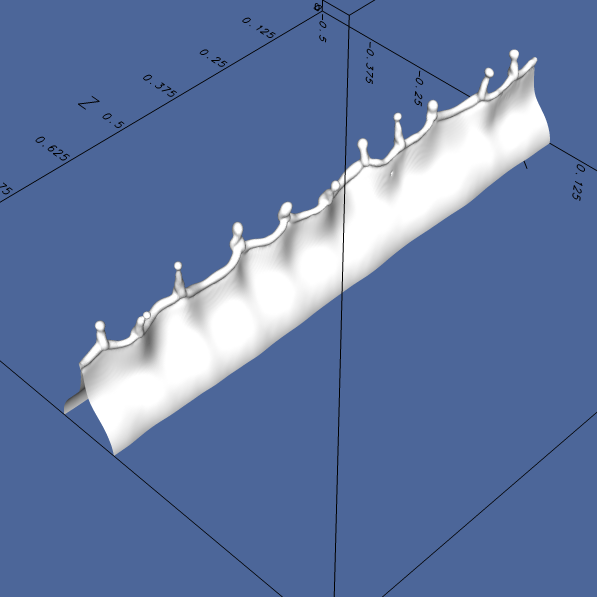}}

	\caption{Isometric snapshots showing the liquid sheet expansion process at $We = 200, \, \varepsilon = 0.06$ and $N_{\rm max} = 25$. From left to right: $t/\tau_{\rm cap} = 0.029$, 0.113 and 0.454.}
	\label{fig:snapshot-iso-We-200}
\end{figure}

\begin{figure}
    \centering
	\subfloat[]{
		\label{fig:snapshot-We-60-0.06-1}
		\includegraphics[width=.32\textwidth]{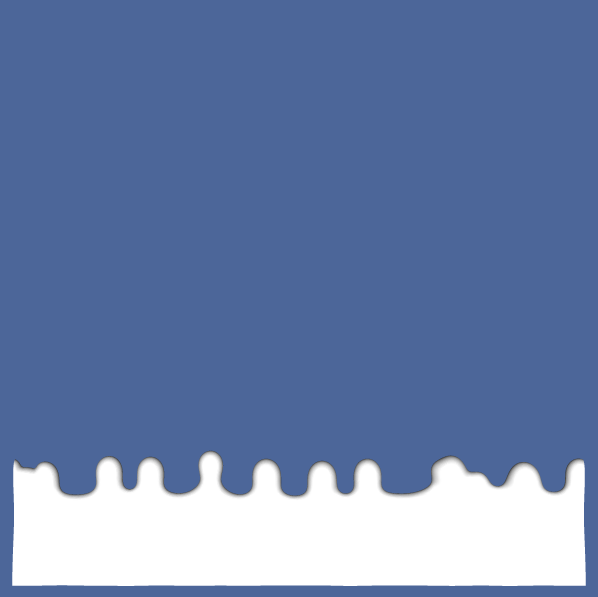}}
	\centering
	\subfloat[]{
		\label{fig:snapshot-We-60-0.06-2}
		\includegraphics[width=.32\textwidth]{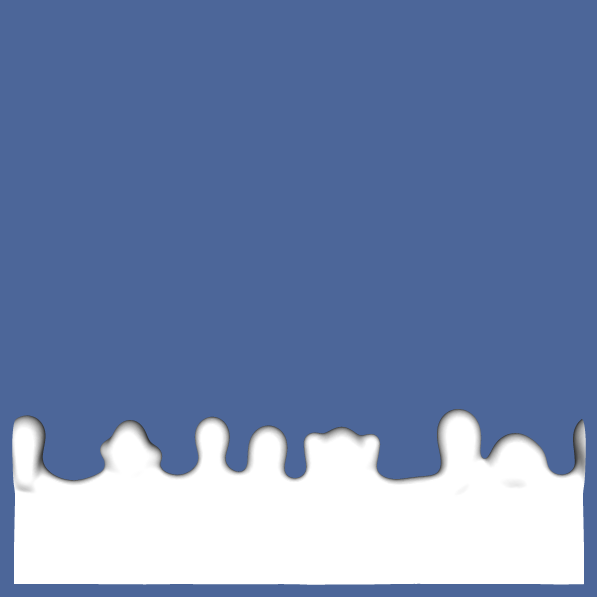}}
    \centering
	\subfloat[]{
		\label{fig:snapshot-We-60-0.06-3}
		\includegraphics[width=.32\textwidth]{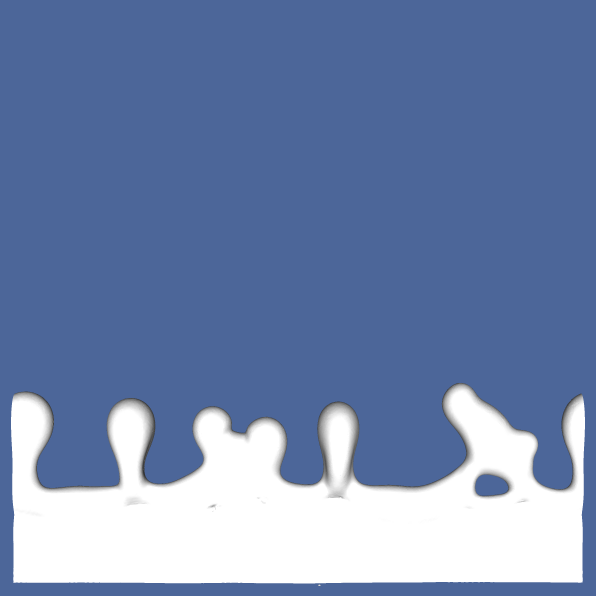}}

    \centering
	\subfloat[]{
		\label{fig:snapshot-We-200-0.06-1}
		\includegraphics[width=.32\textwidth]{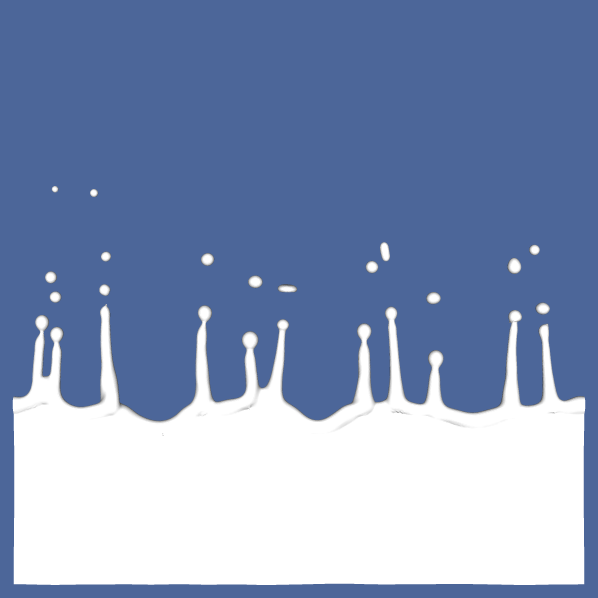}}
	\centering
	\subfloat[]{
		\label{fig:snapshot-We-200-0.06-2}
		\includegraphics[width=.32\textwidth]{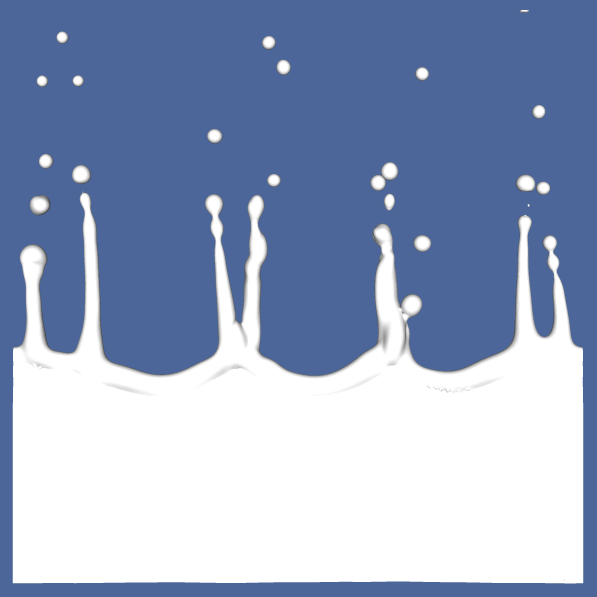}}
    \centering
	\subfloat[]{
		\label{fig:snapshot-We-200-0.06-3}
		\includegraphics[width=.32\textwidth]{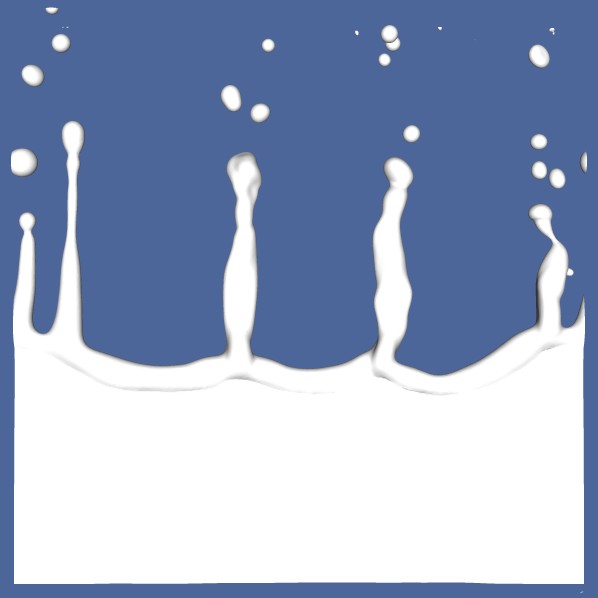}}

    	\centering
	\subfloat[]{
		\label{fig:snapshot-We-60-0.02-1}
		\includegraphics[width=.32\textwidth]{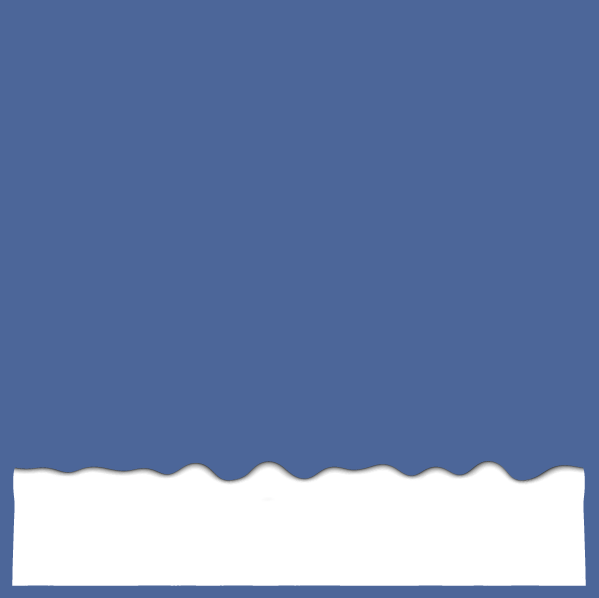}}
	\centering
	\subfloat[]{
		\label{fig:snapshot-We-60-0.02-2}
		\includegraphics[width=.32\textwidth]{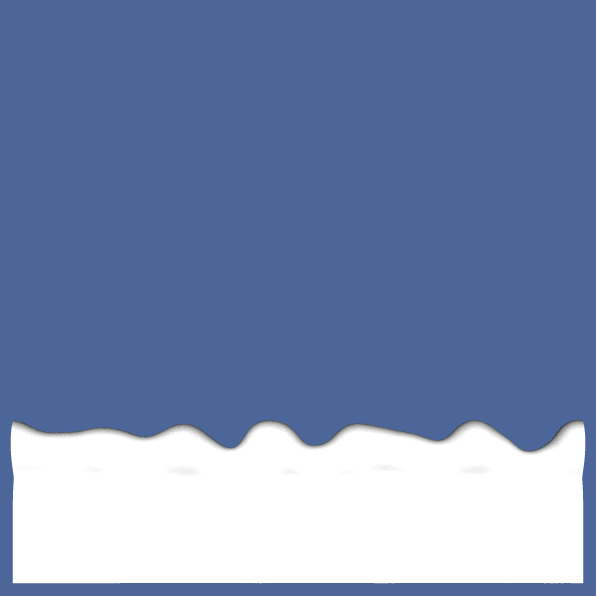}}
    \centering
	\subfloat[]{
		\label{fig:snapshot-We-60-0.02-3}
		\includegraphics[width=.32\textwidth]{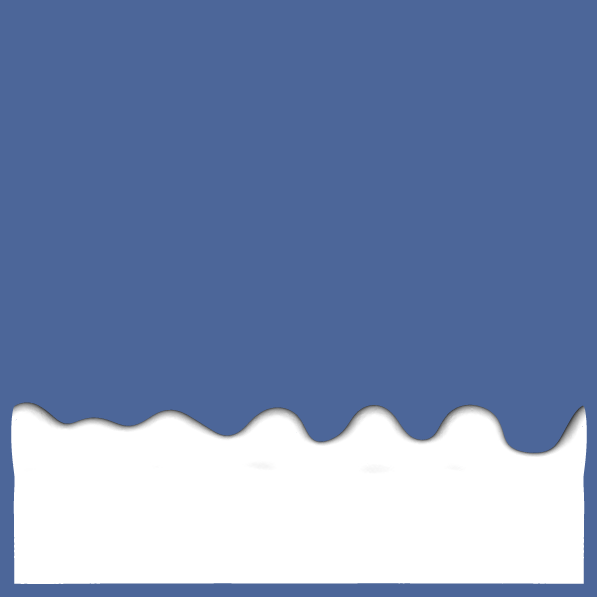}}
	\caption{Snapshots showing the liquid sheet expansion process at $We = 60, \, \varepsilon = 0.06$ (top), $We = 200, \, \varepsilon = 0.06$ (middle), and $We = 60, \, \varepsilon = 0.02$ (bottom). From left to right: $t/\tau_{\rm cap} = 0.91$, 1.82 and 2.73. For all three cases, $N_{\rm max} = 25$.}
	\label{fig:snapshot-overview}
\end{figure}

Here we qualitatively describe the rim splashing process as observed in the simulations before analysing the detailed dynamics at each stage in the following sections. Fig.~\ref{fig:snapshot-iso-We-200} presents the isometric view of the splashing phenomenon at $We = 200$, $\varepsilon = 0.06$ and $N_{\rm max} = 25$, whereas fig.~\ref{fig:snapshot-overview} shows the side view for a few different $(We, \, \varepsilon)$ configurations with $N_{\rm max} = 25$. Tiny air bubbles are entrained within the indentation space between the two rims following the impact \citep{thoraval2013drop, josserand2016droplet, erinin2023plunging_b}, which in our case have no known effects on the subsequent development of liquid-phase flow fields. The rims coalesce rapidly, causing a dramatic increase in the local liquid-phase pressure \citep{neel2020fines}. A very large pressure gradient arises at the surface `indentation' between the two cylinders since the air-phase pressure remains unchanged, leading to the vertical acceleration of fluids near the surface \citep{longuet2001vertical}. The magnitude of the vertical velocity near the contact line can be a few times larger than the initial horizontal velocity $U_0$, causing the contact line to advance upwards rapidly with fluid particles nearby converging to it, while the high-pressure region follows it closely rather than remaining on the axis of symmetry \citep{philippi2016drop}. A thin transverse liquid lamella \citep{neel2020fines} appears as shown in figs.~\ref{fig:snapshot-We-200-0.06-iso-2}, and the location of the lamella foot agrees well with Eq.~\eqref{for:neck_pos_evol} developed in \S\ref{sec:pos-yneck} using potential flow theory \citep{riboux2014experiments}. The liquid lamella is also observed in the experimental study of \cite{goswami2023simultaneous} where two neighbouring drops impact with a solid surface simultaneously and expand into contact. Within the parameter space of the present study, this transverse lamella continues to expand vertically under the pressure difference between the two phases and consumes the two impacting cylinders, evolving into a thin film aligned with the $xy$-plane; in the meantime it is continuously decelerated by capillary force, forming a thick rim at its upper border. 

Without the initial interface perturbation, the liquid rim would remain intact and mostly smooth during the entire simulation period up to $t/\tau_{\rm cap} = 2.73$, where $\tau_{\rm cap} \equiv \sqrt{\rho_l d_0^3/8\sigma}$ is the capillary time scale corresponding to the initial rim diameter. When perturbation is introduced, we observe the amplification of transverse perturbation waves on the rim as shown in fig.~\ref{fig:snapshot-overview}. Depending on the value of $We$, these waves will either slowly increase in amplitude and reach nonlinear development, whose growth rate increases with the initial perturbation amplitude $\varepsilon_0$, as shown in the first and last row of fig.~\ref{fig:snapshot-overview}; or generate slender ligaments which continue to elongate on the top of `cusp' structures \citep{gordillo2014cusps}, as shown in fig.~\ref{fig:snapshot-iso-We-200} and the middle row of fig.~\ref{fig:snapshot-overview}. It is noted that for all three cases presented in fig.~\ref{fig:snapshot-overview}, the number density of transverse ligaments (or the characteristic perturbation wave number when no such ligaments form) decreases over time rather than remaining a constant; a phenomenon which we will analyse in more details in \S\ref{sec:lig-dynamics}. In the meantime, lamella expansion gradually slows down as the vertical position of its bordering rim reaches saturation, and begins to slightly retract at low $We$ values as can be seen in figs.~\ref{fig:snapshot-We-60-0.06-2} and \ref{fig:snapshot-We-60-0.06-3}. As is the paradigm of droplet formation in many previous fragmentation studies \citep{Villermaux2007}, here the transverse ligaments act as the direct and only source of fine drops as the latter intermittently detach from the ligament tips, whose statistics will be presented in \S\ref{sec:frag-stats}. Under the restoring effects of capillary force, most of the fragments undergo prolate-oblate shape oscillations after their detachment, and a small portion of them may cross path and merge to form larger droplets during their flight \citep{tang2022bag}. Note that in our simulations all fragments keep moving upwards before crossing over the top boundary and leaving the simulation domain; while in wave splashing scenarios without wind forcing they will ultimately fall back to the sea surface under gravity and be destroyed \citep{Mostert2021}.

\section{Liquid lamella expansion}
\label{sec:sheet-expansion}
In this section, we study the dynamics of the liquid lamellae consisting of the expanding sheet (\S\ref{subsec:sheet-kinematics}) and its bordering rim (\S\ref{subsec:rim-evolution}). To simplify the problem which features random initial perturbation, here we do not consider the formation of liquid ligaments. Instead, we average the transverse cross section of the coalesced liquid bulk along the $x$ axis following \cite{wang2017drop} so that the lamella expansion process can be described in a quasi-one-dimensional manner, characterised by one-dimensional velocity and thickness profiles $u_y(y,t)$ and $h(y,t)$.

\subsection{Liquid sheet kinematics}
\label{subsec:sheet-kinematics}

The unsteady evolution of the expanding sheet profile is of both fundamental and practical importance for impact problems, especially for predicting their maximum spread radius \citep{yarin2006drop, josserand2016drop, wang2017drop}, and the first step towards modelling the profile evolution of expanding sheets is to understand the fluid motion within them. 

We first derive the velocity profile within liquid lamella sheets in Cartesian coordinates for $We \gg 1$ and $Oh \ll 1$. At early times, the vertically expanding sheet is bounded at its lower end by the line of collision between the two cylinders, which we call the lamella foot. The trajectory of this foot is given by $y_n = 2\sqrt{U_0t/R_0}$, per the analysis given in \S\ref{sec:pos-yneck} and following \cite{gordillo2019theory}. At later times, this lamella foot becomes increasingly indistinct as the colliding cylinders merge. 

We now proceed to discuss the kinematics of the fluid within the vertical lamella sheet itself, subject to these considerations, following the general analytical strategy of \cite{wang2017drop}. After neglecting viscous, compressibility and capillary effects, the quasi-one-dimensional momentum equation in the vertical direction can be written as
\begin{equation}
    \frac{\partial u_y}{\partial t} + u_y \frac{\partial u_y}{\partial y} = 0,   
\end{equation}
which may be presented alternatively in the Lagrangian form along the characteristics $dy/dt = u_y$,
\begin{equation}
    \frac{D u_y}{D t} = 0,
    \label{for:momentum-lagrange}
\end{equation}
where $D/Dt$ is the total derivative. \eqref{for:momentum-lagrange} suggests that the velocity of a fluid particle remains unchanged within the liquid sheet as it travels vertically upwards. We can then integrate along the characteristics and obtain the motion of fluid particles within a Lagrangian frame following it,
\begin{equation}
    y = u_y t + \xi,
    \label{for:motion-lagrange}
\end{equation}
where $u_y$ and $\xi$ are the initial vertical velocity and position of fluid particles. Following \cite{yarin1995impact}, \cite{wang2017drop} assume that the initial velocity is proportional to the initial position, namely $u_y = \beta \xi$, similar to the velocity field of a stagnation-point flow. Consequently, \eqref{for:motion-lagrange} may be rewritten in the Eulerian reference frame as
\begin{equation}
    u_y(y,t) = \frac{y}{t + \frac{1}{\beta}},
    \label{for:sheet-vel-vert-prelim}
\end{equation}
where $1/\beta$ corresponds to the time needed to set up the post-collisional velocity profile within the liquid bulk, which according to \cite{wang2017drop} is at the same order of the collision timescale $d/U_0$. As sheet expansion further progresses so that $t \gg 1/\beta$, \eqref{for:sheet-vel-vert-prelim} asymptotes to 
\begin{equation}
    u_y(y,t) = \frac{y}{t}.
    \label{for:sheet-vel-vertical}
\end{equation}

\begin{figure}
	\centering
	\subfloat[]{
		\label{fig:sheet-vel-prof-We-120}
		\includegraphics[width=.48\textwidth]{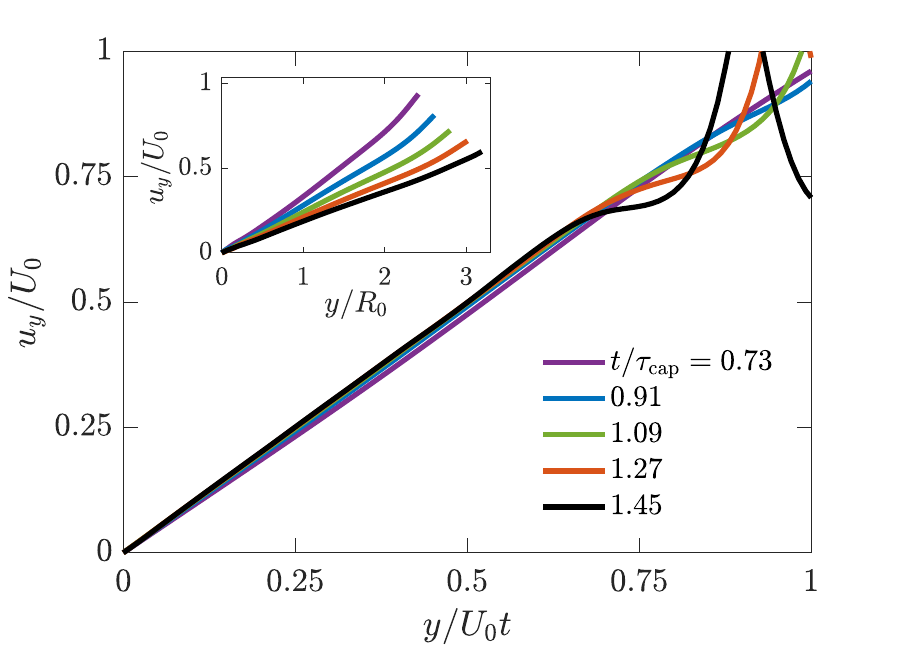}}
    \centering
	\subfloat[]{
		\label{fig:sheet-vel-prof}
		\includegraphics[width=.48\textwidth]{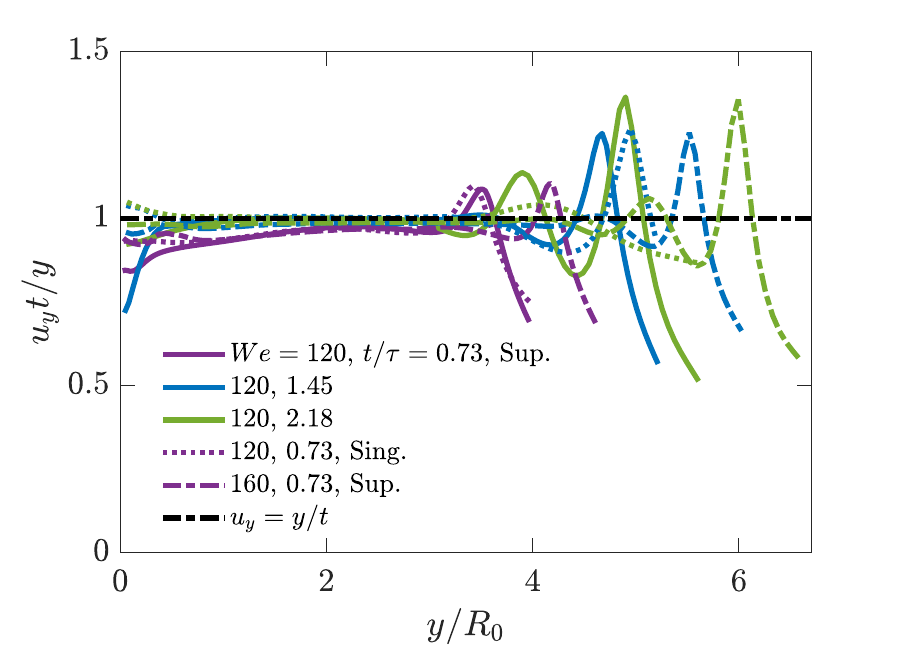}}
	\caption{(a): Liquid sheet velocity profile at $We = 120$, scaled according to \eqref{for:sheet-vel-vertical}; (b): verification of \eqref{for:sheet-vel-vertical} at different values of $We$, time and perturbation waveforms. `Sing.' indicates that the initial perturbation we impose features a single wavenumber $N_{\rm max}$, while `Sup.' denotes a combination of sinusoidal perturbations with wavelengths $\lambda = D/8, \, D/16$ and $D/32$.}
	\label{sheet-v-prof}
\end{figure}

We measure the vertical velocity profile within the expanding sheet from our numerical simulations at different times for $We = 120$, and first plot them in the inset of fig.~\ref{fig:sheet-vel-prof-We-120}. Consistent with our assumption, the vertical velocity $u_y$ is observed to scale linearly with the vertical position $y$, with the slope decreasing over time. In the main plot we rescale the horizontal axis by $U_0 t$, after which the velocity profiles at different times all collapse onto a single straight line $y=x$ for $y \leq 2U_0 t$, agreeing with the theoretical model \eqref{for:sheet-vel-vertical}. For $We = 120$, the collision timescale is $d/U_0 \approx 0.52 \tau_{\rm cap}$, while fig.~\ref{fig:sheet-vel-prof-We-120} indicates that the velocity profile at $t/\tau_{\rm cap} = 0.73$ is already very close to the analytical solution \eqref{for:sheet-vel-vertical}; which suggests that in fact the initialisation timescale $1/\beta \ll d/U_0$, and model \eqref{for:sheet-vel-vertical} is applicable to the early collision stage where $t \approx d/U_0$. Note that as time elapses, the vertical velocity deviates from Eq.~\eqref{for:sheet-vel-vertical} at progressively smaller values of $y/U_0 t$, where the fluid parcels move away from the sheet towards the bordering rim. Indeed, in fig.~\ref{fig:sheet-vel-prof} we also plot the vertical velocity normalised by $y/t$ at different times, $We$ values and perturbation waveforms. All results presented therein feature a range where the normalised velocity $u_y t/y = 1$, suggesting that Eq.~\eqref{for:sheet-vel-vertical} remains valid at different initial configurations within our current parameter space.

Having established the liquid velocity profile within the lamella sheet, we can further solve for its thickness profile utilising the continuity equation, which can be written as follows for a thin sheet expanding in the $y$ direction,
\begin{equation}
    \frac{\partial h}{\partial t} + \frac{\partial (u_y h)}{\partial y} = 0,
    \label{for:sheet-cont-eqn}
\end{equation}
combined with Eq.~\eqref{for:sheet-vel-vertical}, this yields
\begin{equation}
    t \frac{\partial h}{\partial t} + y\frac{\partial h}{\partial y} + h = 0.
\end{equation}
This can be solved by separation of variables in the form of $h(y,t) = f(t)g(y)$ to obtain the evolution of sheet thickness $h$, which is written in a non-dimensional formulation as
\begin{equation}
    \frac{hU_0t}{R_0^2} = f \left(\frac{y}{U_0 t} \right),
    \label{for:sheet-profile}
\end{equation}    
note that this result differs from the axisymmetric configuration where $h U_0^2 t^2 / R_0^3$ evolves self-similarly without an explicit time dependence \citep{wang2017drop, gordillo2019theory}. 

\begin{figure}
	\centering
	\subfloat[]{
		\label{fig:sheet-height-prof-sim-unit}
		\includegraphics[width=.48\textwidth]{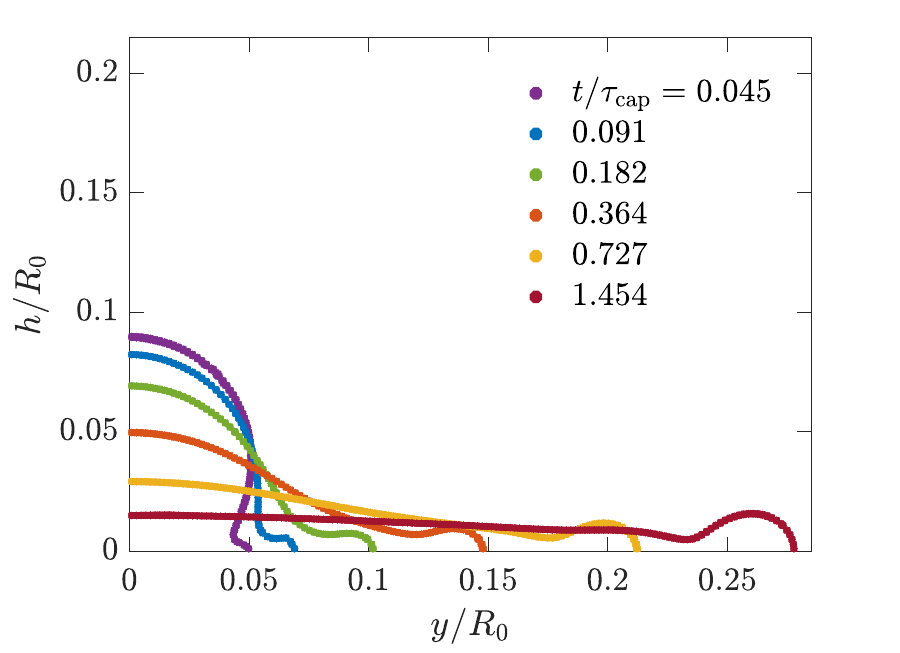}}
	\centering
	\subfloat[]{
		\label{fig:sheet-height-prof}
		\includegraphics[width=.48\textwidth]{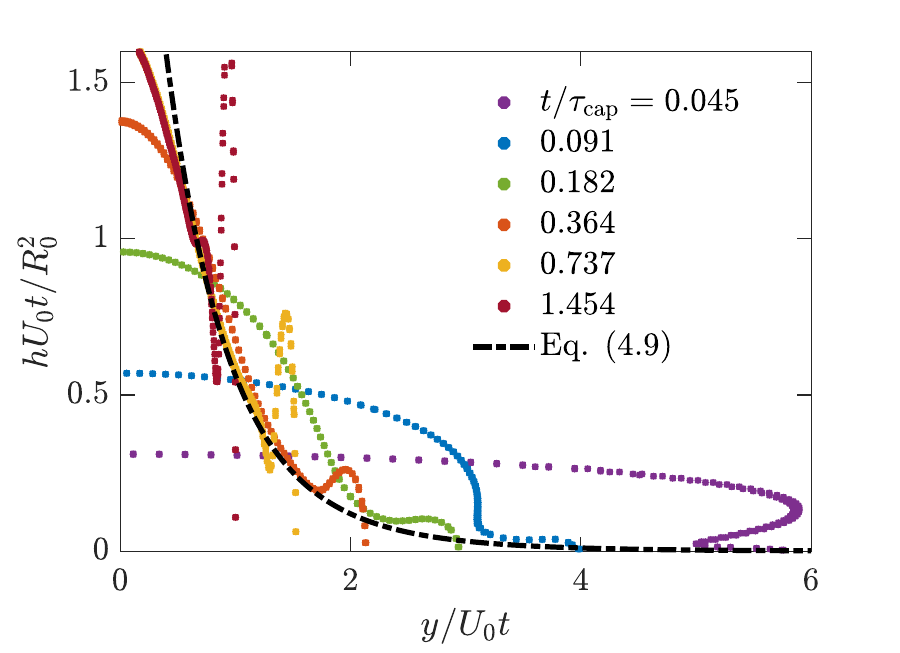}}
	\caption{(a): Liquid sheet profiles at $We = 120$; (b): comparison between interface profiles non-dimensionalised according to \eqref{for:sheet-profile} and the exponential fit \eqref{for:int-exp-fit}.}
	\label{sheet-h-prof}
\end{figure}

We plot the lamella thickness profiles for $We = 120$ in fig.~\ref{fig:sheet-height-prof-sim-unit}, where it can be seen that the `bulge' centred around $y=0$ gradually flattens into an extended thin liquid sheet, pushing the bordering rim further along the vertical direction. The bordering rim is connected to the sheet via a neck, reminiscent of capillary waves upstream of inviscid liquid rims receding at the Taylor-Culick velocity \citep{savva2009viscous}. Fig.~\ref{fig:sheet-height-prof} further shows the profiles rescaled by Eq.~\ref{for:sheet-profile}. The `bulk' region where $y/R_0 \leq \sqrt{2 U_0 t/R_0}$ initially retains its cylindrical shape during the initialisation period $t \sim 1/\beta$, and only comes to agreement with Eq.~\eqref{for:sheet-profile} when the lamella foot disappears and the bulk can no longer be decisively told apart from the lamella. The non-dimensionalised lamella profile in fig.~\ref{fig:sheet-height-prof} is found to be well described by the following functional form $f(x)$,
\begin{equation}
    f(x) = 0.5081 e^{-x} + 2.782 e^{-2x}.
    \label{for:int-exp-fit}
\end{equation}
    
Lastly, we note that although the liquid lamella expands vertically to form a thin film, the latter does not suffer from spontaneous perforation during our simulation period, although this is observed in fig.~14 of \cite{vledouts2016explosive} and marks the onset of bag film fragmentation in droplet aerobreakup problems \citep{tang2022bag, ling2023detailed}, where the liquid films are subject to radial accelerations. Neither does the lamella film experience destabilisation under shear force arising from its interaction with the surrounding gas phase \citep{Villermaux2011, riboux2015diameters}, which may arise at larger $We$ values. We therefore do not take special measures to artificially stabilise \citep{liu2016numerical} or perforate \citep{chirco2021manifold} the expanding lamellae in this study.

\subsection{Bordering rim evolution}
\label{subsec:rim-evolution}
As is discussed in \S\ref{subsec:sheet-kinematics}, while the expanding lamella sheet abides by the velocity profile \eqref{for:sheet-vel-vertical} and the self-similar thickness profile \eqref{for:sheet-profile}, these two models break down for the bordering rim, which demands separate scaling laws to describe its kinemamics. Similar rim structures are ubiquitous in impact problems and act as the crucial link between the expanding sheet and shedding droplets \citep{wang2018universal}, and understanding their motion lays the foundation for further theoretical analysis of ligament merging, which we will perform in \S\ref{subsec:lig-merge}.

\begin{figure}
	\centering
	\subfloat[]{
		\label{fig:rim-pos-evol}
		\includegraphics[width=.48\textwidth]{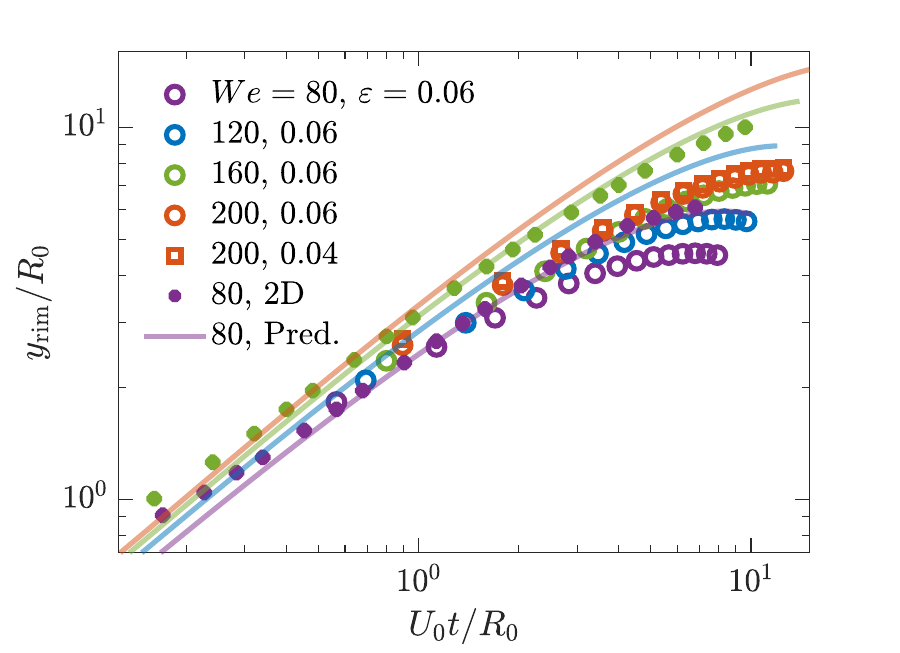}}
	\centering
	\subfloat[]{
		\label{fig:rim-thick-evol}
		\includegraphics[width=.48\textwidth]{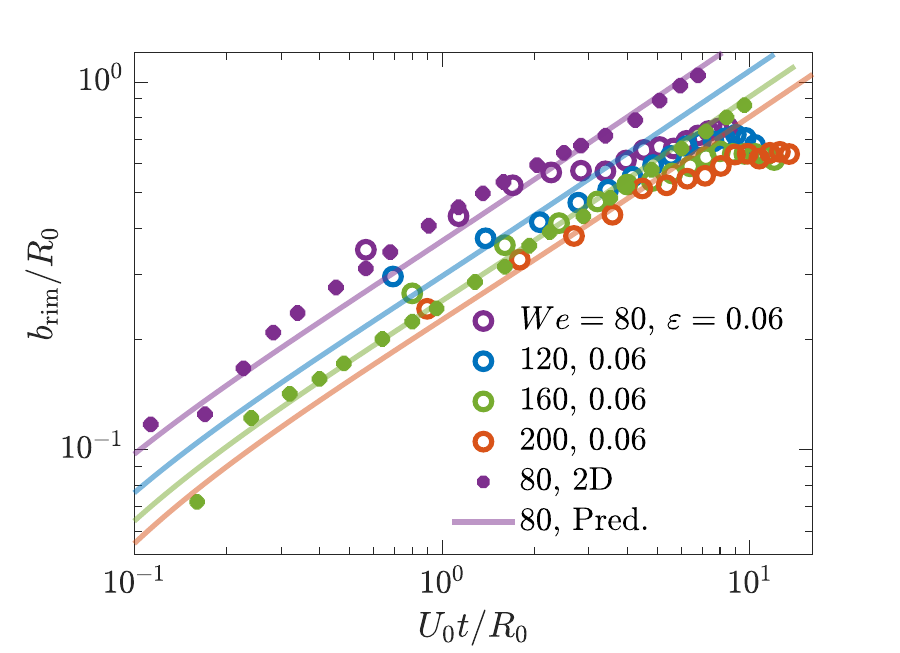}}

    \centering
	\subfloat[]{
		\label{fig:rim-pos-collapse-comp}
		\includegraphics[width=.48\textwidth]{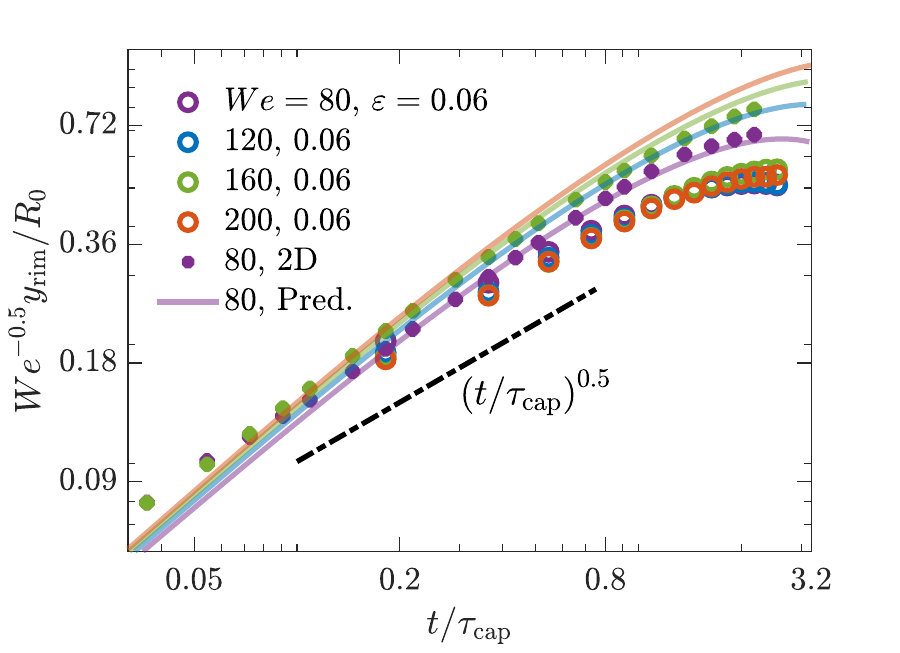}}
	\centering
	\subfloat[]{
		\label{fig:rim-thick-collapse-comp}
		\includegraphics[width=.48\textwidth]{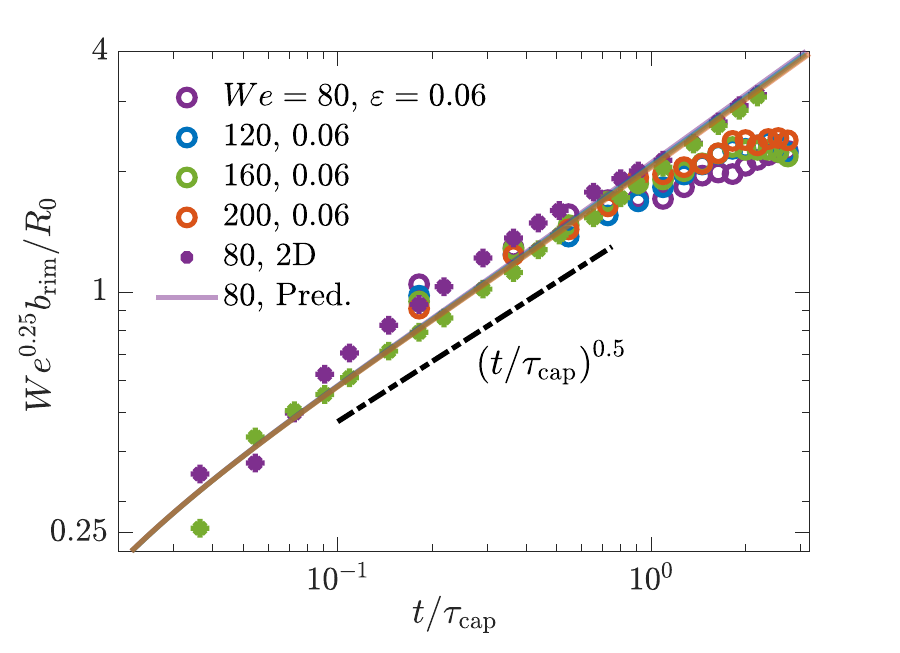}}
	\caption{(a)(b): The evolution of the vertical position $y_{\rm rim}$ (a) and the rim thickness $b_{\rm rim}$ (b) over time, compared with solutions of Eqs.~\eqref{for:rim-dyn-model-mass}-\eqref{for:rim-dyn-model-momentum} at corresponding $We$ values (solid lines). Early-time measurements from two two-dimensional simulations with $We = 80$ and 160 are also included. (c)(d): Results in (a) and (b) rescaled using Eq.~\eqref{for:rim-pos-thick-scaling}.}
	\label{fig:rim-geometry-evol}
\end{figure}

In a quasi-one-dimensional framework, the kinematics of the bordering rim can be characterised by two parameters, namely its average vertical position $y_{\rm rim}$ and diameter $b_{\rm rim}$. We first present their evolution at different $We$ values in figs.~\ref{fig:rim-pos-evol} and \ref{fig:rim-thick-evol}, respectively; where time and length are scaled using $U_0/R_0$ and $R_0$. It is seen that both $y_{\rm rim}$ and $b_{\rm rim}$ first increase with time and eventually saturate; and $y_{\rm rim}$ and $b_{\rm rim}$ show different dependence on $We$, as the former increases and the latter decreases with $We$, although these $We$-dependencies have become very subtle by $We = 200$. Further, the evolution of $y_{\rm rim}$ for $We = 200, \, \varepsilon = 0.06$ and $We = 200, \, \varepsilon = 0.04$ in fig.~\ref{fig:rim-pos-evol} is virtually the same, which shows that the dynamic behaviour of the rim is largely independent of the initial perturbation amplitude $\varepsilon$.

We now seek to further compare our numerical results obtained in figs.~\ref{fig:rim-pos-evol} and \ref{fig:rim-thick-evol} with available theoretical models. Following \cite{gordillo2019theory}, the following mass- and momentum-conservation equations can be proposed in the non-dimensionalised form to describe the dynamic behaviour of the advancing lamella rim in the inviscid limit,
\begin{gather}
    \frac{\pi}{8} \frac{d b_{\rm rim}^2}{dt} = [u_y(y_{\rm rim}, t) - v_{\rm rim}]h(y_{\rm rim}, t), \label{for:rim-dyn-model-mass} \\ 
    \frac{d y_{\rm rim}}{dt} = v_{\rm rim}, \\
    \frac{\pi}{8} \frac{d}{dt} (b_{\rm rim}^2 v_{\rm rim}) = u_y(y_{\rm rim}, t){[u_y(y_{\rm rim}, t) - v_{\rm rim}]} h(y_{\rm rim}, t) - \frac{8}{We}, \label{for:rim-dyn-model-momentum}
\end{gather}
where the cylinder radius $R_0$, initial impact velocity $U_0$ and their quotient $R_0/U_0$ are chosen as the reference scales for length, velocity and time. The numerical solutions of the ordinary differential equation (ODE) system \eqref{for:rim-dyn-model-mass}-\eqref{for:rim-dyn-model-momentum} at various $We$ values, with initial conditions as defined in \S\ref{sec:pos-yneck}, are presented in figs.~\ref{fig:rim-pos-evol} and \ref{fig:rim-thick-evol} respectively as transparent solid lines. Since most of our measurements from three-dimensional simulations are taken when $U_0 t/R_0 > 1$, we also present results of two-dimensional simulations at $We = 80$ and 160 using solid dots, which extend to much earlier times. An excellent agreement between the predictions of \eqref{for:rim-dyn-model-mass}-\eqref{for:rim-dyn-model-momentum} and the two-dimensional results is observed. Three-dimensional measurements of $y_{\rm rim}$ and $b_{\rm rim}$ are found to saturate earlier and become smaller than their two-dimensional counterparts and theoretical predictions as time elapses. These observations can be explained by taking into account the formation of ligaments on the lamella rim. This mechanism arises due to highly nonlinear transverse perturbations imposed on the liquid rim, which is not present in two-dimensional simulations. At sufficiently large times, the growth of ligaments causes a substantial mass flux away from the liquid rim, which becomes more prominent at higher $We$ values as is shown in fig.~\ref{fig:snapshot-overview}. Other factors may also play a role, e.g., the initial overlapping between the two rims in three-dimensional simulations.


In figs.~\ref{fig:rim-pos-collapse-comp} and \ref{fig:rim-thick-collapse-comp} we present the evolution of $y_{\rm rim}$ and $b_{\rm rim}$ again, where time is now non-dimensionalised using the capillary timescale $\tau_{\rm cap}$. The growth of both $y_{\rm rim}$ and $b_{\rm rim}$ in three-dimensional simulations is consistent with a power law of $\sqrt{t/\tau_{\rm cap}}$ up to $t/\tau_{\rm cap} \approx 1.4$, after which deviation from this power law is observed. It is also found that prefactors of $\sqrt{We}$ (for $y_{\rm rim}$) and ${We}^{-1/4}$ (for $b_{\rm rim}$) can collapse the evolution data reasonably well, especially for $b_{\rm rim}$ as a single master curve is clearly observed in fig.~\ref{fig:rim-thick-collapse-comp}, leading to the following scaling arguments,
\begin{equation}
    \frac{y_{\rm rim}}{R_0} \propto \sqrt{We} \sqrt{\frac{t}{\tau_{\rm cap}}}, \quad \frac{b_{\rm rim}}{R_0} \propto {We}^{-1/4} \sqrt{\frac{t}{\tau_{\rm cap}}},
    \label{for:rim-pos-thick-scaling}
\end{equation}  
which we will use for further theoretical analysis of the ligament merging phenomenon in \S\ref{subsec:lig-merge}, while a complete determination for \eqref{for:rim-pos-thick-scaling} valid for very late time remains for future work. The scaling of $We$ in \eqref{for:rim-pos-thick-scaling} agree with the experimental results of \cite{Villermaux2011} and \cite{wang2018universal} for drops impacting a small surface; although the dependence on time is different, as their models aim at describing the entire sheet expansion-retraction motion. Nonetheless, similar empirical $\sqrt{t}$ scalings have been proposed by \cite{mongruel2009early}, \cite{thoroddsen2012micro} and \cite{visser2015dynamics} for quantifying the radial position of the expanding lamellae in the inertial regime, indicating that this simple form of time dependence can still describe the rim kinematics reasonably well between its formation and the onset of the retraction motion. Note that (4.9) implies the evolution of the liquid momentum carried by the rim $p_{\rm rim} \equiv \pi \rho_l D b_{\rm rim}^2 \Dot{y}_{\rm rim}$ is independent of the values of $We$, even though its average vertical velocity $\Dot{y}_{\rm rim}$ and volume $\Omega_{\rm rim} \equiv \pi b_{\rm rim}^2 D$ do depend on $We$. This contrasts with the axisymmetric results of \cite{wang2022mass} where the rim volume remains independent of $We$ due to their axisymmetric configuration.

\section{Transverse liquid ligaments}
\label{sec:lig-dynamics}

\subsection{Formation and growth}
\label{subsec:lig-formation}
\begin{figure}
	\centering
	\subfloat[]{
		\label{fig:snapshot-lig-gen-1}
		\includegraphics[width=.32\textwidth]{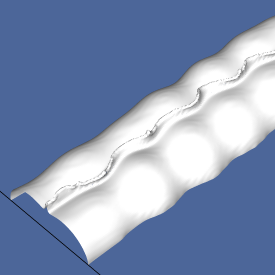}}
	\centering
	\subfloat[]{
		\label{fig:snapshot-lig-gen-2}
		\includegraphics[width=.32\textwidth]{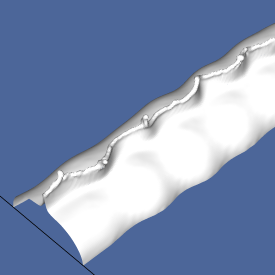}}
    \centering
	\subfloat[]{
		\label{fig:snapshot-lig-gen-3}
		\includegraphics[width=.32\textwidth]{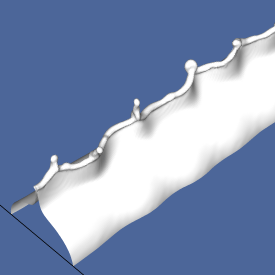}}
	\caption{Snapshots taken from a simulation case at $We = 200, \, \varepsilon = 0.06$ and $N_{\rm max} = 25$ showing ligaments generated from the `indentation' region between two colliding rims. From left to right: $t/\tau_{\rm cap} = 0.045, \, 0.091$ and 0.136.}
	\label{fig:snapshot-lig-gen}
\end{figure}

In the scenario considered by \cite{wang2021growth}, liquid ligaments grow slowly out of the corrugated bordering rim along its azimuthal direction, which they ascribed to a combination of local geometry, pulling effects of inertial force associated with rim deceleration and the global liquid-phase mass conservation. For this study, and given our perturbation profile, we find the transverse ligaments form very early for $We \geq 120$, nearly at the same time when the lamella is born out of the indentation region between the two cylinders, as presented in the simulation snapshots of fig.~\ref{fig:snapshot-lig-gen}. A closer look at fig.~\ref{fig:snapshot-lig-gen-2} reveals that the ligaments are produced preferentially from the concave regions along the two perturbed cylinders, suggesting that the ligament formation process is closely associated with the initial rim perturbation waveform. Indeed, a previous investigation by \cite{gordillo2020impulsive} suggests that the ligaments originate from the non-uniform initial distribution of normal interface velocity, which is in turn determined by the upstream liquid velocity and the curved initial cavity profile at the moment of impact. Note also that the nascent ligaments do not always project exactly vertically along the vertical $y$-axis; rather, they can display complex twisting and surface oscillation motions, and may grow in an oblique direction. At the same time, the liquid sheet beneath its bordering rim also features a wavy surface not perfectly aligned with the $xy$-plane. These are most likely due to the difference in the initial interface perturbations imposed on the two colliding rims, which gives rise to oscillatory motions under capillary effects.

\begin{figure}
	\centering
	\subfloat[]{
		\label{fig:lig-length-growth}
		\includegraphics[width=.48\textwidth]{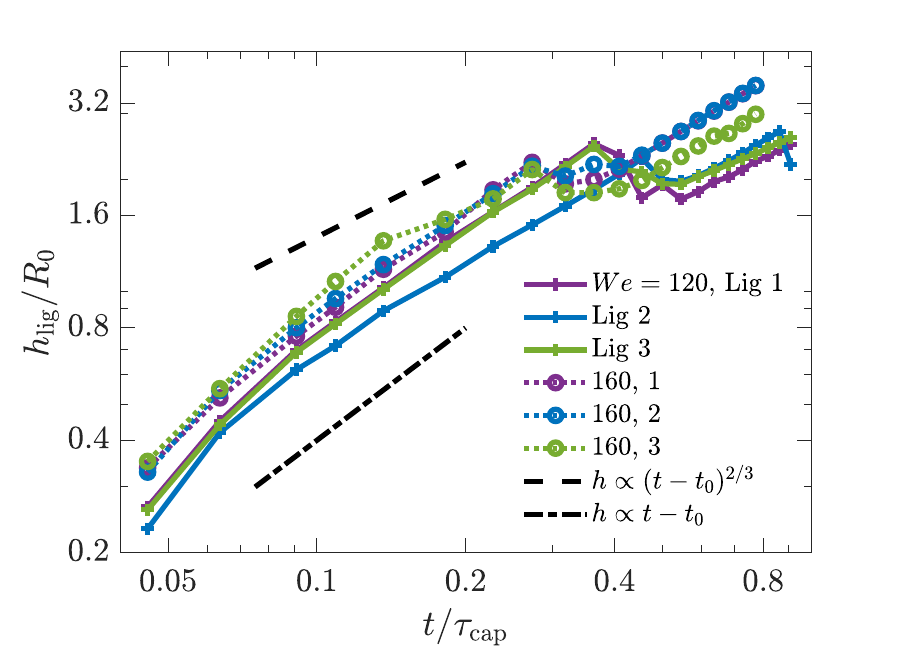}}
	\centering
	\subfloat[]{
		\label{fig:lig-vel-struct}
		\includegraphics[width=.48\textwidth]{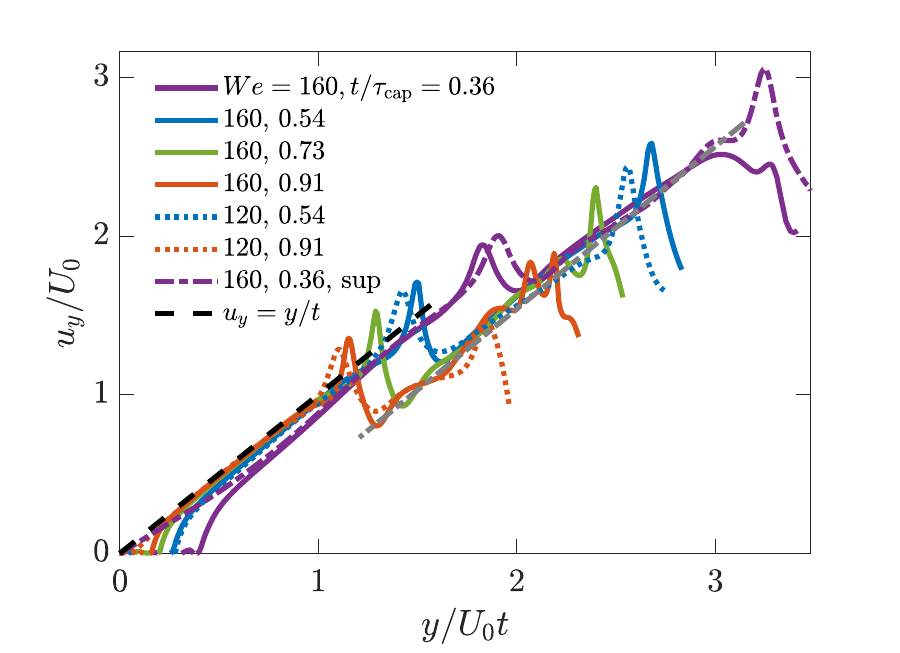}}
	\caption{(a): The evolution of liquid ligament length measured at different $We$ values, compared with the $t^{2/3}$ scaling law of \cite{lai2018bubble} and a linear growth model. (b): Vertical component of liquid velocity $u_y$ measured within liquid sheets and ligaments, showing the ballistic region within the ligament proposed by \cite{gekle2010generation}. `Sup' denotes that the initial rim perturbation is a superposition of sinusoidal signals with wavelengths $\lambda = D/8, \, D/16$ and $D/32$.}
	\label{fig:lig-struct-evol}
\end{figure}

We now move on to discuss the subsequent growth of the height of these ligaments after their formation, which is shown in fig.~\ref{fig:lig-length-growth} for rim collision at $We = 120$ and 160, where we present the height evolution of three individual ligaments at each $We$. The shedding of the first fine drop from these ligaments is characterised by a kink around $t = 0.4 \tau_{\rm cap}$. While this initial pinch-off may happen at even earlier times as $We$ increases, as shown in fig.~7 of \cite{wang2018unsteady}; this is still much later than the onset of the micro-splashing phenomena investigated by \cite{thoroddsen2012micro}, which happens at $t/\tau_{\rm cap} = O(10^{-4})$ (see e.g., their fig. 5b). Before this first pinch-off event, the height $h_{\rm lig}$ of different ligaments increases following a similar trend, and the growth rate at $We = 160$ is higher than that at $We = 120$. Given that the initial phase of growth of these ligaments is governed primarily by inertio-capillary effects, we compare it in fig.~\ref{fig:lig-length-growth} with the self-similar power law of $h \propto t^{2/3}$, proposed by \cite{lai2018bubble} in their investigation of inertio-capillary-dominated collapse of small surface bubbles. It is found that the height increase of the majority of transverse ligaments (except Ligament 1 at $We = 120$) agrees better with the linear growth model. This is most likely because the formation of fast jets observed by \cite{lai2018bubble} is preceded by focusing of interfacial capillary waves at the bottom of the bubble cavity while the liquid bulk remains quiescent; whereas here the bulk velocity plays a vital role in driving the closure of rim indentation, and may thus modify the rate of jet growth. In addition, a recent study by \cite{gordillo2023jets} suggests that the exponent of power laws dictating the evolution of jet radius and speed is dependent on the initial geometry of the collapsing air cavity, which may also account for the difference between our results and those of \cite{lai2018bubble} since the concave regions on our perturbed cylindrical surfaces do not feature a uniform radius of curvature.

To better understand the fluid motion within the growing ligaments, we measure the liquid-phase vertical velocity $u_y$ from the bottom of the expanding sheet up to the tip of a single ligament, and plot it in fig.~\ref{fig:lig-vel-struct} as a function of the vertical coordinate $y$. Two linear scaling regimes are observed; the first one is well described by $u_y = y/t$, which corresponds back to the velocity profile \eqref{for:sheet-vel-vertical} we established for the expanding sheet. As the fluid particles move higher up and away from the sheet, its velocity first increases and then abruptly decreases as it enters the neck and rim region respectively; a similar abrupt deceleration is also noted by \cite{wang2022mass} for drop collision problems. Interestingly, the combination of the neck and rim causes a constant decrease in the vertical velocity of approximately $0.7U_0$ for different $We$ values, times and perturbation waveforms. This pattern is not explicable from the scaling model \eqref{for:rim-pos-thick-scaling} since according to it, the fluid velocity should be equal to the average rim velocity $\Dot{y}_{\rm rim}$ after deceleration, which is always one half of the sheet velocity $y_{\rm rim}/t$ before deceleration. The implication is that the fluid velocity at the ligament root is always faster than the average rim velocity, which \cite{wang2021growth} ascribed to the additional acceleration due to the interface curvature at the rim-ligament junction. After this constant offset at the bordering rim which is most likely a viscous effect \citep{ghabache2014liquid}, $u_y$ grows linearly again with the vertical position $y$ with the same slope as the sheet region. This second linear scaling regime most likely corresponds to the ballistic region (although in the present study gravity is not included) in Worthington jets identified by \cite{gekle2010generation}, where the liquid particles travel at constant speed upwards before being slowed down once again at the bulb by capillary effects.

\begin{figure}
	\centering
	\subfloat[]{
		\label{fig:diss-contour-1}
		\includegraphics[width=.48\textwidth]{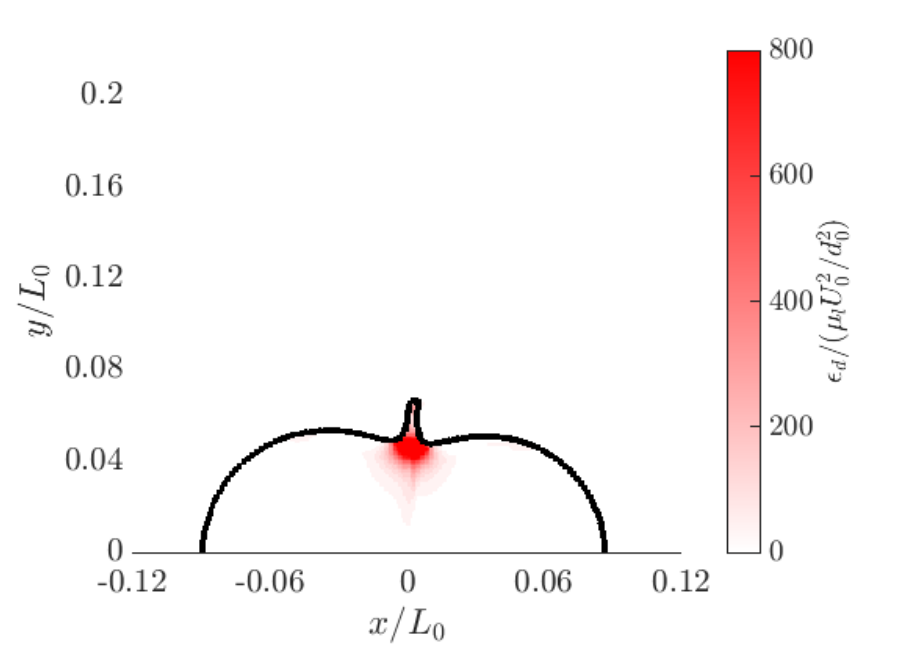}}
	\centering
	\subfloat[]{
		\label{fig:diss-contour-2}
		\includegraphics[width=.48\textwidth]{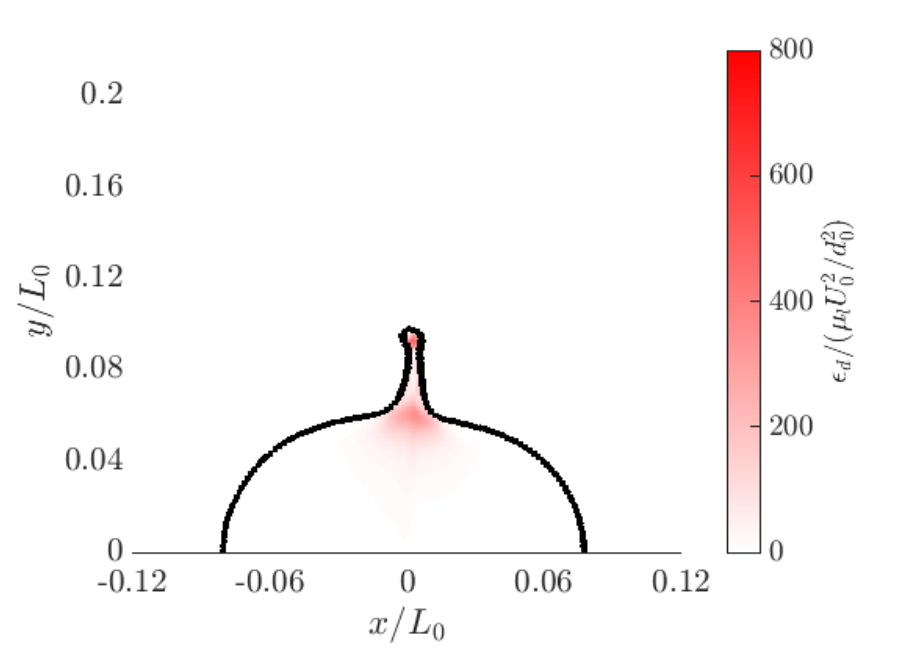}}
  
        \centering
	\subfloat[]{
		\label{fig:diss-contour-3}
		\includegraphics[width=.48\textwidth]{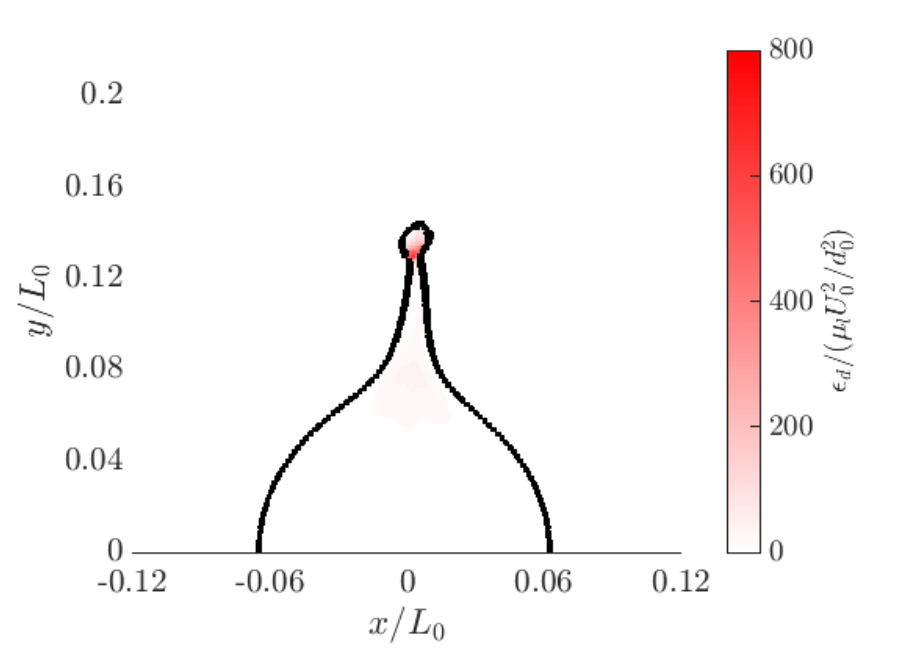}}
	\centering
	\subfloat[]{
		\label{fig:diss-contour-4}
		\includegraphics[width=.48\textwidth]{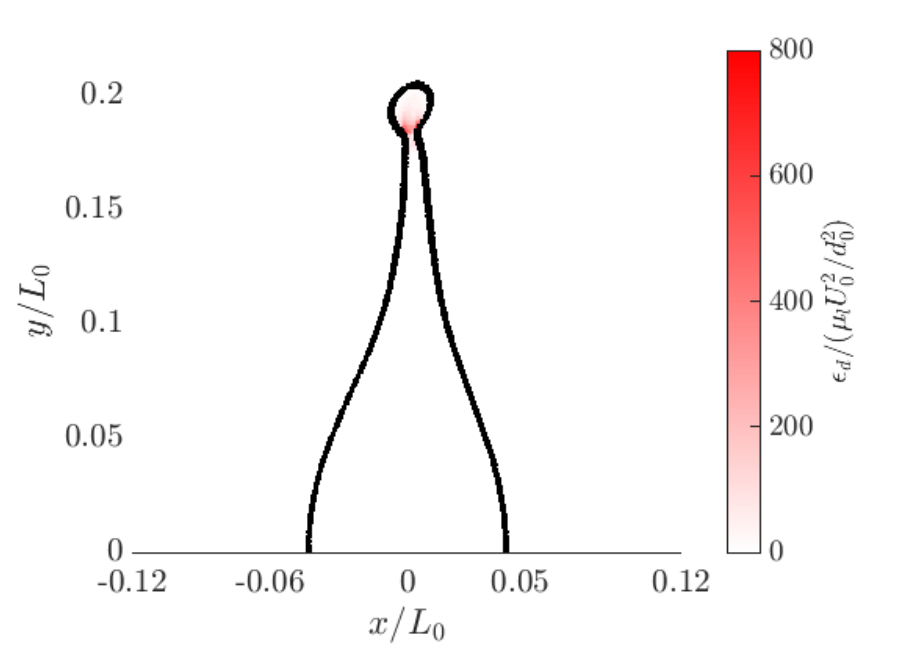}}
	\caption{Contour plots visualising the 2D distribution of instantaneous liquid-phase dissipation rate $\epsilon_d$ within the centre-plane $x/D = 0.5$ for $t/\tau_{\rm cap}=0.045$ (a), 0.091 (b), 0.182 (c) and 0.364 (d), where $We = 200$ and $\varepsilon_0 = 0.06$.}
	\label{fig:diss-contour}
\end{figure}

While \cite{wang2022mass} were able to demonstrate theoretically that energy dissipation at the bordering rim is responsible for the local rapid deceleration of fluid particles, which we also observed in fig.~\ref{fig:lig-vel-struct}, the detailed nonlinear dissipation mechanism is not captured by their one-dimensional model. As discussed in \S\ref{sec:introduction}, there is still a dissipation deficit of $15\%$ not covered by their calculations occurring at early-time. To help elucidate this discrepancy, we present contour plots in fig.~\ref{fig:diss-contour} for a simulation case at $We = 200$, showing the distribution of the liquid-phase viscous dissipation rate $\epsilon_d \equiv \mu_l (\partial u_i / \partial x_j) (\partial u_j / \partial x_i)$ within the centre-plane $x=0$ for $t/\tau_{\rm cap} \leq 0.36$, covering the early deformation period $t/\tau_{\rm cap} \leq 0.2$ where \cite{wang2022mass} observed the dissipation deficit (see their fig.~19). It is found that when the lamella is born at very early time and its bordering rim has not yet fully developed (fig.~\ref{fig:diss-contour-1}), there is an extremely high concentration of $\epsilon_d$ located at the lamella foot, agreeing with the simulation results of \cite{wildeman2016spreading} for drops impacting a smooth surface (see their fig.~4) and \cite{fudge2023drop} for drops impacting a liquid pool (see their fig.~8b). As time elapses, the bordering rim takes shape with its neck featuring relatively high concentration of $\epsilon_d$; whereas the dissipation at the lamella foot weakens and eventually becomes negligible by $t/\tau_{\rm cap} = 0.364$ (fig.~\ref{fig:diss-contour-4}), matching the saturation trend shown in fig.~19 of \cite{wang2022mass} for the dissipation deficit. This decay pattern of liquid-phase dissipation may be explained as follows. The liquid particles feeding the lamella at early time mostly comes from a very narrow boundary straddling the corners on either side of the lamella foot \citep{riboux2014experiments}; and they generate vorticity \citep{batchelor2000, li2018early} and experience capillary deceleration while traversing the highly-curved free surface, leading to very large values of viscous dissipation. Since this early-time deceleration is geometry-induced, \cite{fudge2023drop} further hypothesised that the magnitude of the velocity gradient remains largely unchanged at different flow configurations so that the early-time dissipation rate is proportional to the liquid viscosity $\mu_l$; although this is not directly verified in the present work. As the liquid sheet extends further upwards, the interface curvature at the lamella foot decreases and capillary deceleration becomes much weaker, hence the decrease in the dissipation rate $\epsilon_d$. Eventually, as the lamella foot is no longer discernable from the flattened liquid bulk and the capillary effects become negligible, the inviscid sheet velocity profile \eqref{for:sheet-vel-vertical} will be established.

These observations of the liquid-phase dissipation evolution therefore suggest that the unidentified dissipation deficit found by \cite{wang2022mass} is most likely linked with the early-time lamella formation which also features strong viscous effects; and according to \cite{wildeman2016spreading} and \cite{naraigh2023analysis}, this kind of dissipation is a type of `general head loss' imposed by the deformation mode of the impacting object and independent of the detailed impact parameters. The head loss in impact problems dissipates a fixed fraction of the initial kinetic energy via recirculating flows \citep{wildeman2016spreading, naraigh2023analysis}, which matches the observations of \cite{wang2022mass}. It is also noted that besides the early-time lamella foot and the late-time rim neck, the flow field elsewhere within the coalesced liquid bulk shown in \ref{fig:diss-contour} is nearly inviscid, supporting our derivation of the centerline velocity profile within the liquid sheet \eqref{for:sheet-vel-vertical} based on inviscid flow assumptions. While we have not fully explored the influence of $Oh$ on the lamella expansion process, it can be expected that in the inviscid limit where $Oh \ll 1$, its influence will be confined to the narrow `viscous' regions identified in this section, and thus will not significantly affect the deformation of the droplet bulk, or the ligament dynamics and fragmentation mechanisms to be discussed below.

\subsection{Ligament merging phenomenon}
\label{subsec:lig-merge}

\begin{figure}
	\centering
	\subfloat[]{
		\label{fig:snapshot-lig-death-1}
		\includegraphics[width=.32\textwidth]{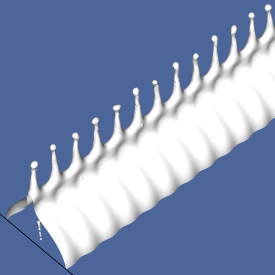}}
	\centering
	\subfloat[]{
		\label{fig:snapshot-lig-death-2}
		\includegraphics[width=.32\textwidth]{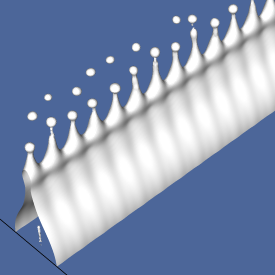}}
    \centering
	\subfloat[]{
		\label{fig:snapshot-lig-death-3}
		\includegraphics[width=.32\textwidth]{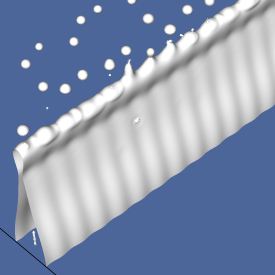}}

    \centering
	\subfloat[]{
		\label{fig:snapshot-lig-death-comp-1}
		\includegraphics[width=.32\textwidth]{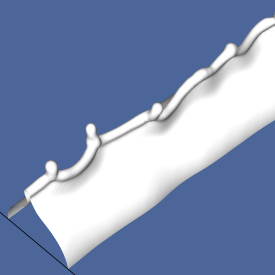}}
	\centering
	\subfloat[]{
		\label{fig:snapshot-lig-death-comp-2}
		\includegraphics[width=.32\textwidth]{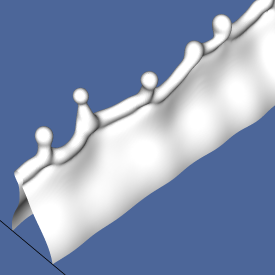}}
    \centering
	\subfloat[]{
		\label{fig:snapshot-lig-death-comp-3}
		\includegraphics[width=.32\textwidth]{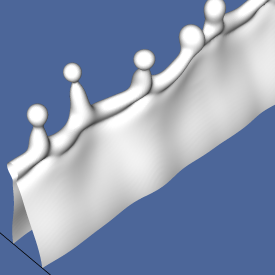}}
	\caption{Snapshots at $We = 120, \, \varepsilon = 0.06$ and $N_{\rm max} = 25$ showing ligament evolution from monochromatic initial perturbations (upper row) and filtered white-noise perturbations (lower row). Re-absorption of ligaments back into the rim is observed for the monochromatic perturbation case after two cycles of drop shedding. From left to right: $t/\tau_{\rm cap} = 0.2, \, 0.4$ and 0.6.}
	\label{fig:snapshot-lig-death}
\end{figure}

As is noted in \S\ref{sec:overview}, the number of transverse ligaments on the rim will decrease over time as they merge with their neighbours. This merging process is observed only when the initial perturbation is not monochromatic such that the nascent ligaments are not equidistantly spaced, and it turns out crucial in maintaining the growth of ligaments and the continuation of fragment shedding. Figs.~\ref{fig:snapshot-lig-death-1}-\ref{fig:snapshot-lig-death-3} show the development of ligaments formed out of monochromatic perturbation waveforms, and it is observed that after two rounds of pinching-off events at their tips, the ligaments can no longer sustain their own growth and are re-absorbed back into the underlying liquid rim; whereas ligaments formed out of filtered white noise perturbations at the same values of $We, \, \varepsilon$ and $N_{\rm max}$ merge with their neighbours and survive end-pinching, continuing to grow and shed fragments until the end of simulation, as observed in figs.~\ref{fig:snapshot-lig-death-comp-1}-\ref{fig:snapshot-lig-death-comp-3}.

\begin{figure}
	\centering
	\subfloat[]{
		\label{fig:snapshot-lig-merge-1}
		\includegraphics[width=.32\textwidth]{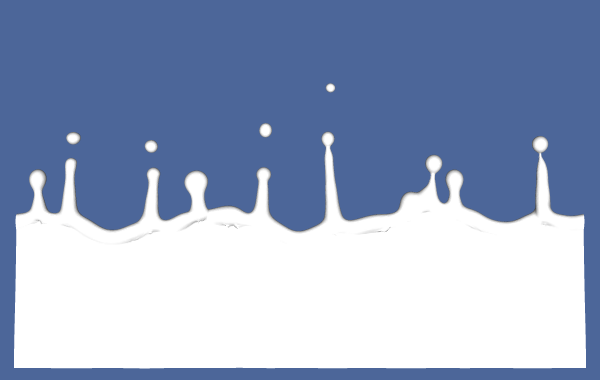}}
	\centering
	\subfloat[]{
		\label{fig:snapshot-lig-merge-2}
		\includegraphics[width=.32\textwidth]{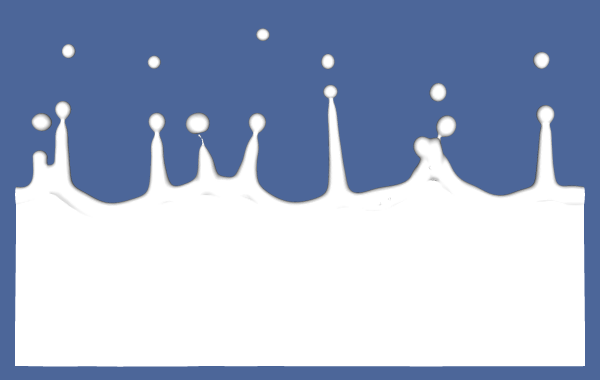}}
    \centering
	\subfloat[]{
		\label{fig:snapshot-lig-merge-3}
		\includegraphics[width=.32\textwidth]{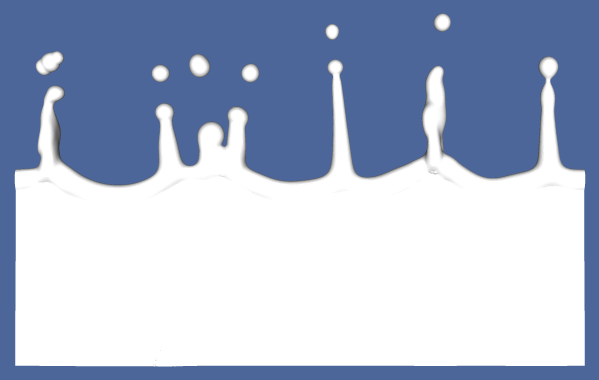}}
	\caption{Snapshots taken from a simulation case at $We = 160, \, \varepsilon = 0.04$ and $N_{\rm max} = 25$ showing ligaments merging on the corrugated rim bordering the expanding sheet, while shedding fragments via the end-pinching mechanism.}
	\label{fig:snapshot-lig-merge}
\end{figure}

The ligament merging process is shown in more details in the snapshots of fig.~\ref{fig:snapshot-lig-merge}, where the merging ligaments are observed to be located on rim `cusp' structures extruding from the liquid sheet beneath, an indication of the non-uniform incoming mass distribution along the bordering rim \citep{gordillo2014cusps, wang2018unsteady}. When ligament merging occurs, the roots of two neighbouring ligaments approach each other; liquid is then drawn upwards from the underlying cusp in between the two ligaments, causing them to coalesce into a thicker and more corrugated ligament; while the length of the fused ligament does not differ significantly from those of its parents. Ligament merging is thus capable of delaying the next end-pinching event since the fused ligament takes longer to be stretched, thus allowing incoming fluid from the rim to sustain their growth and evade re-absorption into the rim when the depleted ligament becomes too short, as observed in fig.~\ref{fig:snapshot-lig-death}. Note that the monotonically decreasing trend of ligament numbers observed by us and \cite{wang2021growth} differs from the early experimental work of \cite{thoroddsen1998evolution} on drop impact, where they also observed splitting of liquid fingers aside from their merging behaviour, so that the total finger number remains approximately constant.

To the knowledge of the authors, a scaling law for the ligament numbers accounting for their merging dynamics is not yet available. The early work of \cite{marmanis1996scaling} found that the number of liquid fingers at the maximum spread radius for high-speed drop impacts scales with $Re^{3/4}$ ($Re$ being the impact Reynolds number), without accounting for their evolution over time; whereas the recent $We^{3/8}$ scaling model proposed by \cite{wang2021growth} is based on the understanding that the ligaments are formed from a subset of rim corrugations arising from a combined Rayleigh-Taylor (RT) and Rayleigh-Plateau (RP) instability; a physical mechanism which is most likely not yet active within our parameter space as we find the formation of ligaments more closely linked with the initial interface perturbation geometry. The early theoretical analysis of \cite{yarin1995impact} predicted that perturbed free rims will spontaneously develop `cusp' structures due to nonlinearity, where two neighbouring rim sections impinge and give rise to free jets. This physical picture closely resembles our present observations, although \cite{yarin1995impact} did not proceed to develop scaling models for the splashing fragments. We therefore seek to derive a new ligament number scaling model accounting for the merging dynamics in this section, and compare it with the simulation results.

\begin{figure}
    \centering
	\subfloat[]{
		\label{fig:frag-lig-ratio}
		\includegraphics[width=.48\textwidth]{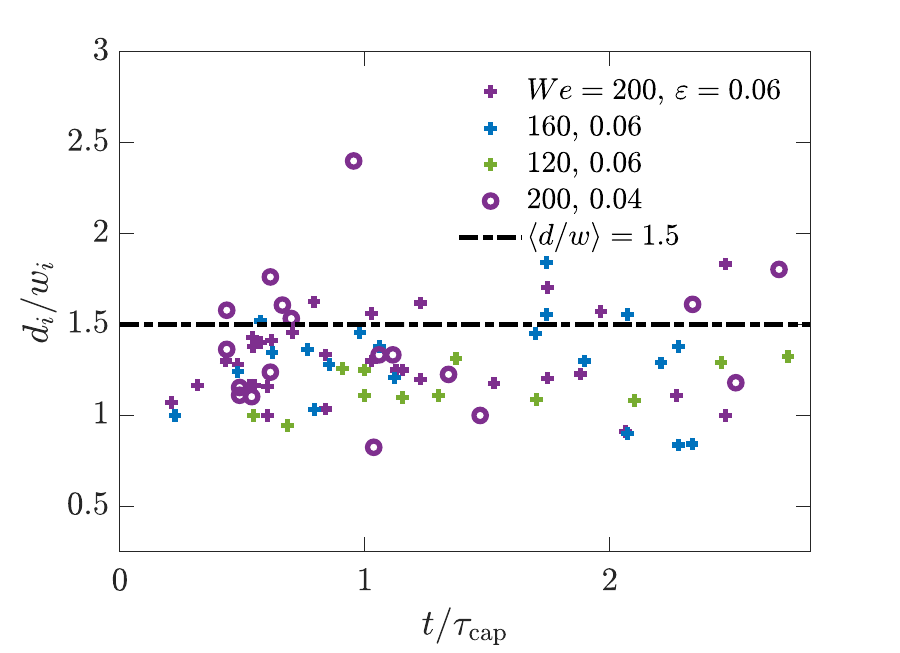}}
	\centering
	\subfloat[]{
		\label{fig:d-frag-evol}
		\includegraphics[width=.48\textwidth]{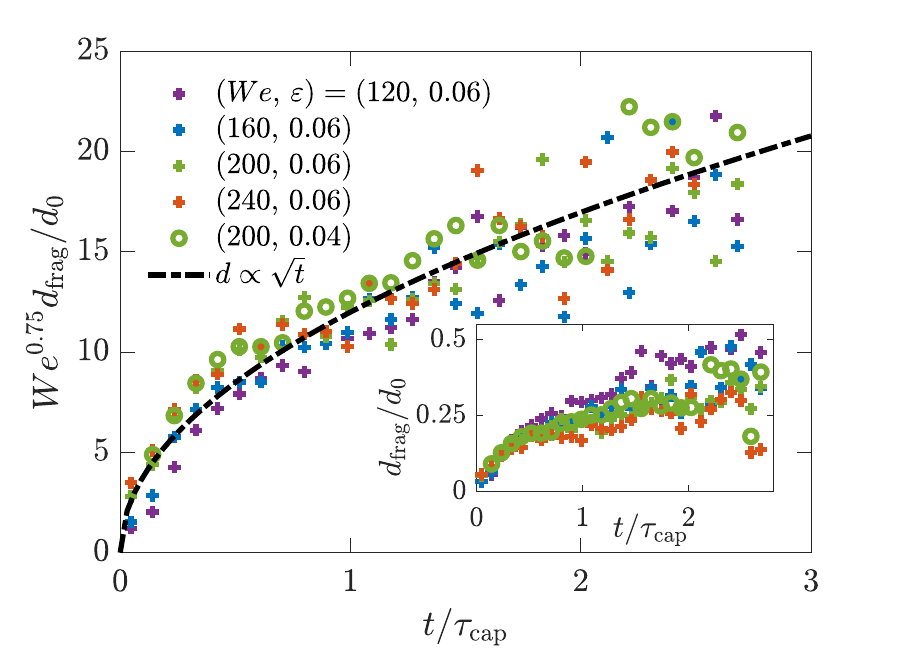}}
	\caption{(a): Measurement of the ratio between the diameter $d_i$ of the detaching fragment and the width $w_i$ of its originating ligament at different ejection times. (b): The evolution of the fragment diameter $d_{\rm frag}$ of ejected fragments. The results in (b) have been ensemble-averaged across three realisations for each $(We, \, \varepsilon)$ configuration, and rescaled by $We^{-0.75}$ in the main plot.}
	\label{fig:nd-frag-evol}
\end{figure}

As the first step towards quantifying ligament dynamics, we seek to determine the evolution of the average ligament width $w_{\rm lig}$ over time by inspecting the evolution of fragment diameter $d_{\rm frag}$, which can be easily reconstructed using the droplet-tracking algorithm of \cite{chan2021identifying}. The connection between these two quantities is established in fig.~\ref{fig:frag-lig-ratio}, where we plot the ratio between $d_{\rm frag}$ and $w_{\rm lig}$ at the instant of pinch-off. The ligament diameters are measured at the cross section corresponding to one half of the total ligament length. Most of the measured data are found to scatter between 1 and 2, centred around 1.4 which is close to the average value of 1.5 as found by \cite{wang2018unsteady}. These are below the theoretical value of 1.89 as predicted by the RP instability \citep{gordillo2010generation}, indicating that end-pinching is indeed the dominant fragment production mechanism for the present study, and that the diameter of fragments $d_{\rm frag}$ remains in proportion to the width of their parent ligaments $w_{\rm lig}$.

The inset of fig.~\ref{fig:d-frag-evol} shows the diameters of the fragments $d_{\rm frag}$ versus their time of formation, where the data for each configuration of $(We, \, \varepsilon)$ have been averaged across three individual realisations to reduce the range of scatter. It is found that larger fragments are generally produced at later times and smaller $We$ values. The scatter in data increases over time, which is most likely due to the increase in the diameter difference and the surface corrugation of remaining ligaments as they merge with one other. The main plot suggests that the evolution of individual fragment diameter in fig.~\ref{fig:d-frag-evol} roughly scales with $We^{-3/4} \sqrt{t/\tau_{\rm cap}}$, while noting that our fitted prefactor ${We}^{-3/4}$ is not conclusive and remains to be further validated at higher $We$ values. Since the fragment size remains in proportional to the width of their parent ligament according to fig.~\ref{fig:frag-lig-ratio}, it can be inferred that
\begin{equation}
    w_{\rm lig} \propto d_{\rm frag} \propto \sqrt{\frac{t}{\tau_{\rm cap}}}.
    \label{for:w-lig-scaling}
\end{equation}

\begin{figure}
	\centering
	\includegraphics[width=.98\textwidth]{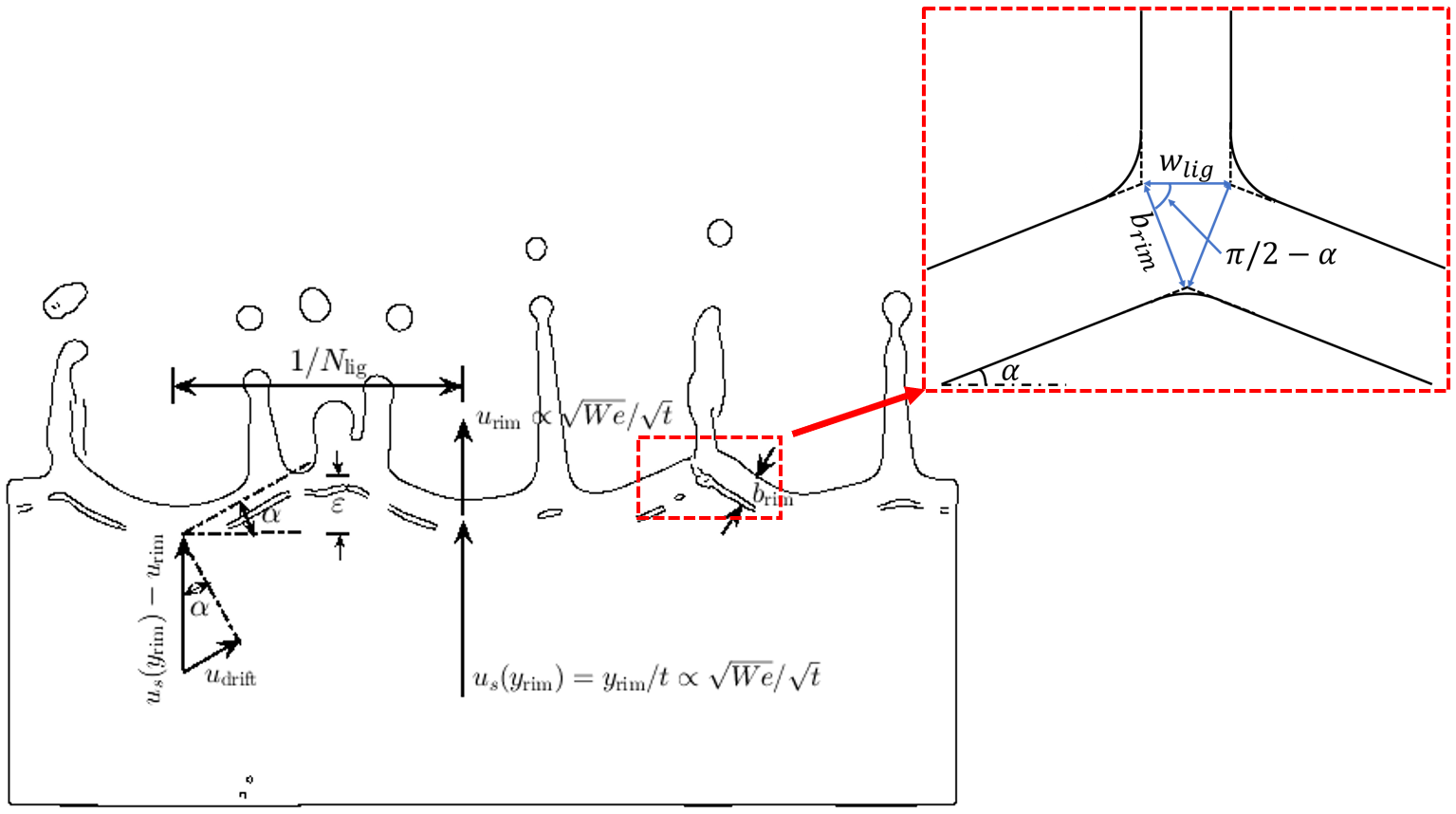}  
	\caption{Main plot: Sketch showing the quantities defined in \S\ref{subsec:lig-merge} for developing the ligament merging model \eqref{for:lig-num-scaling}. Inset: Sketch showing the local geometry of the junction region at the ligament base.}
	\label{fig:lig-merge-sketch}
\end{figure}

\cite{gordillo2014cusps} and \cite{wang2018unsteady} proposed the following drift velocity of ligaments $u_{\rm drift}$ on top of a liquid cusp to characterise their migration,
\begin{equation}
    u_{\rm drift} = [u_s(y_{\rm rim}) - u_{\rm rim}] \sin \alpha,
    \label{for:drift-vel}
\end{equation}
where $u_s(y_{\rm rim}) = y_{\rm rim}/t$ is the velocity at which liquids enters the rim from the expanding liquid sheet, calculated from the self-similar expanding sheet velocity profile \eqref{for:sheet-vel-vertical}; and $u_{\rm rim}$ is the vertical rim velocity determined by differentiating the scaling law for $y_{\rm rim}$ in \eqref{for:rim-pos-thick-scaling}. It is noted that both $u_s(y_{\rm rim})$ and $u_{\rm rim}$ scale as $\sqrt{We} \cdot t^{-1/2}$, and $\alpha$ is the angle between the local rim and the horizontal plane as shown in fig.~\ref{fig:lig-merge-sketch}. Overall, \eqref{for:drift-vel} suggests that the migration of ligaments is driven by the tangential projection of the net incoming fluid velocity along the corrugated rim.

As $u_{\rm drift}$ drives the ligament migration and causes the ligament number density $N_{\rm lig}$ to decrease, the average transverse ligament spacing $1/N_{\rm lig}$ on the rim consequently increases. Therefore,
\begin{equation}
    u_{\rm drift} \propto \frac{d}{dt} \left(\frac{1}{N_{\rm lig}} \right).
    \label{for:u-drift-n-lig}
\end{equation}

The averaged value of $\sin \alpha$ can be determined by inspecting the geometry of the junction region between the ejected ligaments and the lamella rim, which is shown in the inset of fig.~\ref{fig:lig-merge-sketch}. Making use of the law of cosines:
\begin{equation}
    \frac{w_{\rm lig}}{2b_{\rm rim}} = \cos \left( \frac{\pi}{2} - \alpha \right) = \sin \alpha.
\end{equation}
This combined with Eqs.~\eqref{for:w-lig-scaling} and \eqref{for:rim-pos-thick-scaling} yields
\begin{equation}
    \sin \alpha = C(We),
    \label{for:alpha-prediction}
\end{equation}
suggesting that the rim slope $\alpha$ remains unchanged over time and depends only upon the impact Weber number. Taking into account that $\tan \alpha \approx \varepsilon_{\rm rim} N_{\rm lig}$, $\alpha$ remaining constant also indicates that as the average ligament spacing $1/N_{\rm lig}$ becomes larger over time owing to the ongoing merging of ligaments, the rim corrugation $\varepsilon_{\rm rim}$ also increases proportionally to maintain a constant local slope.

By further incorporating \eqref{for:u-drift-n-lig} and \eqref{for:drift-vel}, the following model predicting the evolution of the ligament number density $N_{\rm lig}$ can be derived,
\begin{equation}
    N_{\rm lig} \propto {\left(\frac{t}{\tau_{\rm cap}} \right)}^{-1/2}.
    \label{for:lig-num-scaling}
\end{equation}
\begin{figure}
	\centering
	\subfloat[]{
		\label{fig:n_lig_N_max_sweep}
		\includegraphics[width=.48\textwidth]{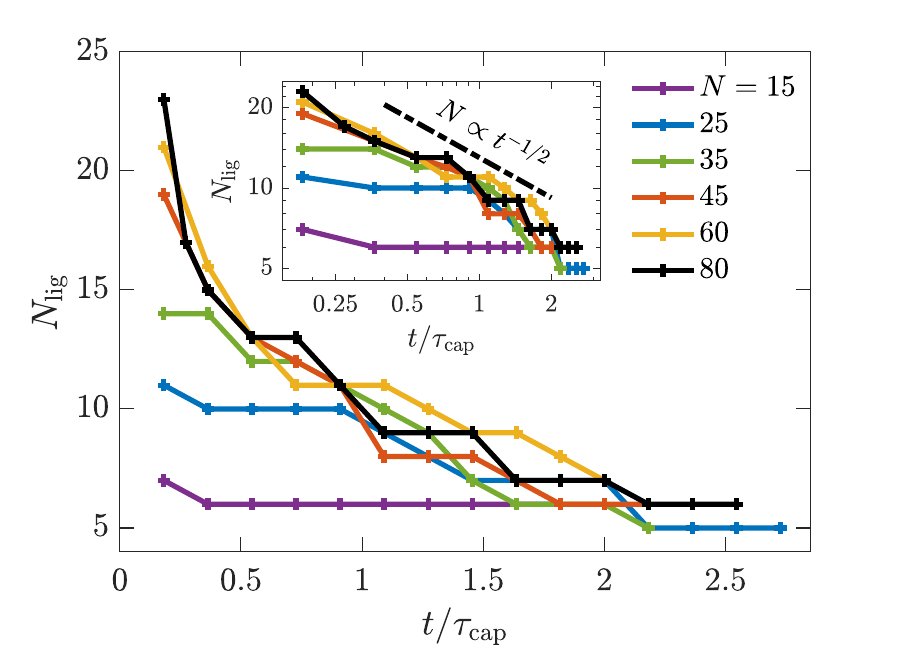}}
	\centering
	\subfloat[]{
		\label{fig:n_lig_We_sweep}
		\includegraphics[width=.48\textwidth]{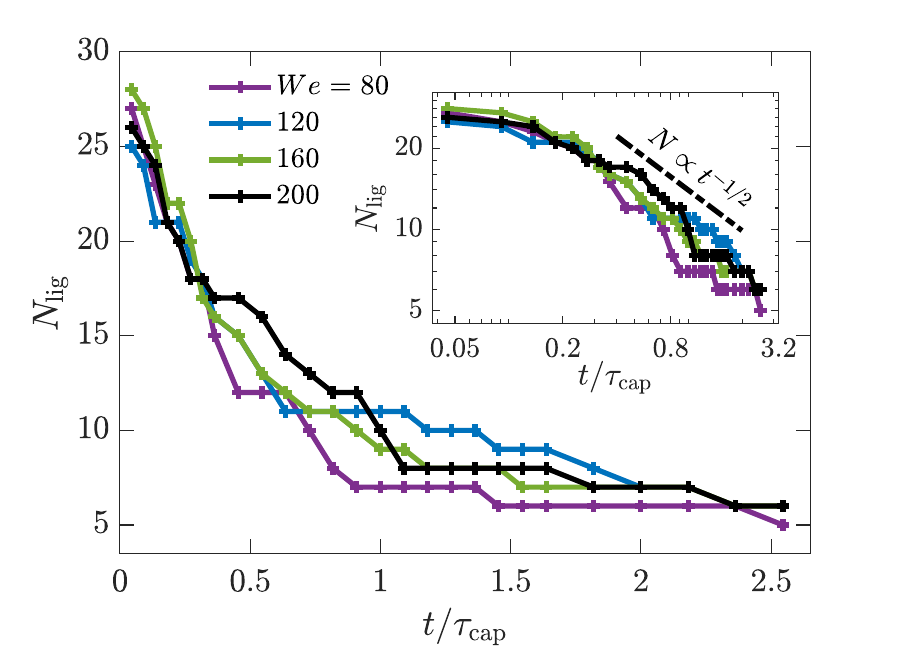}}
	\caption{Evolution of the ligament number density $N_{\rm lig}$ at different values of $N_{\rm max}$ with $We = 120$ (a) and different $We$ with $N_{\rm max} = 60$ (b). The insets compare the evolution of $N_{\rm lig}$ with our model \eqref{for:lig-num-scaling}.}
	\label{fig:n-lig-evol}
\end{figure}

In the main plots of fig.~\ref{fig:n-lig-evol} we show the decay of the ligament number density $N_{\rm lig}$ at different values of $N_{\rm max}$ and $We$. Fig.~\ref{fig:n_lig_N_max_sweep} shows that while increasing $N_{\rm max}$ leads to the formation of more ligaments at early time, $N_{\rm lig}$ appears to reach saturation and does not increase proportionally with $N_{\rm max}$ when $t/\tau_{\rm cap} = 0.18$; and it is likely that viscous damping effects are particularly strong when the lamella is ejected at the early impact stage, which may smooth out short-wavelength components in the initial perturbation spectra (see our fig.~\ref{fig:diss-contour} and relevant discussions). Larger $N_{\rm max}$ also leads to faster decay of $N_{\rm lig}$ as expected, since the average distance between neighbouring ligaments becomes smaller; and for all values of $N_{\rm max}$, fig.~\ref{fig:n_lig_N_max_sweep} suggests that the evolution of $N_{\rm lig}$ becomes largely similar for $t/\tau_{\rm cap} > 1.5$, where the remaining ligaments take much longer to migrate and merge. Fig.~\ref{fig:n_lig_We_sweep} shows that the decay of $N_{\rm lig}$ does not appear to depend strongly on $We$ and thus lends some support to the prediction of \eqref{for:lig-num-scaling} that it is $We$-independent, although ensemble averaging would be needed to ascertain this due to the randomness in the initial perturbation waveform. The insets of fig.~\ref{fig:n-lig-evol} show the evolution of $N_{\rm lig}$ again in log-log axes, which is also compared with model \eqref{for:lig-num-scaling}. The inset of fig.~\ref{fig:n_lig_N_max_sweep} indicates that the measured results collapse well and agree with \eqref{for:lig-num-scaling} for $N_{\rm max} \geq 35$ throughout the period of measurement, whereas at $N_{\rm max} = 25$ the decay of $N_{\rm lig}$ is initially slower, and only matches the prediction of $t^{-1/2}$ for $t/\tau_{\rm cap} \geq 1$. The measurement of $N_{\rm lig}$ extends to earlier times in fig.~\ref{fig:n_lig_We_sweep}, and its inset suggests that the evolution of $N_{\rm lig}$ at different $We$ is initially slower and matches model \eqref{for:lig-num-scaling} only after $t/\tau_{\rm cap} \geq 0.2$. This slower decay at early times observed in both insets arises most likely because the rim slope $\alpha$ takes a finite period of time to develop before reaching the steady-state value given by \eqref{for:alpha-prediction}. Overall, these results indicate that model \eqref{for:lig-num-scaling} offers a good working description of the ligament merging phenomenon occurring within our parameter space. However, the approximations we make for its derivation restricts its application to the scenario when $N_{\rm lig}$ is large and the slope $\alpha$ of the corrugated rim has reached the steady-state value predicted by \eqref{for:alpha-prediction}.

\section{Droplet generation and characteristics}
\label{sec:frag-stats}

Since the expanding sheet remains intact during our simulation period, fragments are formed solely through the breakup of liquid ligaments originating from the bordering rim. The majority of fragments are produced via the end-pinching mechanism, a process that coincides with the ligament merging phenomena and is already visible in the snapshots presented in figs.~\ref{fig:snapshot-overview} and \ref{fig:snapshot-lig-merge}. Namely, the ligament tips are decelerated by capillary force and produce surface corrugations, which in turn creates a local pressure gradient within the ligament neck that drains the liquid towards the tip and triggers pinch-off. After this, the enlarged ligament tip detaches as a primary drop \citep{gordillo2010generation}, whose diameter is proportional to the parent ligament width as shown in fig.~\ref{fig:frag-lig-ratio}. Occasionally, satellite drops are formed from the small amount of remnant liquid within the neck before it can be fully reabsorbed into the ligament after end-pinching, as also shown in fig.~5b of \cite{wang2018unsteady}. In contrast with the primary drops, the production of these satellite drops is sensitive to initial liquid-phase velocity perturbations, thus introducing randomness to the jet breakup process. However, they do not affect the size and velocity of the subsequent primary drops ejected, as shown by \cite{berny2022size} in the instance of jet-droplet production by bubble-bursting. Possibly due to the difference in their generation mechanisms, our ligaments can grow much longer than those \cite{wang2018unsteady} observed in their experiments; and at late times some particularly long ligaments may break up due to the RP instability and shed multiple primary fragments at a time, as can be seen in the rightmost ligament in fig.~\ref{fig:snapshot-We-200-0.06-2} and fig.~2f of \cite{liu2022speed}. This appears inconsistent with the conclusion of \cite{wang2021growth} that a ligament can only pinch off to produce one primary drop at a time under the chaotic dripping regime.

\begin{figure}
	\centering
	\subfloat[]{
		\label{fig:n-frag-evol}
		\includegraphics[width=.48\textwidth]{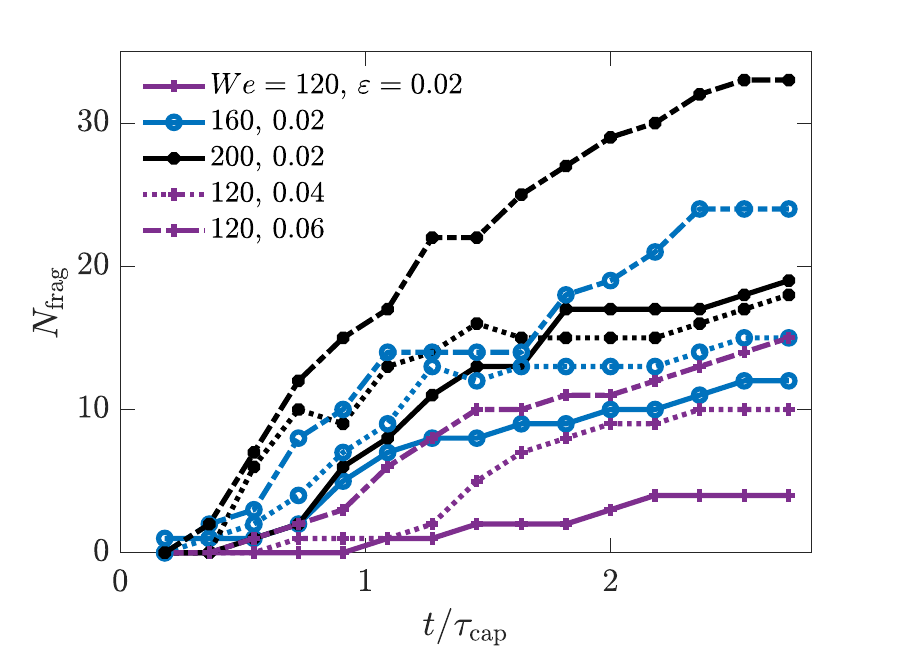}}
	\centering
	\subfloat[]{
		\label{fig:u-t-rel}
		\includegraphics[width=.48\textwidth]{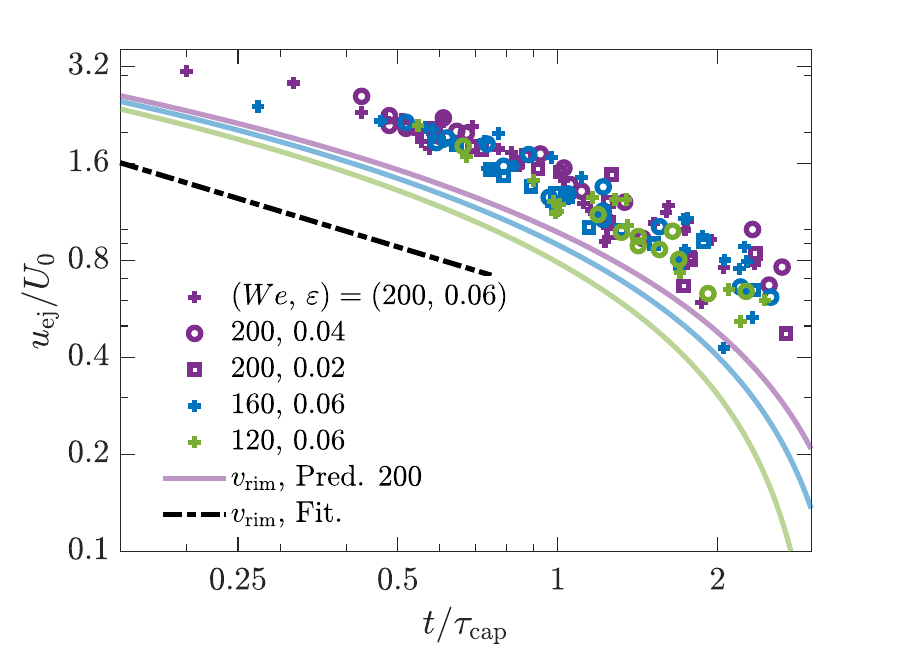}}
	\caption{The evolution of the total number density $N_{\rm frag}$ (a) and the ejection velocity (b) of primary fragments, compared with the rim velocity $u_{\rm rim}$ derived from solving Eqs.~\eqref{for:rim-dyn-model-mass}-~\eqref{for:rim-dyn-model-momentum} (solid transparent lines) and \eqref{for:rim-pos-thick-scaling} (dash-dotted line).}
	\label{fig:frag-sheet-rel}
\end{figure}

We first analyse the time evolution of various fragment properties during the rim collision process. Fig.~\ref{fig:n-frag-evol} shows the evolution of the total number density of primary fragments $N_{\rm frag}$, where it is observed to generally increase over time towards saturation, despite infrequent decreases due to coalescence of primary fragments with another fragment or a neighbouring ligament. It is also noted that $N_{\rm frag}$ increases with both $We$ and $\varepsilon$, as larger $We$ and $\varepsilon$ encourages the growth and subsequent pinch-off of liquid ligaments. 

The decrease of drop production rate $\Delta N_{\rm frag}/\Delta t$ can be explained as follows. The ongoing ligament merging phenomenon causes the ligament number density to decrease, hence fewer fragments can detach at the same time. In the meanwhile, merged ligaments become more corrugated and thicker, thus end-pinching events occur at larger ligament widths as time elapses. Consequently, the time interval between successive end-pinching events also becomes longer, since the necking timescale
\begin{equation}
    t_{\rm neck} \equiv 1.13 \sqrt{\frac{\rho_l w_{\rm lig}^3}{\sigma}}
    \label{for:necking-time}
\end{equation}
increases with $w_i$ according to the experimental results of \cite{wang2018unsteady}. 

Fig.~\ref{fig:u-t-rel} shows the evolution of the fragment ejection speed $u_{\rm ej}$, the speed of the drop at the moment it detaches from its parent ligament. The fastest drops found in our simulations feature $u_{\rm ej}$ slightly over $3U_0$, which is comparable to the fragment ejection velocity in the prompt splashing phenomena \citep{burzynski2020splashing}. Regardless of the detailed geometrical features of the parent ligaments, when scaled with the initial collision speed $U_0$, the decay of $u_{\rm ej}$ over time does not significantly depend on either $We$ or $\varepsilon_0$. We further compare the measured $u_{\rm ej}$ values with the predictions of both the rim dynamic equations \eqref{for:rim-dyn-model-mass}-\eqref{for:rim-dyn-model-momentum} and the scaling model \eqref{for:rim-pos-thick-scaling}. Both capture the early-time evolution of $u_{\rm ej}$ up to $t/\tau_{\rm cap} \approx 0.5$, after which the decrease of $u_{\rm ej}$ slightly steepens and shows a better agreement with the predictions of \eqref{for:rim-dyn-model-mass}-\eqref{for:rim-dyn-model-momentum}. This is most likely due to the late-time capillary deceleration not well-represented by \eqref{for:rim-pos-thick-scaling}. We note that while the rim velocity predicted by \eqref{for:rim-dyn-model-mass}-\eqref{for:rim-dyn-model-momentum} is closer to $u_{\rm ej}$, they tend to over-predict the three-dimensional simulation results, as already observed in fig.~\ref{fig:rim-geometry-evol}; suggesting the existence of a gap between the actual rim velocity and the fragment ejection velocity. This agrees with the earlier results of \cite{liu2022speed}, where their fig.~5 also shows fragment speeds remaining higher than the rim velocity. Similar measurements of the fragment ejection velocity have also been reported by \cite{thoroddsen2012micro} for micro-splashing in drop impact problems at much larger values of $We$, which were well explained by the theory of \cite{riboux2015diameters}. However, in \cite{riboux2015diameters} the lamella rim fragments under Rayleigh-Taylor and capillary instabilities, and therefore the velocity and size of ejected droplets are directly determined by the rim. In \cite{thoroddsen2012micro} fragmentation also occurs shortly after the emergence of the lamella sheet, based on which \cite{riboux2015diameters} modelled the droplet ejection velocity using flow field information at the lamella foot. These differ from our scenario where the ejection of droplets are governed by the end-pinching of ligaments erupting on the lamella rim \citep{wang2018unsteady}. Development of theoretical models capable of predicting fragment ejection speed in our case thus requires detailed knowledge of ligament growth and merging dynamics, which is out of the scope of the current work.

\begin{figure}
	\centering
	\subfloat[]{
		\label{fig:drop-pdf-evol}
		\includegraphics[width=.48\textwidth]{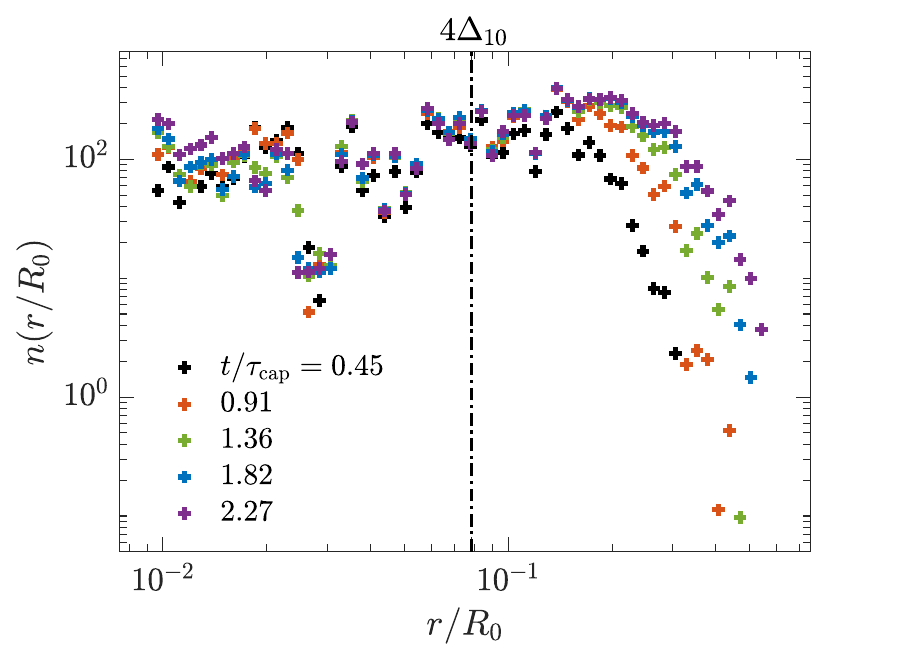}}
	\centering
	\subfloat[]{
		\label{fig:drop-pdf-neel-comp}
		\includegraphics[width=.48\textwidth]{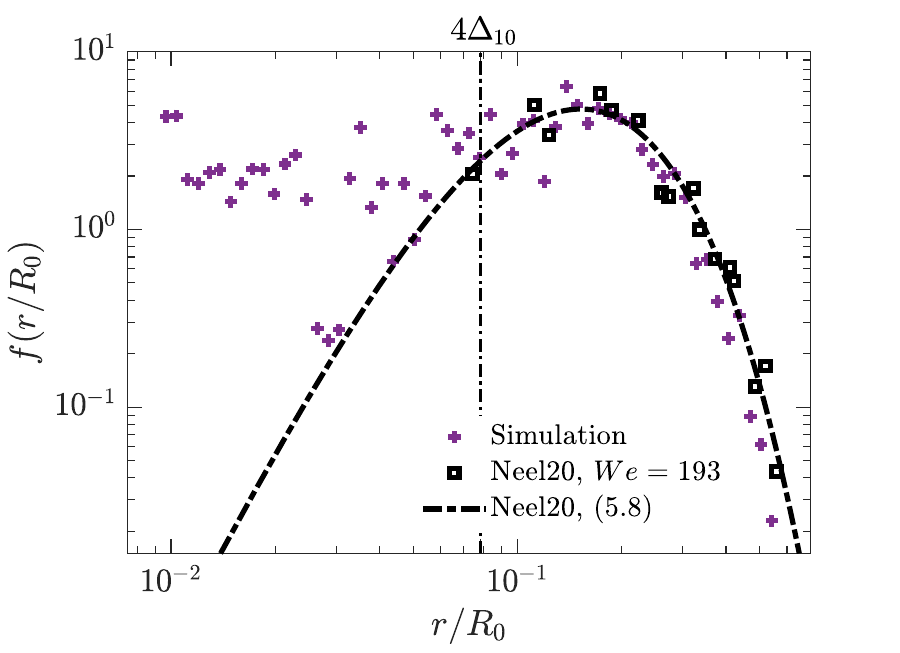}}

        \centering
	\subfloat[]{
		\label{fig:drop-pdf-We-sweep}
		\includegraphics[width=.48\textwidth]{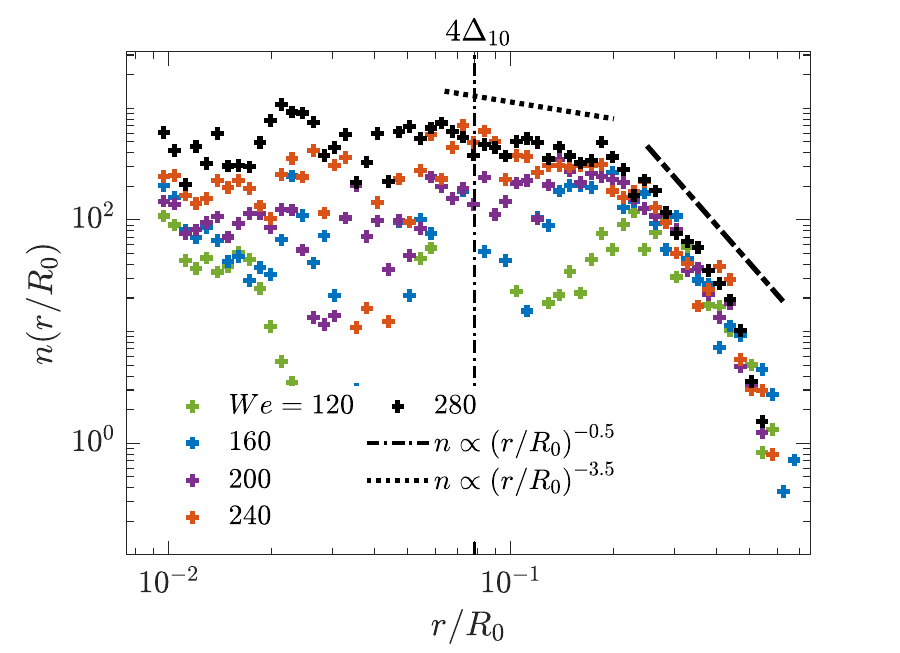}}
	\centering
	\subfloat[]{
		\label{fig:drop-pdf-epsilon-nmax-sweep}
		\includegraphics[width=.48\textwidth]{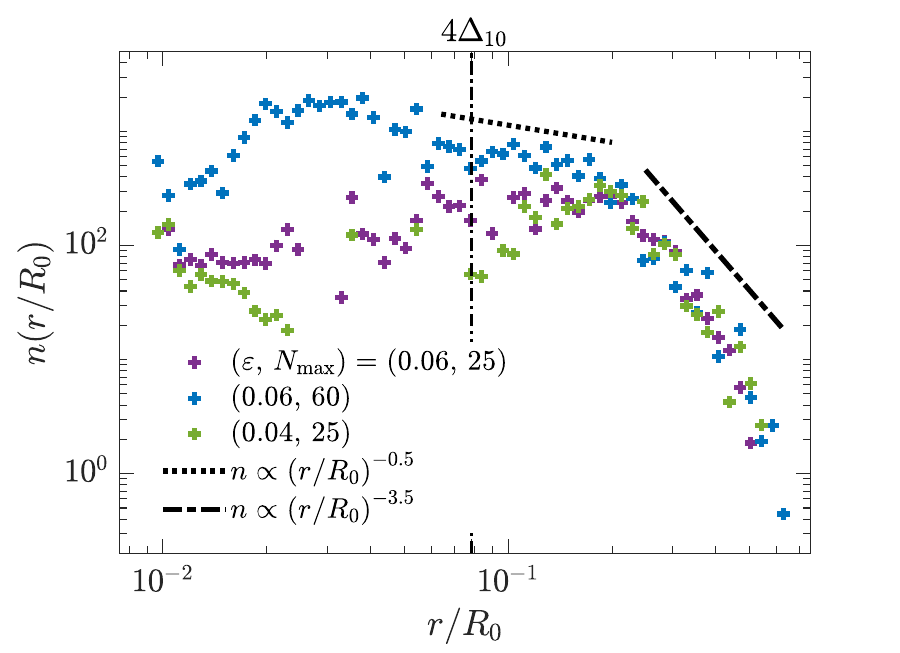}}
	\caption{(a): The evolution of time- and ensemble-averaged size distribution function $n(r/R_0)$ of all fragments produced by colliding rims at $We = 200$, $\varepsilon = 0.06$, and $N_{\rm max} = 25$. (b): The fragment size probability distribution function $f(r/R_0)$ compared with the experimental data and model of \cite{neel2020fines}. (c,d): The influence of $We$ (c) and $\varepsilon$ and $N_{\rm max}$ (d) on the fragment size distribution function, where $\varepsilon = 0.06$ and $N_{\rm max} = 25$ for all simulation results presented in (c), and $We = 200$ for those presented in (d).}
	\label{fig:drop-pdf}
\end{figure}

\begin{figure}
	\centering
	\subfloat[]{
		\label{fig:vel-y-pdf-evol}
		\includegraphics[width=.48\textwidth]{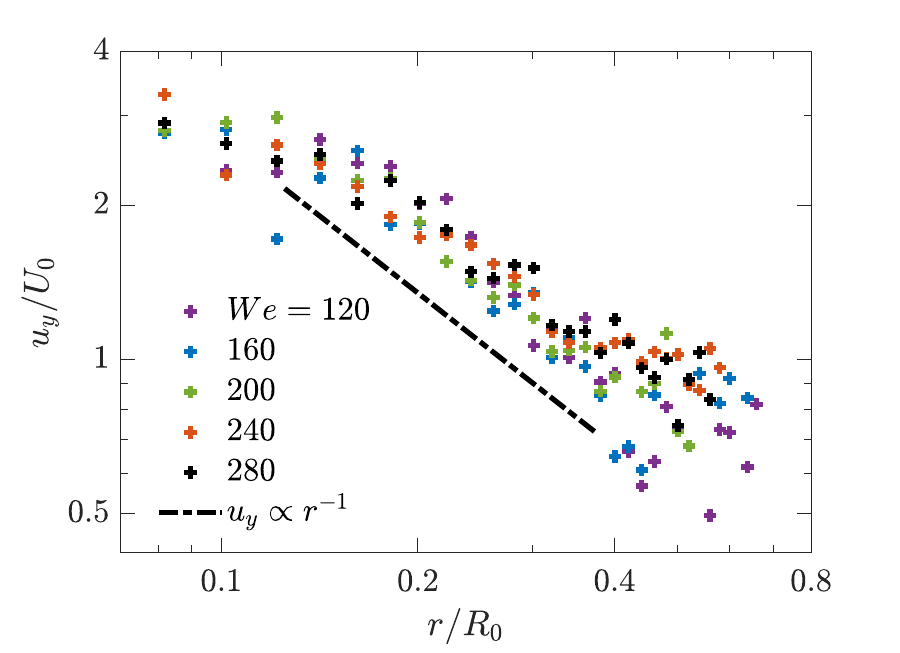}}
	\centering
	\subfloat[]{
		\label{fig:vel-xz-pdf-evol}
		\includegraphics[width=.48\textwidth]{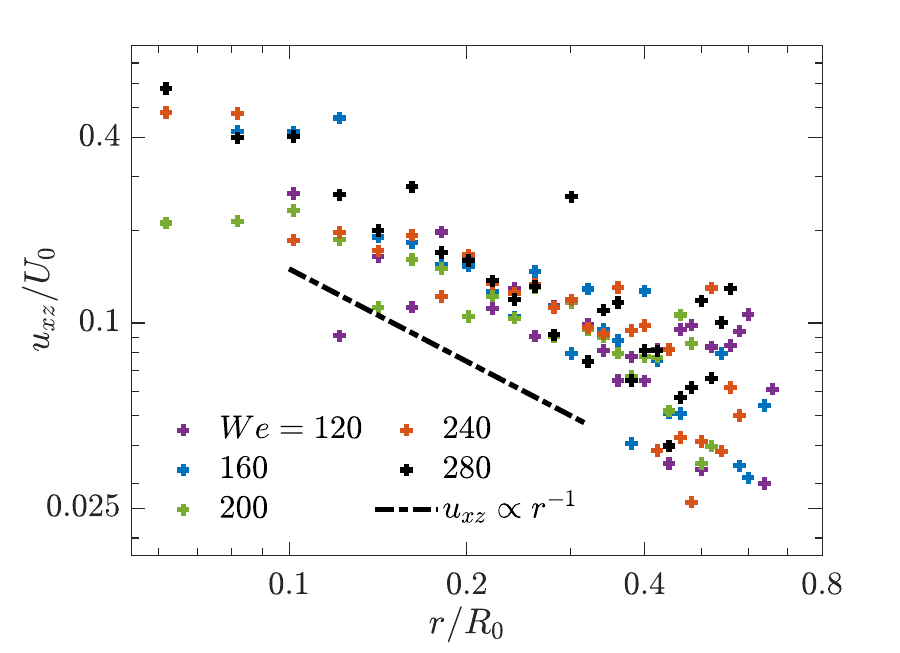}}

        \centering
	\subfloat[]{
		\label{fig:vel-dens-evol}
		\includegraphics[width=.48\textwidth]{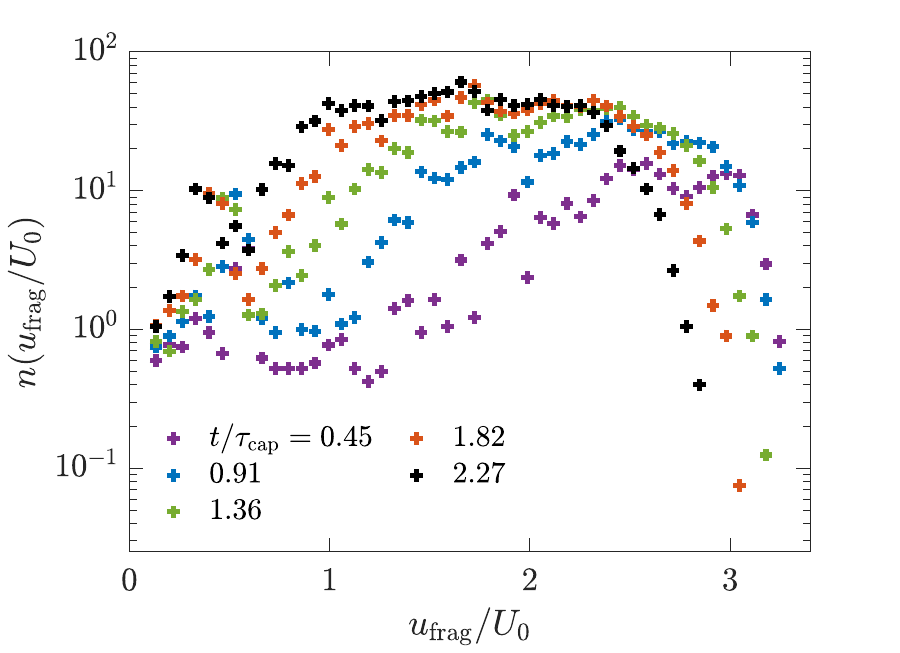}}
        \centering
	\subfloat[]{
		\label{fig:vel-dens-We-sweep}
		\includegraphics[width=.48\textwidth]{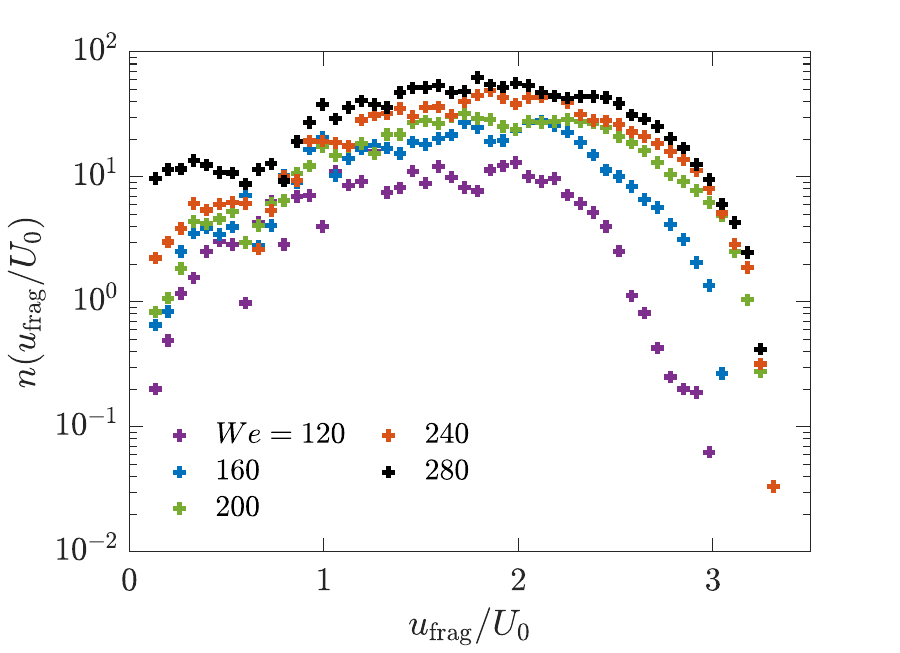}}
	\caption{(a,b): Ensemble-averaged vertical (a) and in-plane (b) components of fragment ejection velocity $u_y$ and $u_{xz}$ calculated at various $We$ values, with $N_{\rm max} = 25$. (c): Evolution of the fragment velocity distribution over time, obtained at $We = 200, \, \varepsilon = 0.06$ and $N_{\rm max} = 25$. (d): The probability distribution functions of fragment velocities at different $We$ values.}
	\label{fig:vel-dist}
\end{figure}

Since rim collision is a transient process where both the total number and the individual size of fragments increase over time, as shown in Fig.~\ref{fig:nd-frag-evol}, it is of interest to determine how the fragment size and velocity distribution functions evolve with time. We first show the fragment size distributions $n(r/R_0)$ in fig.~\ref{fig:drop-pdf-evol} at different times $t_c/\tau_{\rm cap}$, which are computed by sampling over a time window of $t_c - 0.45\tau_{\rm cap} \leq t \leq t_c + 0.45\tau_{\rm cap}$. To ensure statistical convergence of the data, three individual realisations are computed for each initial configuration $(We, \, \varepsilon, \, N_{\rm max})$; and all results presented in figs.~\ref{fig:drop-pdf} and~\ref{fig:vel-dist} have been averaged across these ensembles. It can be seen that initially fragments with $r \leq 0.2R_0$ are produced at early time. While the number density $n$ within this range remains largely unchanged over time and does not appear to depend strongly on $r$, this should be treated with caution due to a lack of fully established grid independence of fragment statistics at small sizes, as discussed in \S\ref{sec:statistics-converge}. As time elapses, the number of larger fragments increases and causes the falling tail of the distribution to move further to the right, indicating that larger drops are fewer and produced later in time. This can be attributed to the ongoing ligament merging process, as it increases the thickness of individual ligaments.

We now seek to compare our numerical results with the experimental data and the theoretical model of \cite{neel2020fines}. It is noted therein that the variation in the fragment size distribution for the rim collision problem arises from two sources, namely those of the transverse ligament size and of the fragments produced from the breakup of a single ligament. A linear superposition of these two effects yields the following size distribution function,
\begin{equation}
    p \left( \zeta = \frac{r}{\Bar{r}} \right) = \frac{2{(mn)^{\frac{m+n}{2}}}}{\Gamma(m) \Gamma(n)} \zeta^{\frac{m+n}{2} - 1} K_{m-n} (2\sqrt{mn\zeta}),
    \label{for:neel20-5.8}
\end{equation}
where $K_{m-n}$ is a modified $(m-n)$-th order Bessel function of the second kind; and $m$ and $n$ reflect the roughness of the distribution of ligament widths and corrugation amplitudes on individual ligaments. \cite{neel2020fines} fit their experimental data at $We = 193$ using \eqref{for:neel20-5.8} with $m=40$, $n=5$, which we can reproduce in fig.~\ref{fig:drop-pdf-neel-comp} together with our fragment size probability density function (pdf) at $We = 200$. Note that the fragment sizes were originally normalised by \cite{neel2020fines} using the average fragment diameter $\Bar{d}$ in their fig.~15b, which is shown in their fig.~13b to saturate at large $We$ values and remain proportional to their interstitial sheet thickness $h$. The sheet thickness is in turn related to the pre-collisional rim radius $R_0$ via the rim collision Weber number, given in their work as $We = 16 R_0/h$. This allows us to estimate their average fragment diameter as
\begin{equation}
    \Bar{d} = \chi h = \frac{16 \chi}{We} R_0.
\end{equation}
We find that setting the coefficient $\chi$ to 5 leads to an excellent match between our simulation results with $(We, \, \varepsilon, \, N_{\rm max}) = (200, \, 0.06, \, 25)$ and the re-normalised data of \cite{neel2020fines} for $r \geq 4\Delta_{10}$. Fig.~13b of \cite{neel2020fines} suggests a $\chi$ value of approximately 25 for their controlled rim production setup, which is larger than our fitted value by a factor of 5. This might be because rim fragmentation has completed in the experiments, and the larger fragments produced at later times increases the value of $\Bar{d}$. Our numerical results differ from \cite{neel2020fines} for $r \leq 4\Delta_{10}$, where their model \eqref{for:neel20-5.8} exhibits a fall-off not found in our data. This more uniform portion of fragment size distribution for small $r$ values is also found in fig.~23b of \cite{lhuissier2012bursting}, where they attributed it to the transverse impact between adjacent rim ligaments. The differences between their experimental and our numerical configurations may also play an important role, as their expanding rims feature toroidal shapes and therefore coalesce within a finite period of time. Last but not least, the two liquid rims are connected by an interstitial thin film in the configuration of \cite{neel2020fines}, whereas in our case the rims come into contact directly. 

The dependence of the time-averaged fragment size distribution on the controlling parameters $We, \, \varepsilon$ and $N_{\rm max}$ is further shown in figs.~\ref{fig:drop-pdf-We-sweep} and \ref{fig:drop-pdf-epsilon-nmax-sweep}. It can be seen that the number density $n$ of small fragments with $r \leq 4\Delta_{10}$ continues to increase with $We$, $\varepsilon$ and $N_{\rm max}$, consistent with our observations in fig.~\ref{fig:n-frag-evol}. More specifically, here the sensitivity of $n$ to $\varepsilon$ and $N_{\rm max}$, as shown in fig.~\ref{fig:drop-pdf-epsilon-nmax-sweep}, further supports our understanding that the differences between the numerical and experimental initial conditions causes the difference between the corresponding results. The number density of intermediate fragments with $4\Delta_{10} \leq r \leq 0.2 R_0$ also increases with $We$ in fig.~\ref{fig:drop-pdf-We-sweep}; but different from that of smaller fragments, it appears to asymptote to ${(r/R_0)}^{-1/2}$ for $We \geq 240$ or $N_{\rm max} \geq 60$. The tail of the distribution functions consisting of even larger fragments with $r \geq 0.2 R_0$ remains approximately independent of the controlling parameters, and its decaying trend agrees well with a power law of ${(r/R_0)}^{-7/2}$. While this finding differs from that of \cite{neel2020fines}, where $m$ and $n$ decrease with increasing $We$ and cause the slope of the tail to decrease correspondingly, this may again be due to differences in detail of the initial conditions. Overall, figs.~\ref{fig:drop-pdf-We-sweep} and~\ref{fig:drop-pdf-epsilon-nmax-sweep} suggest that as $We$ increases, the fragment size distribution within $r \geq 4\Delta$ asymptotes to a regime independent of the controlling parameter, and well described by a power-law decay with a break in slope; which may originate from the insensitivity of ligament merging and breakup phenomena to the initial perturbation configurations. It is noted that a similar transition between two power law regimes has been observed in the droplet size distributions associated with wave breaking by \cite{Mostert2021, erinin2023plunging_b}, although their distributions feature steeper slopes, with the power law transitioning from $r^{-2}$ to $r^{-6}$.

Here we show that the power-law scaling ${(r/R_0)}^{-1/2}$ we observed in figs.~\ref{fig:drop-pdf-We-sweep} and \ref{fig:drop-pdf-epsilon-nmax-sweep} at small fragment sizes can be derived from the ligament merging dynamics previously established in this work. The ligament merging timescale is defined as the ratio between the average ligament spacing $L_0/N_{\rm lig}$ and the drift velocity $u_{\rm drift}$, which can be evaluated based on Eq.~\eqref{for:drift-vel},
\begin{equation}
    \Delta t_{\rm merge} \equiv \frac{L_0}{N_{\rm lig} u_{\rm drift}} \propto N_{\rm lig}^{-2}.
\end{equation}

The rate of droplet shedding is controlled by the ligament necking process, thus $\Delta t_{\rm shed} = t_{\rm neck}$ as given in Eq.~\eqref{for:necking-time} \citep{wang2018unsteady}. Consequently, the total number of fragments shed from all ligaments between two consecutive merging events can be estimated as
\begin{equation}
    N_{\rm frag} = N_{\rm lig} \frac{\Delta t_{\rm merge}}{\Delta t_{\rm shed}} \propto N_{\rm lig}^{-1} w^{-3/2},
    \label{for:n-frag-scaling-intermediate}
\end{equation}
where $w$ is the average width of ligaments.

As we observed in figs.~\ref{fig:frag-lig-ratio} and \ref{fig:d-frag-evol}, $w \propto r \propto \sqrt{t/\tau_{\rm cap}}$, whereas Eq.~\eqref{for:lig-num-scaling} suggests $N_{\rm lig} \propto {(t/\tau_{\rm cap})}^{-1/2}$. Substituting these two scalings into Eq.~\eqref{for:n-frag-scaling-intermediate} leads to
\begin{equation}
    N_{\rm frag} \propto {(t/\tau_{\rm cap})}^{-1/4} \propto {(r/R_0)}^{-1/2}.
    \label{for:n-frag-scaling}
\end{equation}

These derivations suggest that the ${(r/R_0)}^{-1/2}$ scaling at small fragment sizes can be explained as a competition between ligament merging and end-pinching, whereas the ${(r/R_0)}^{-7/2}$ scaling found at larger fragment size ranges still awaits further analysis. It is likely that this steeper dependence on $r$ originates from the presence of corrugations on individual ligaments, or the variation of ligament width across the bordering rim \citep{neel2020fines}; neither of which has been considered in the derivations above. 

Lastly, we discuss the fragment velocity statistics. Figs.~\ref{fig:vel-y-pdf-evol} and~\ref{fig:vel-xz-pdf-evol} present the ensemble-averaged vertical (fig.~\ref{fig:vel-y-pdf-evol}) and in-plane (fig.~\ref{fig:vel-xz-pdf-evol}) components $u_y$ and $u_{xz}$ of the fragment ejection velocity, which are plotted as functions of the fragment radius $r$. Since most of the initial liquid momentum is deflected in the vertical direction over the collision process, $u_y$ remains a few times to a decade larger than $u_{xz}$. Fig.~\ref{fig:vel-y-pdf-evol} shows that the $u_y$ values for fragments within the size range of $0.1 \leq r/R_0 \leq 0.4$ collapse reasonably well when scaled by the initial rim velocity $U_0$, and is well predicted by a power-law decay model $u_y \propto {(r/R_0)}^{-1}$. The velocity distribution at larger fragment sizes deviate from this power-law scaling, most likely due to a combination of late-time effects including amplified ligament corrugations, rim deceleration and the onset of RP instability on the ligaments. The in-plane velocity component values $u_{xz}$ measured in fig.~\ref{fig:vel-xz-pdf-evol} are more scattered compared with fig.~\ref{fig:vel-y-pdf-evol}, but the distribution can still be roughly described by the same power-law decay model $u_{xz} \propto {(r/R_0)}^{-1}$. These observations further corroborate the fragment size distribution model \eqref{for:n-frag-scaling} we proposed, since the power-law exponent of $-1$ for $u_y$ can be derived based on our observations in figs.~\ref{fig:u-t-rel} and Eq.~\eqref{for:w-lig-scaling},
\begin{equation}
     u_y \propto {(t/\tau_{\rm cap})}^{-1/2} = {\left[(t/\tau_{\rm cap})^{1/2} \right]}^{-1} \propto {(r/R_0)}^{-1}.
\end{equation}
As the only source of the liquid in-plane motion is the transverse drifting of ligaments, the in-plane velocity component $u_{xz}$ can be estimated by the drifting velocity $u_{\rm drift}$. Combining Eqs.~\eqref{for:drift-vel} and \eqref{for:lig-num-scaling} similarly leads to
\begin{equation}
     u_{xz} \propto u_{\rm drift} \propto N_{\rm lig} \propto {(t/\tau_{\rm cap})}^{-1/2} \propto {(r/R_0)}^{-1}.
\end{equation}
The good agreement between the velocity scaling models derived above and our numerical results once again highlights the importance of the rim ligament merging phenomenon in determining the size and velocity statistics of splashing fragments.

Figs.~\ref{fig:vel-dens-evol} and~\ref{fig:vel-dens-We-sweep} show the number density of fragments as a function of their travelling speed $u_{\rm frag}$. Fig.~\ref{fig:vel-dens-evol} suggests that while most of the fragments produced at early time feature a skewed velocity distribution peaking at $u_{\rm frag} \approx 2.8U_0$, the maximum fragment speed decreases and more fragments travelling at lower speeds are recorded as time elapses, and the distribution function has developed a plateau by $t/\tau_{\rm cap} = 1.36$. This suggests that as the ligaments continue to grow in length and break up, the droplets produced come to span uniformly across a large range of travelling speeds, with fewer drops featuring very slow or particularly fast speeds. Fig.~\ref{fig:vel-dens-We-sweep} shows the velocity distribution at different $We$ values averaged over the entire collision process. It is found that as $We$ increases the velocity distribution becomes broader; and for $We$ beyond 240, the right tail of the velocity distribution appears to reach a $We$-independent regime, similar to our observation in fig.~\ref{fig:drop-pdf-We-sweep} for the size distribution of fragments; and the fastest speeds recorded is around $u_{\rm frag} \approx 3.2 U_0$. While a direct comparison with breaking wave statistics \citep{Mostert2021, erinin2023plunging_b} is out of the scope of the current work, it is noted that the shapes of velocity distribution functions obtained here in figs.~\ref{fig:vel-dens-evol} and~\ref{fig:vel-dens-We-sweep} differ from their counterparts in wave breaking phenomena (see, e.g. fig. 16d of \cite{Mostert2021} and fig. 12 of \cite{erinin2023plunging_b}), as the latter are skewed and narrower than our distributions. This may be due to the presence of gravity in wave breaking, which may arrest the ligament fragmentation process and define a short timescale for completing the splashing phenomenon, thereby reducing the total number of fragments. These splash fragments are themselves also decelerated by gravity. In the velocity distribution, this would appear as a narrowing of the distribution, with lower velocities at the peak and the higher velocities represented by a long tail. Other fragmentation mechanisms in the wave-breaking phenomena may also alter the shape of velocity distribution, e.g., the bursting of surface bubbles \citep{lhuissier2012bursting, Berny2021}, whose contribution to the production of droplets is known to be significant after the wave splashing phase.

\section{Conclusions}
\label{sec:conclusions}

We have investigated the collision and subsequent fragmentation of perturbed liquid rims, focusing on the range of $120 \leq We \leq 280$ that allows for the growth and merging of transverse ligaments and production of fine drops from such ligaments via the pinch-off mechanism. We look into different parts of the post-collisional liquid bulk as it evolves over time, and our key findings are summarised as below:

Firstly, following the quasi-one-dimensional approach of \cite{wang2017drop}, we derive analytical solutions of the liquid velocity and free surface profiles for the vertically expanding lamella sheet, which are shown to be in good agreement with the numerical results. Capillary effects have been neglected in this model for sheet evolution, but prove significant for the dynamics of the bordering rim. We then compared the growth of its vertical position and thickness with the theoretical model of \cite{gordillo2019theory}, and develop scaling laws collapsing the data.

Secondly, we analyse the behaviour of transverse ligaments on top of the lamella rim. The ligaments produce fragments primarily via the end-pinching mechanism, and when the initial perturbation waveform is polychromatic, they migrate on the rim and merge with each other to form thicker and more corrugated ligaments, thus preventing their absorption into the rim and sustaining the fragmentation process. A novel scaling model is derived for predicting the evolution of ligament number density based on the migration speed model of \cite{wang2018unsteady}.

Lastly, we present the size and velocity statistics associated with the rim collision phenomenon. An excellent agreement between our fragment size distribution and the experimental results of \cite{neel2020fines} is found within the range of grid convergence ($r \geq 4\Delta_{10}$). The fragment size distribution becomes insensitive to the initial configurations when $We$ or $N_{\rm max}$ further increases, which can be described using a power law with a break in slope. A theoretical model is proposed predicting the power-law distribution observed for $r \leq 0.2R_0$. Over time, the fragment speed $u_{\rm frag}$ develops a largely uniform spread over the range of $0 \leq u_{\rm frag} \leq 3.2 U_0$ as their parent ligaments continue to grow vertically and decelerate to form slower drops.

The implications of the present work are manifold. Firstly, it sheds new light on the fluid physics involved in a liquid impact problem that has not received much attention prior to the recent works of \cite{neel2020fines} and \cite{agbaglah2021breakup}. Furthermore, the results we obtained are also of reference value for ongoing research works on spherical drop impacts, especially the influence of initial perturbations on the fragmentation process, which may also be present during the early-time prompt splashing phenomena observed in previous experimental studies \citep{burzynski2020splashing, wang2023analysis}. Lastly, this work serves as a stepping stone towards understanding the secondary splashing phenomenon observed in wave breaking events and the associated fragment statistics \citep{Mostert2021, erinin2023plunging_b}, and provides the basis for investigating the influence of other physical mechanisms not covered in the present work, e.g., viscosity, gravity and air-phase turbulence effects.

\section{Declaration of Interests}
The authors report no conflict of interest.

\section{Acknowledgments}
\label{sec:acknowledge}
We thank D. Champlin, former at the Missouri University of Science and Technology, for preliminary data related to this study. Insights from Prof. Alfonso A. Castrej{\'o}n-Pita have helped in improving the theoretical analysis of this work. The authors would like to thank EPSRC for the computational time made available on the UK supercomputing facility ARCHER2 via the UK Turbulence Consortium (EP/R029326/1). Use of the University of Oxford Advanced Research Computing (ARC) facility is also acknowledged. K. Tang is supported by a Research Studentship at the University of Oxford.

\appendix
\section{Grid convergence and $Oh$-dependency of fragment statistics}
\label{sec:statistics-converge}

\begin{figure}
	\centering
	\subfloat[]{
		\label{fig:drop-pdf-grid-converge-0.0025}
		\includegraphics[width=.48\textwidth]{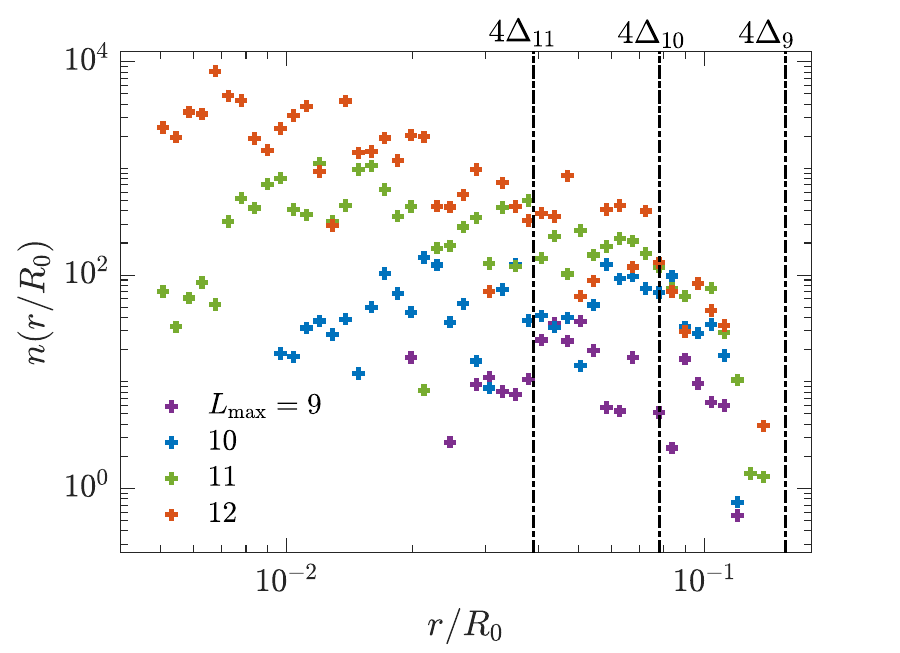}}
    \centering
	\subfloat[]{
		\label{fig:vel-pdf-grid-converge-0.0025}
		\includegraphics[width=.48\textwidth]{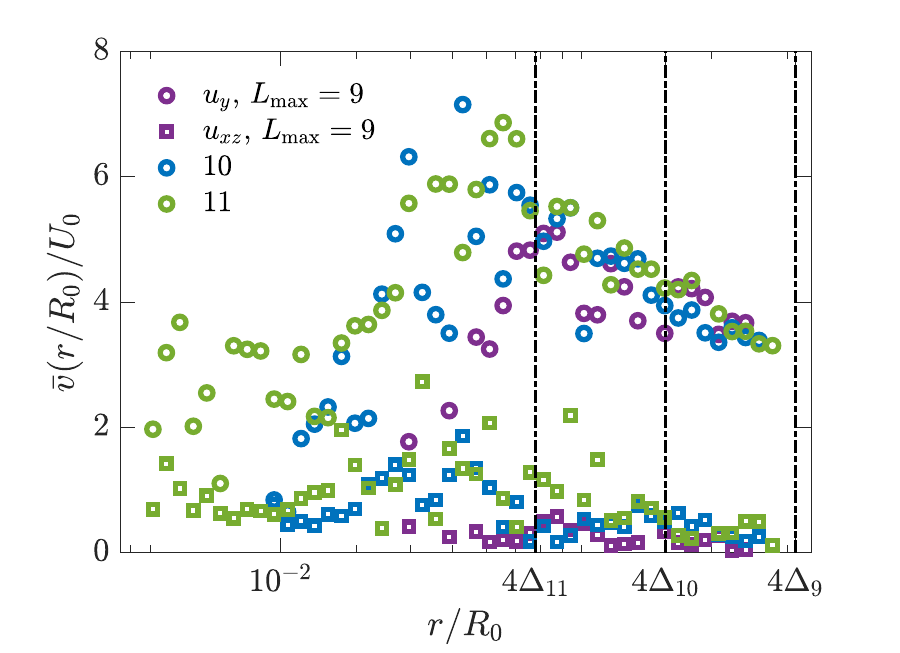}}

    \centering
	\subfloat[]{
		\label{fig:drop-pdf-grid-converge-0.02}
		\includegraphics[width=.48\textwidth]{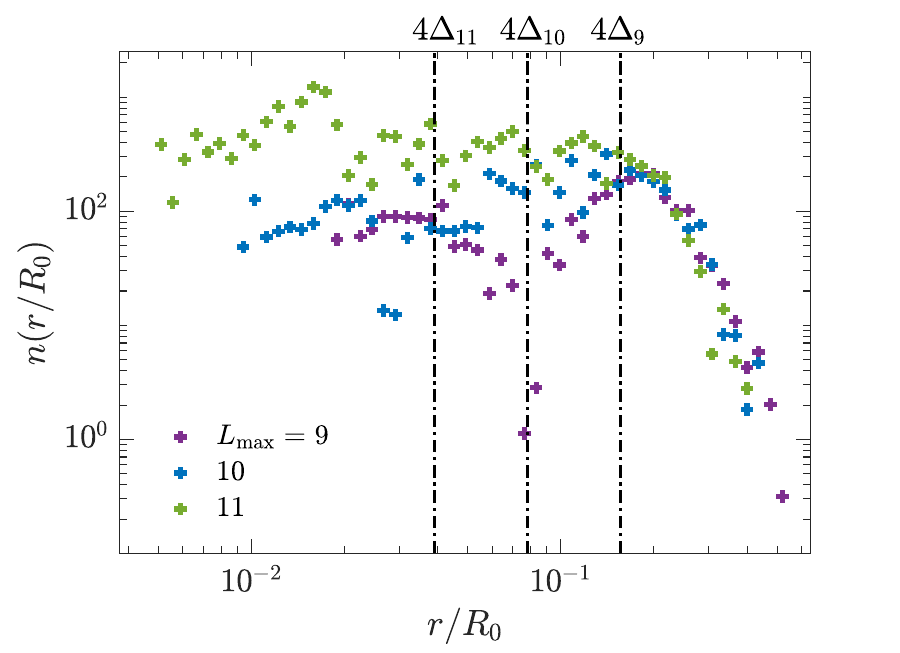}}
	\centering
	\subfloat[]{
		\label{fig:vel-pdf-grid-converge-0.02}
		\includegraphics[width=.48\textwidth]{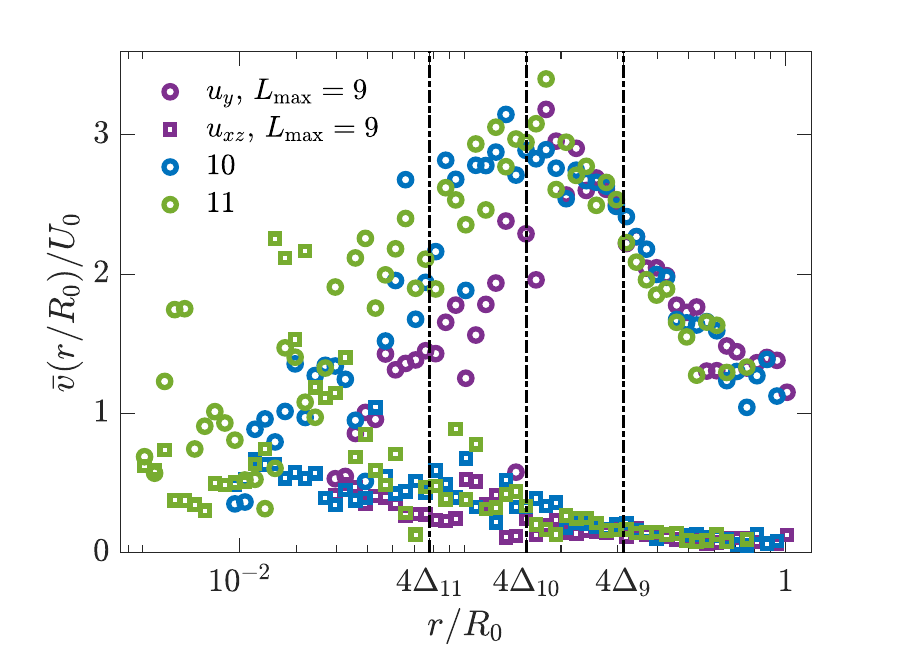}}

    \centering
	\subfloat[]{
		\label{fig:drop-pdf-grid-converge-Oh-sweep}
		\includegraphics[width=.48\textwidth]{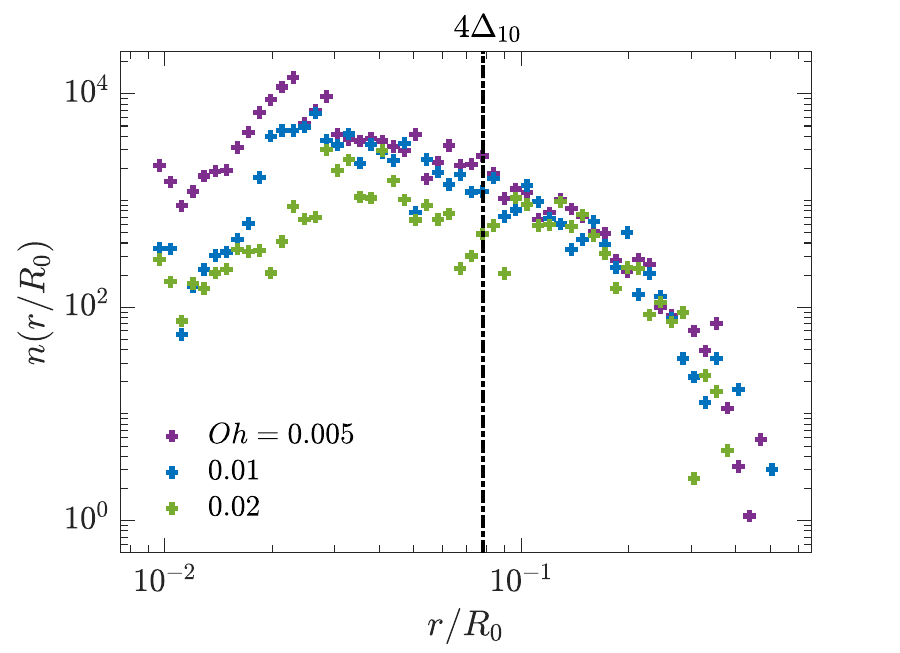}}
	\centering
	\subfloat[]{
		\label{fig:vel-pdf-grid-converge-Oh-sweep}
		\includegraphics[width=.48\textwidth]{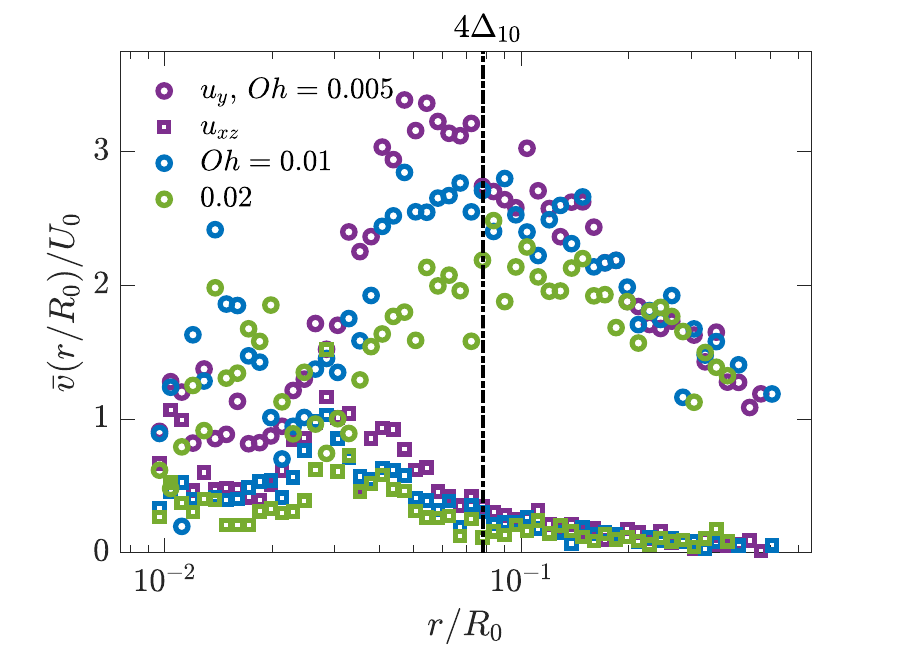}}
	\caption{The size (a,c) and velocity (b,d) distributions of fragments $n(r/R_0)$ and $\Bar{v}(r/R_0)$ at $We = 200$, $\varepsilon = 0.06$ and $N_{\rm max} = 25$ at $t/\tau_{\rm cap} = 0.23$ (a,b) and 1.82 (c,d), binned and averaged across three realisations with the same initial configurations at $L_{\rm max} = 9$, 10 and 11. Fig.~\ref{fig:drop-pdf-grid-converge-0.0025} also includes statistics produced from a single realisation at $L_{\rm max} = 12$. (e,f): The size (e) and velocity (f) distributions of fragments at $We = 280$ and different $Oh$ values at $t/\tau_{\rm cap} = 1.82$.}
	\label{fig:drop-pdf-converge}
\end{figure}

In this section, we discuss the numerical convergence for the fragment statistics of rim collision reported in this work. For this purpose, we present the time- and ensemble-averaged fragment size and velocity distributions $n(r/R_0)$ and $\Bar{v}(r/R_0)$ for $We = 200$, $\varepsilon = 0.06$ and $N_{\rm max} = 25$ at $t/\tau_{\rm cap} = 0.23$ in fig.~\ref{fig:drop-pdf-converge}, where the upper and lower rows show respectively the results at early ($t/\tau_{\rm cap} = 0.23$) and late ($t/\tau_{\rm cap} = 1.82$) times. The recorded fragment data at a given time are first collected across different realisations with the same initial configuration, and then binned according to the fragment radius $r$. The count in each bin thus produces the fragment size distribution $n(r/R_0)$. We then average the speed of fragments within each bin to obtain the distribution of fragment velocity $\Bar{v}(r/R_0)$. Three identical ensemble realisations are obtained at maximum resolution level $L_{\rm max} = 9$, 10 and 11, whereas in fig.~\ref{fig:drop-pdf-grid-converge-0.0025} we also include data produced from a single realisation at $L_{\rm max} = 12$. It has been known from previous numerical studies \citep{riviere2021sub, Mostert2021, tang2022bag, wang2023analysis} that grid independence for fragmentation problems can be challenging to obtain, especially for small fragments near the grid size $\Delta$; and a radius threshold of $r \geq 4\Delta$ has been recommended beyond which the fragment size distributions are considered fully converged. We therefore also add vertical dashed lines in each subplot of fig.~\ref{fig:drop-pdf-converge} showing the values of $4\Delta$ corresponding to each resolution level to facilitate comparison with these threshold values. 

The early-time fragment size distributions presented in fig.~\ref{fig:drop-pdf-grid-converge-0.0025} show large ranges of scatter due to relatively small number of fragments produced from each ensemble realisation, with the size of the tiniest fragments becoming increasingly smaller as the resolution level $L_{\rm max}$ increases. Nevertheless, it is observed that the tail of the distribution functions obtained at $L_{\rm max} = 10$ and 11 agrees when $r/R_0 \geq 0.1R_0$, which roughly corresponds to the threshold value of $4\Delta_{10}$. We note that the grid convergence behaviour of large fragments at early times are further improved when $N_{\rm max}$ is increased, although these results are not included in the present work. The distributions of velocity components $u_y$ and $u_{xz}$ in fig.~\ref{fig:vel-pdf-grid-converge-0.0025} show better agreement across all three resolution levels down to $r = 4\Delta_{10}$. Below this radius threshold, large scatters in the velocity data at $L_{\rm max} = 10$ and 11 are observed, although they appear to agree in trend with each other and differ from the results at $L_{\rm max} = 9$. When rim collision proceeds to later times, more fragments are produced and the range of scatter in the size and velocity distributions becomes smaller, as shown in figs.~\ref{fig:drop-pdf-grid-converge-0.02} and \ref{fig:vel-pdf-grid-converge-0.02}. Nevertheless, the fragment size range where grid convergence of fragment size and velocity distributions is fully established remains largely unchanged from the early time results, namely, the right tail of the fragment size (fig.~\ref{fig:drop-pdf-grid-converge-0.02}) and velocity distribution (fig.~\ref{fig:vel-pdf-grid-converge-0.02}) are fully converged for $r/R_0 \geq 4\Delta_9$, and the difference between results at $L_{\rm max} = 10$ and 11 become significant for $r/R_0 \leq 4\Delta_{10}$. These results suggest that for a given grid resolution level $L_{\rm max}$, $4\Delta_{L_{\rm max}}$ can be regarded as the lower limit of fragment radius above which the fragment size and velocity distributions can be considered fully grid converged, while the statistics of fragments below this threshold are still grid-dependent and should be treated with caution \citep{Mostert2021}.

Figs.~\ref{fig:drop-pdf-grid-converge-Oh-sweep} and \ref{fig:vel-pdf-grid-converge-Oh-sweep} present the fragment size and velocity distributions at a few different $Oh$ values at $t/\tau_{\rm cap} = 1.82$. Fig.~\ref{fig:drop-pdf-grid-converge-Oh-sweep} shows that for $r \geq 4\Delta_{10}$ where grid convergence of fragment statistics is fully established, the fragment size distribution is not sensitive to viscous effects. On the other hand, the velocity of fragments with radii near $4\Delta_{10}$ as shown in fig.~\ref{fig:vel-pdf-grid-converge-Oh-sweep} decreases slightly with increasing $Oh$ values, which might be because of more significant viscous dissipation at the lamella feet and rim necks, as discussed in \S\ref{subsec:lig-formation}.

\section{Theoretical analysis of the lamella foot advancement}
\label{sec:pos-yneck}

Here we analytically solve for the evolution of the advancement of the lamella foot position $y_n$ after the two cylinders come into contact. This is similar to the axisymmetric analysis due to \cite{riboux2014experiments}, having here been adapted for the planar problem; and it is expected that similar to their case, $y_n \propto \sqrt{U_0 t/R_0}$. Taking into account the geometrical symmetry present in the problem, we consider the equivalent configuration where a single liquid rim impacts on a flat plane. Neglecting viscous effects and air entrainment, the flow field within the liquid phase is prescribed by the Laplace equation within the Cartesian coordinate $yOz$,
\begin{equation}
    \nabla^2 \phi = \frac{\partial^2 \phi}{\partial y^2} + \frac{\partial^2 \phi}{\partial z^2} = 0,
    \label{for:laplace-cart}
\end{equation}
where $\phi$ is the velocity potential. From now on, we select the rim radius $R_0$, the collision velocity $U_0$ and their quotient $R_0/U_0$ as reference length, velocity and time scales to non-dimensionalise all physical properties within this section.

\begin{figure}
	\centering
	\subfloat[]{
		\label{fig:cyl_cross_sec}
		\includegraphics[height=.4\textwidth]{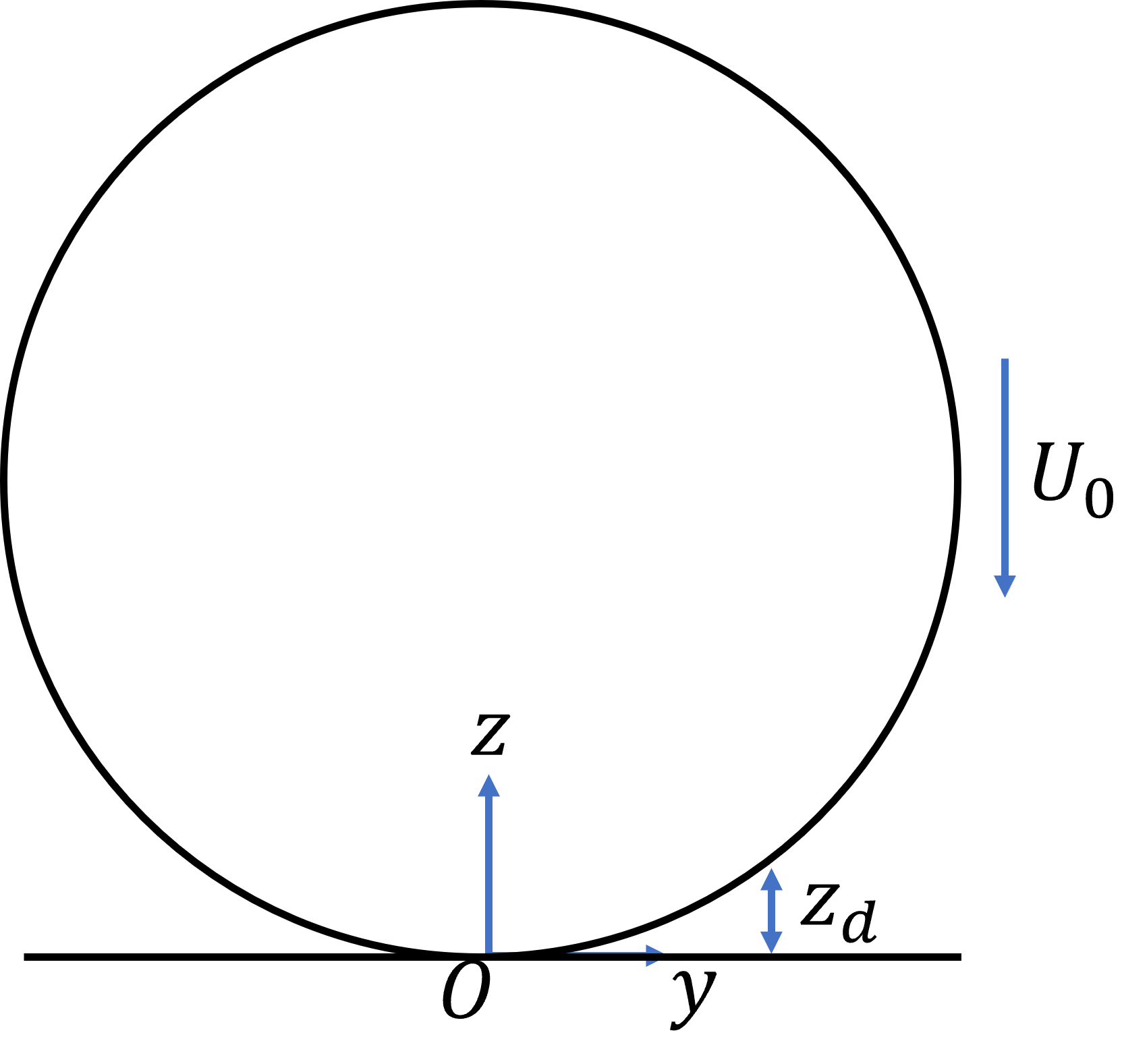}}
	\centering
	\subfloat[]{
		\label{fig:laplace_sketch}
		\includegraphics[height=.4\textwidth]{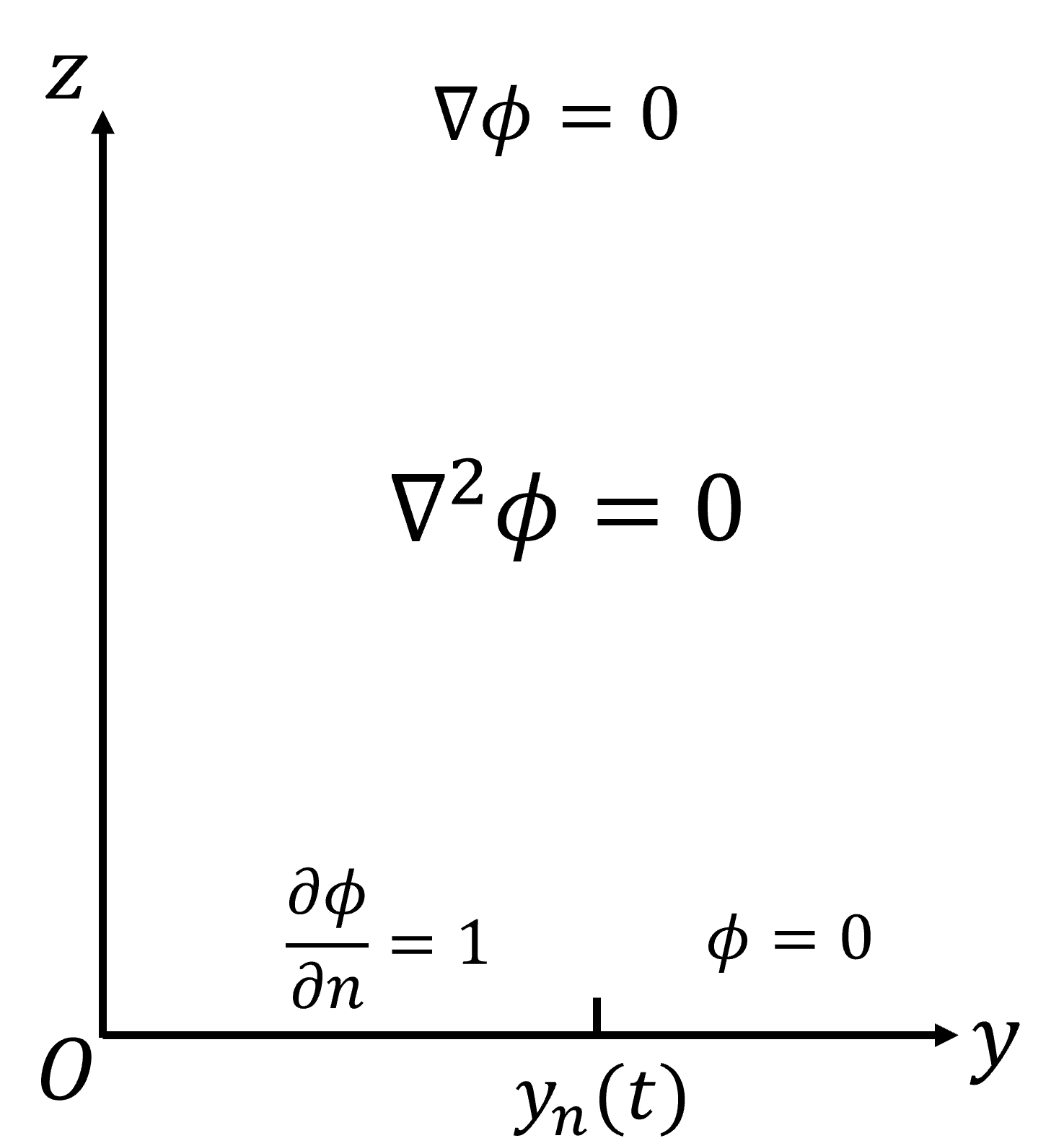}}
	\caption{(a): Sketch showing the cross-sectional view of the liquid cylinder collision problem. (b): Sketch showing the boundary conditions \eqref{for:laplace_bc_1}, \eqref{for:laplace_bc_2} and \eqref{for:laplace_bc_3} under which we solve the Laplace equation \eqref{for:laplace-cart}.}
	\label{fig:lamella_neck_sketch}
\end{figure}

Now consider the boundary conditions for the domain of interest, namely the contact region $z \ll 1$ within the liquid phase close to the bottom plane. We fix our reference frame on the liquid bulk descending at non-dimensionalised unit velocity $-1$, thus we impose an opposite normal velocity at the bottom,
\begin{equation}
    \frac{\partial \phi}{\partial n} = 1, \quad |y| \leq y_n(t), \, z = 0.
    \label{for:laplace_bc_1}
\end{equation}
Within the same reference frame, the liquid phase velocity decays to 0 far away from the contact region, thus
\begin{equation}
    \nabla \phi = 0, \quad z \gg 1.
    \label{for:laplace_bc_2}
\end{equation}
The final boundary condition comes from considering the flow condition on the drop surface (but outside the contact area) for $1 \gg y \geq y_n(t)$ and $z=z_d\ (y)$. At very early time immediately after the impact, the rim bulks largely retain their cylindrical shape, as shown in fig.~\ref{fig:cyl_cross_sec}. Thus, the drop shape in the vicinity of the stagnation point can be approximated as
\begin{equation}
    z_d = 1 - \sqrt{1-y^2} \approx \frac{y^2}{2} + o(y^2) \approx 0
\end{equation}
up to first order in $y$. At high $We$ values capillary effects can be neglected, and the unsteady Bernoulli Equation can be applied along the drop surface,
\begin{equation}
    \frac{\partial \phi}{\partial t} + \frac{u^2}{2} = \left[ \frac{\partial \phi}{\partial t} + \frac{u^2}{2} \right]_{y=y_n(t)} \approx \frac{1}{2}.
    \label{for:unsteady-bernoulli}
\end{equation}
For small values of $t$, unsteady effects dominate and Eq.~\eqref{for:unsteady-bernoulli} can be further simplified using $\phi(t=0)=0$ at the free surface,
\begin{equation}
    \phi \approx \phi(t=0) - \frac{1}{2} \left(1-{|\nabla \phi|}^2 \right) t \approx \phi(t=0) = 0, \quad |y| \geq y_n(t), \, z = 0.
    \label{for:laplace_bc_3}
\end{equation}

The solution of Eq.~\eqref{for:laplace-cart} subject to boundary conditions~\eqref{for:laplace_bc_1}, \eqref{for:laplace_bc_2} and~\eqref{for:laplace_bc_3} thus corresponds to the flow field induced by an infinitely long lamina with width $2y_n$ moving perpendicular to its surface. The following solution written in elliptical coordinates is provided in Art. 71, 3° of \cite{lamb1932hydrodynamics}:
\begin{equation}
\psi=ae^{-\xi} \cos \eta,
\end{equation}
where $y=a\cosh \xi \cos \eta, \, z=a \sinh \xi \sin \eta$. Utilising $\eta=0$ at $z=0$, we derive the total vertical velocity $u_z$ for $y>y_n(t), \, z=0$ within the laboratory frame,
\begin{equation}
  u_z(z=0) = -1 - \frac{\partial \psi}{\partial z} = -1 - \frac{2}{{\left[ \frac{y}{y_n} + \sqrt{{\left( \frac{y}{y_n} \right)}^2-1} \right]}^2-1}.
  \label{for:un_total}
\end{equation}
Now apply Wagner’s condition \citep{wagner1932uber, riboux2014experiments} to Eq.~\eqref{for:un_total}; namely, the lamella foot position $y_n(t)$ is fixed by the time when a point on the drop interface with initial coordinates $y=y_n (t), \, z_d=y_n^2/2$ reaches $z=0$,
\begin{equation}
    \frac{y_n^2}{2} - t - \int_0^{y_n} \frac{2}{{\left[ \frac{y_n}{\kappa} + \sqrt{{\left( \frac{y_n}{\kappa} \right)}^2-1} \right]}^2-1}    \frac{d\tau}{d\kappa} d\kappa = 0.
\end{equation}
Where $\kappa$ is a dummy variable indicating the lamella foot location for $\tau<t$. Taking $\kappa = y_n(t) \sin{\lambda}$, and since we already know $\kappa \propto \sqrt{t}$, we assume that $d\tau/d\kappa= C\kappa$. Thus one arrives at
\begin{equation}
    \frac{y_n^2}{2} - t - C y_n^2 \int_0^{\pi/2} (\sin \lambda -\sin \lambda \cos \lambda ) d\lambda = 0,
\end{equation}
which leads to
\begin{equation}
    \frac{dt}{dy_n}= y_n (1-C) = Cy_n.
\end{equation}
Thus $C=1/2$, and by integration one finds $y_n = 2\sqrt{t}$, or written dimensionally as
\begin{equation}
    y_n/R_0 = 2\sqrt{U_0 t / R_0}.
    \label{for:neck_pos_evol}
\end{equation}

\begin{figure}
	\centering
	\subfloat[]{
		\label{fig:lamella_foot_profile}
		\includegraphics[width=.48\textwidth]{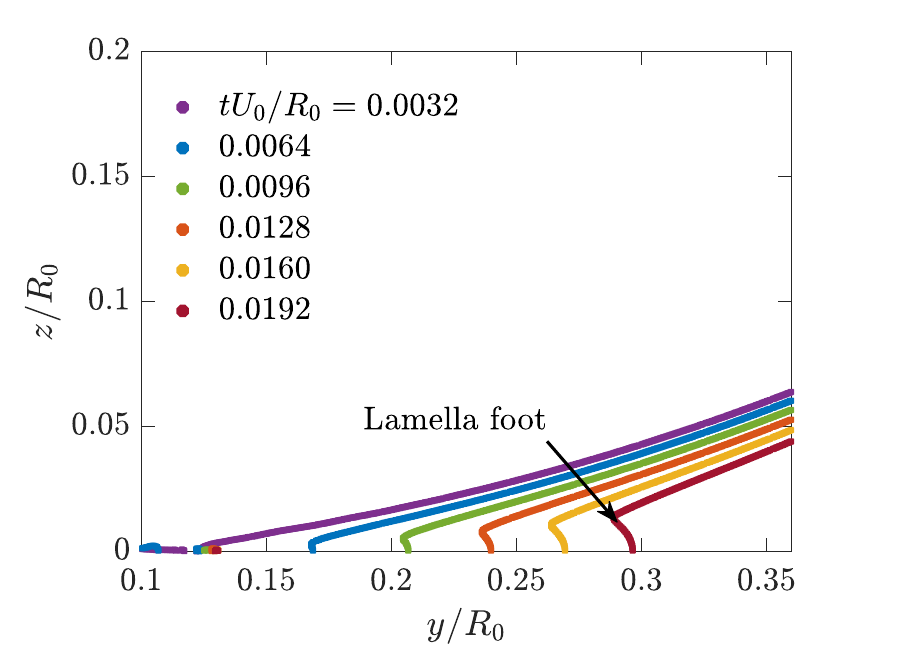}}
	\centering
	\subfloat[]{
		\label{fig:lamella_advance_comp}
		\includegraphics[width=.48\textwidth]{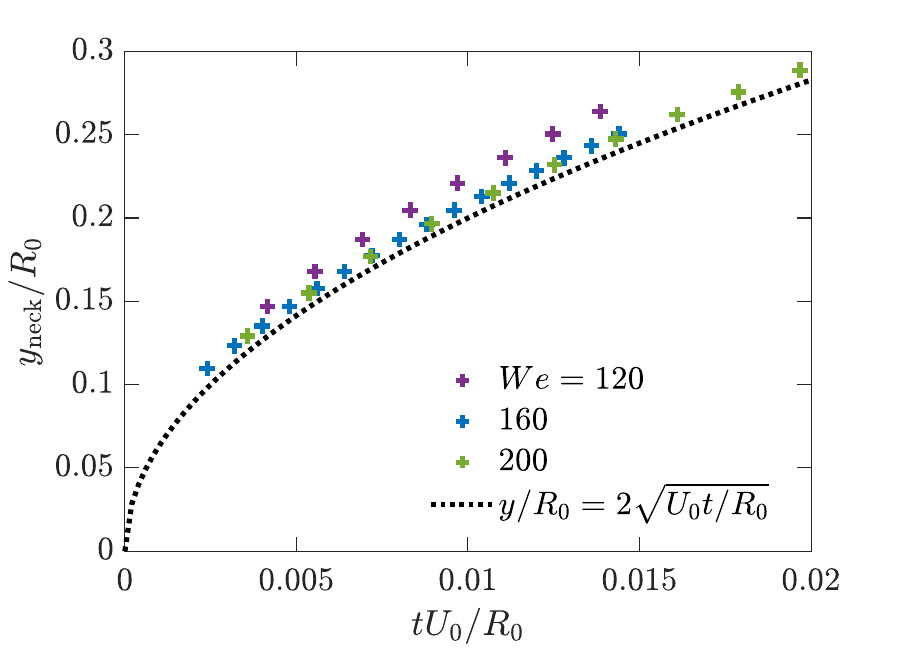}}
	\caption{(a): Two-dimensional simulation results at $L_{\rm max} = 15$ for $We = 160$ showing the evolution of the contact region. (b): Comparison between simulation results at different values of $We$ and the theoretical prediction \eqref{for:neck_pos_evol} at very early time.}
	\label{fig:lamella_neck_comp}
\end{figure}

We conduct two-dimensional numerical simulations at a very high resolution level $L_{\rm max} = 15$ to investigate the evolution of the interface profile close to the contact region, which we show in fig.~\ref{fig:lamella_foot_profile}. It can be observed that for $tU_0/R_0 \geq 0.0096$, a bulge on the interface profile appears at $z=0$ representing the nascent lamella, whereas the local minimum in $y$ denotes the lamella foot. Fig.~\ref{fig:lamella_advance_comp} further compares the evolution of the lamella foot location $y_n$ measured from the numerical simulations with the theoretical prediction \eqref{for:neck_pos_evol}, where a good agreement is reached for $We \geq 160$, indicating that the potential flow analysis employed in this section indeed captures the lamella ejection at very early time. 

Solving Eqs.~\eqref{for:rim-dyn-model-mass}-\eqref{for:rim-dyn-model-momentum} requires the initial values of $(b_{\rm rim}, \, y_{\rm rim}, \, v_{\rm rim})$ at the moment of lamella foot formation $t_e$. \eqref{for:neck_pos_evol} suggests that
\begin{equation}
    y_{\rm rim} (t_e) = 2\sqrt{t_e}, \quad v_{\rm rim} (t_e) = 1/\sqrt{t_e}.
    \label{for:rim-dyn-ics}
\end{equation}
We also expect that the nascent rim thickness $b_{\rm rim}(t_e)$ can be approximated using the lamella height function $h[y_{\rm rim}(t_e), \, t_e]$. Thus the ejection time $t_e$ remains the only unknown in the initial conditions, although our numerical simulations suggest that it is close to zero. To regularize the initial conditions, we choose a well-resolved and sufficiently small time $t_e' = 4.5\times 10^{-3} \tau_{\rm cap}$ when solving Eqs.~\eqref{for:rim-dyn-model-mass}-\eqref{for:rim-dyn-model-momentum}. We have confirmed that changing values of $t_e'$ does not have significant influences on the solution of Eqs.~\eqref{for:rim-dyn-model-mass}-\eqref{for:rim-dyn-model-momentum}.

\bibliographystyle{jfm}
\bibliography{references}

\begin{thebibliography}{90}
\expandafter\ifx\csname natexlab\endcsname\relax\def\natexlab#1{#1}\fi
\def\au#1{#1} \def\ed#1{#1} \def\yr#1{#1}\def\at#1{#1}\def\jt#1{\textit{#1}} \def\bt#1{#1}\def\bvol#1{\textbf{#1}} \def\vol#1{#1} \def\pg#1{#1} \def\publ#1{#1}\def\arxiv#1{#1}\def\org#1{#1}\def\st#1{\textit{#1}}

\bibitem[Agbaglah {\em et~al.\/}(2015)Agbaglah, Thoraval, Thoroddsen, Zhang, Fezzaa \& Deegan]{agbaglah2015drop}
{\sc \au{Agbaglah, G.}, \au{Thoraval, M.-J.}, \au{Thoroddsen, S.~T.}, \au{Zhang, L.~V.}, \au{Fezzaa, K.} \& \au{Deegan, R.~D.}} \yr{2015}  \at{Drop impact into a deep pool: vortex shedding and jet formation}.  \jt{Journal of Fluid Mechanics}  \bvol{764},  \pg{R1}.

\bibitem[Agbaglah(2021)]{agbaglah2021breakup}
{\sc \au{Agbaglah, G.~G.}} \yr{2021}  \at{Breakup of thin liquid sheets through hole--hole and hole--rim merging}.  \jt{Journal of Fluid Mechanics}  \bvol{911},  \pg{A23}.

\bibitem[Andreas {\em et~al.\/}(1995)Andreas, Edson, Monahan, Rouault \& Smith]{andreas1995spray}
{\sc \au{Andreas, E.~L.}, \au{Edson, J.~B.}, \au{Monahan, E.~C.}, \au{Rouault, M.~P.} \& \au{Smith, S.~D.}} \yr{1995}  \at{The spray contribution to net evaporation from the sea: A review of recent progress}.  \jt{Boundary-Layer Meteorology}  \bvol{72},  \pg{3--52}.

\bibitem[Batchelor(2000)]{batchelor2000}
{\sc \au{Batchelor, G.~K.}} \yr{2000} {\em An Introduction to Fluid Dynamics\/}.  \publ{Cambridge University Press}.

\bibitem[Berny {\em et~al.\/}(2022)Berny, Deike, Popinet \& S{\'e}on]{berny2022size}
{\sc \au{Berny, A.}, \au{Deike, L.}, \au{Popinet, S.} \& \au{S{\'e}on, T.}} \yr{2022}  \at{Size and speed of jet drops are robust to initial perturbations}.  \jt{Physical Review Fluids}  \bvol{7}~(1),  \pg{013602}.

\bibitem[Berny {\em et~al.\/}(2021)Berny, Popinet, S{\'e}on \& Deike]{Berny2021}
{\sc \au{Berny, A.}, \au{Popinet, S.}, \au{S{\'e}on, T.} \& \au{Deike, L.}} \yr{2021}  \at{Statistics of jet drop production}.  \jt{Geophysical Research Letters}  \bvol{48}~(10),  \pg{e2021GL092919}.

\bibitem[Brackbill {\em et~al.\/}(1992)Brackbill, Kothe \& Zemach]{brackbill1992continuum}
{\sc \au{Brackbill, J.~U.}, \au{Kothe, D.~B.} \& \au{Zemach, C.}} \yr{1992}  \at{A continuum method for modeling surface tension}.  \jt{Journal of Computational Physics}  \bvol{100}~(2),  \pg{335--354}.

\bibitem[Burzynski {\em et~al.\/}(2020)Burzynski, Roisman \& Bansmer]{burzynski2020splashing}
{\sc \au{Burzynski, D.~A.}, \au{Roisman, I.~V.} \& \au{Bansmer, S.~E.}} \yr{2020}  \at{On the splashing of high-speed drops impacting a dry surface}.  \jt{Journal of Fluid Mechanics}  \bvol{892},  \pg{A2}.

\bibitem[Castrej{\'o}n-Pita {\em et~al.\/}(2021)Castrej{\'o}n-Pita, Betton, Campbell, Jackson, Morgan, Tuladhar, Vadillo \& Castrejon-Pita]{castrejon2021formulation}
{\sc \au{Castrej{\'o}n-Pita, A.~A.}, \au{Betton, E.~S.}, \au{Campbell, N.}, \au{Jackson, N.}, \au{Morgan, J.}, \au{Tuladhar, T.~R.}, \au{Vadillo, D.~C.} \& \au{Castrejon-Pita, J.~R.}} \yr{2021}  \at{Formulation, quality, cleaning, and other advances in inkjet printing}.  \jt{Atomization and Sprays}  \bvol{31}~(4).

\bibitem[Castrej{\'o}n-Pita {\em et~al.\/}(2015)Castrej{\'o}n-Pita, Castrej{\'o}n-Pita, Thete, Sambath, Hutchings, Hinch, Lister \& Basaran]{castrejon2015plethora}
{\sc \au{Castrej{\'o}n-Pita, J.~R.}, \au{Castrej{\'o}n-Pita, A.~A.}, \au{Thete, S.~S.}, \au{Sambath, K.}, \au{Hutchings, I.~M.}, \au{Hinch, J.}, \au{Lister, J.~R.} \& \au{Basaran, O.~A.}} \yr{2015}  \at{Plethora of transitions during breakup of liquid filaments}.  \jt{Proceedings of the National Academy of Sciences}  \bvol{112}~(15),  \pg{4582--4587}.

\bibitem[Chan {\em et~al.\/}(2021)Chan, Dodd, Johnson \& Moin]{chan2021identifying}
{\sc \au{Chan, W. H.~R.}, \au{Dodd, M.~S.}, \au{Johnson, P.~L.} \& \au{Moin, P.}} \yr{2021}  \at{Identifying and tracking bubbles and drops in simulations: A toolbox for obtaining sizes, lineages, and breakup and coalescence statistics}.  \jt{Journal of Computational Physics}  \bvol{432},  \pg{110156}.

\bibitem[Cheng {\em et~al.\/}(2022)Cheng, Sun \& Gordillo]{cheng2022drop}
{\sc \au{Cheng, X.}, \au{Sun, T.-P.} \& \au{Gordillo, L.}} \yr{2022}  \at{Drop impact dynamics: Impact force and stress distributions}.  \jt{Annual Review of Fluid Mechanics}  \bvol{54},  \pg{57--81}.

\bibitem[Chirco {\em et~al.\/}(2022)Chirco, Maarek, Popinet \& Zaleski]{chirco2021manifold}
{\sc \au{Chirco, L.}, \au{Maarek, J.}, \au{Popinet, S.} \& \au{Zaleski, S.}} \yr{2022}  \at{Manifold death: a volume of fluid implementation of controlled topological changes in thin sheets by the signature method}.  \jt{Journal of Computational Physics}  \bvol{467},  \pg{111468}.

\bibitem[Cimpeanu \& Papageorgiou(2018)]{cimpeanu2018three}
{\sc \au{Cimpeanu, R.} \& \au{Papageorgiou, D.~T.}} \yr{2018}  \at{Three-dimensional high speed drop impact onto solid surfaces at arbitrary angles}.  \jt{International Journal of Multiphase Flow}  \bvol{107},  \pg{192--207}.

\bibitem[Culick(1960)]{culick1960comments}
{\sc \au{Culick, F. E.~C.}} \yr{1960}  \at{Comments on a ruptured soap film}.  \jt{Journal of Applied Physics}  \bvol{31}~(6),  \pg{1128--1129}.

\bibitem[Deike(2022)]{deike2022mass}
{\sc \au{Deike, L.}} \yr{2022}  \at{Mass transfer at the ocean--atmosphere interface: the role of wave breaking, droplets, and bubbles}.  \jt{Annual Review of Fluid Mechanics}  \bvol{54},  \pg{191--224}.

\bibitem[Eggers {\em et~al.\/}(1999)Eggers, Lister \& Stone]{eggers1999coalescence}
{\sc \au{Eggers, J.}, \au{Lister, J.~R.} \& \au{Stone, H.~A.}} \yr{1999}  \at{Coalescence of liquid drops}.  \jt{Journal of Fluid Mechanics}  \bvol{401},  \pg{293--310}.

\bibitem[Erinin {\em et~al.\/}(2023{\natexlab{{\em a\/}}})Erinin, Liu, Wang, Liu \& Duncan]{erinin2023plunging_b}
{\sc \au{Erinin, M.~A.}, \au{Liu, C.}, \au{Wang, S.~D.}, \au{Liu, X.} \& \au{Duncan, J.~H.}} \yr{2023{\natexlab{{\em a\/}}}}  \at{Plunging breakers. {P}art 2. {D}roplet generation}.  \jt{Journal of Fluid Mechanics}  \bvol{967},  \pg{A36}.

\bibitem[Erinin {\em et~al.\/}(2023{\natexlab{{\em b\/}}})Erinin, Liu, Wang \& Duncan]{erinin2023plunging_a}
{\sc \au{Erinin, M.~A.}, \au{Liu, X.}, \au{Wang, S.~D.} \& \au{Duncan, J.~H.}} \yr{2023{\natexlab{{\em b\/}}}}  \at{Plunging breakers. {P}art 1. {A}nalysis of an ensemble of wave profiles}.  \jt{Journal of Fluid Mechanics}  \bvol{967},  \pg{A35}.

\bibitem[Fudge {\em et~al.\/}(2023)Fudge, Cimpeanu, Antkowiak, Castrej{\'o}n-Pita \& Castrej{\'o}n-Pita]{fudge2023drop}
{\sc \au{Fudge, B.~D.}, \au{Cimpeanu, R.}, \au{Antkowiak, A.}, \au{Castrej{\'o}n-Pita, J.~R.} \& \au{Castrej{\'o}n-Pita, A.~A.}} \yr{2023}  \at{Drop splashing after impact onto immiscible pools of different viscosities}.  \jt{Journal of Colloid and Interface Science}  \bvol{641},  \pg{585--594}.

\bibitem[Gao {\em et~al.\/}(2021)Gao, Deane \& Shen]{gao2021bubble}
{\sc \au{Gao, Q.}, \au{Deane, G.~B.} \& \au{Shen, L.}} \yr{2021}  \at{Bubble production by air filament and cavity breakup in plunging breaking wave crests}.  \jt{Journal of Fluid Mechanics}  \bvol{929},  \pg{A44}.

\bibitem[Garc{\'\i}a-Geijo {\em et~al.\/}(2021)Garc{\'\i}a-Geijo, Quintero, Riboux \& Gordillo]{garcia2021spreading}
{\sc \au{Garc{\'\i}a-Geijo, P.}, \au{Quintero, E.~S.}, \au{Riboux, G.} \& \au{Gordillo, J.~M.}} \yr{2021}  \at{Spreading and splashing of drops impacting rough substrates}.  \jt{Journal of Fluid Mechanics}  \bvol{917},  \pg{A50}.

\bibitem[Gekle \& Gordillo(2010)]{gekle2010generation}
{\sc \au{Gekle, S.} \& \au{Gordillo, J.~M.}} \yr{2010}  \at{Generation and breakup of {W}orthington jets after cavity collapse. {P}art 1. {J}et formation}.  \jt{Journal of Fluid Mechanics}  \bvol{663},  \pg{293--330}.

\bibitem[Ghabache {\em et~al.\/}(2014)Ghabache, S{\'e}on \& Antkowiak]{ghabache2014liquid}
{\sc \au{Ghabache, {\'E}.}, \au{S{\'e}on, T.} \& \au{Antkowiak, A.}} \yr{2014}  \at{Liquid jet eruption from hollow relaxation}.  \jt{Journal of fluid mechanics}  \bvol{761},  \pg{206--219}.

\bibitem[Gordillo \& Blanco-Rodr{\'\i}guez(2023)]{gordillo2023jets}
{\sc \au{Gordillo, J.~M.} \& \au{Blanco-Rodr{\'\i}guez, F.~J.}} \yr{2023}  \at{Theory of the jets ejected after the inertial collapse of cavities with applications to bubble bursting jets}.  \jt{Physical Review Fluids}  \bvol{8}~(7),  \pg{073606}.

\bibitem[Gordillo \& Gekle(2010)]{gordillo2010generation}
{\sc \au{Gordillo, J.~M.} \& \au{Gekle, S.}} \yr{2010}  \at{Generation and breakup of worthington jets after cavity collapse. part 2. tip breakup of stretched jets}.  \jt{Journal of Fluid Mechanics}  \bvol{663},  \pg{331--346}.

\bibitem[Gordillo {\em et~al.\/}(2014)Gordillo, Lhuissier \& Villermaux]{gordillo2014cusps}
{\sc \au{Gordillo, J.~M.}, \au{Lhuissier, H.} \& \au{Villermaux, E.}} \yr{2014}  \at{On the cusps bordering liquid sheets}.  \jt{Journal of Fluid Mechanics}  \bvol{754},  \pg{R1}.

\bibitem[Gordillo {\em et~al.\/}(2020)Gordillo, Onuki \& Tagawa]{gordillo2020impulsive}
{\sc \au{Gordillo, J.~M.}, \au{Onuki, H.} \& \au{Tagawa, Y.}} \yr{2020}  \at{Impulsive generation of jets by flow focusing}.  \jt{Journal of Fluid Mechanics}  \bvol{894},  \pg{A3}.

\bibitem[Gordillo {\em et~al.\/}(2019)Gordillo, Riboux \& Quintero]{gordillo2019theory}
{\sc \au{Gordillo, J.~M.}, \au{Riboux, G.} \& \au{Quintero, E.~S.}} \yr{2019}  \at{A theory on the spreading of impacting droplets}.  \jt{Journal of Fluid Mechanics}  \bvol{866},  \pg{298--315}.

\bibitem[Goswami \& Hardalupas(2023)]{goswami2023simultaneous}
{\sc \au{Goswami, A.} \& \au{Hardalupas, Y.}} \yr{2023}  \at{Simultaneous impact of droplet pairs on solid surfaces}.  \jt{Journal of Fluid Mechanics}  \bvol{961},  \pg{A17}.

\bibitem[He {\em et~al.\/}(2019)He, Xia \& Zhang]{he2019non}
{\sc \au{He, C.}, \au{Xia, X.} \& \au{Zhang, P.}} \yr{2019}  \at{Non-monotonic viscous dissipation of bouncing droplets undergoing off-center collision}.  \jt{Physics of Fluids}  \bvol{31}~(5),  \pg{052004}.

\bibitem[He {\em et~al.\/}(2022)He, Yue \& Zhang]{he2022spin}
{\sc \au{He, C.}, \au{Yue, L.} \& \au{Zhang, P.}} \yr{2022}  \at{Spin-affected reflexive and stretching separation of off-center droplet collision}.  \jt{Physical Review Fluids}  \bvol{7}~(1),  \pg{013603}.

\bibitem[Hopper(1993{\natexlab{{\em a\/}}})]{hopper1993acoalescence}
{\sc \au{Hopper, R.~W.}} \yr{1993{\natexlab{{\em a\/}}}}  \at{Coalescence of two viscous cylinders by capillarity: {P}art {I}. {T}heory}.  \jt{Journal of the American Ceramic Society}  \bvol{76}~(12),  \pg{2947--2952}.

\bibitem[Hopper(1993{\natexlab{{\em b\/}}})]{hopper1993bcoalescence}
{\sc \au{Hopper, R.~W.}} \yr{1993{\natexlab{{\em b\/}}}}  \at{Coalescence of two viscous cylinders by capillarity: {P}art {II}. {S}hape evolution}.  \jt{Journal of the American Ceramic Society}  \bvol{76}~(12),  \pg{2953--2960}.

\bibitem[Jackiw \& Ashgriz(2022)]{jackiw2022prediction}
{\sc \au{Jackiw, I.~M.} \& \au{Ashgriz, N.}} \yr{2022}  \at{Prediction of the droplet size distribution in aerodynamic droplet breakup}.  \jt{Journal of Fluid Mechanics}  \bvol{940}.

\bibitem[Josserand {\em et~al.\/}(2016)Josserand, Ray \& Zaleski]{josserand2016droplet}
{\sc \au{Josserand, C.}, \au{Ray, P.} \& \au{Zaleski, S.}} \yr{2016}  \at{Droplet impact on a thin liquid film: anatomy of the splash}.  \jt{Journal of Fluid Mechanics}  \bvol{802},  \pg{775--805}.

\bibitem[Josserand \& Thoroddsen(2016)]{josserand2016drop}
{\sc \au{Josserand, C.} \& \au{Thoroddsen, S.~T.}} \yr{2016}  \at{Drop impact on a solid surface}.  \jt{Annual Review of Fluid Mechanics}  \bvol{48},  \pg{365--391}.

\bibitem[Kiger \& Duncan(2012)]{kiger2012air}
{\sc \au{Kiger, K.~T.} \& \au{Duncan, J.~H.}} \yr{2012}  \at{Air-entrainment mechanisms in plunging jets and breaking waves}.  \jt{Annual Review of Fluid Mechanics}  \bvol{44},  \pg{563--596}.

\bibitem[Lai {\em et~al.\/}(2018)Lai, Eggers \& Deike]{lai2018bubble}
{\sc \au{Lai, C.-Y.}, \au{Eggers, J.} \& \au{Deike, L.}} \yr{2018}  \at{Bubble bursting: Universal cavity and jet profiles}.  \jt{Physical Review Letters}  \bvol{121}~(14),  \pg{144501}.

\bibitem[Lamb(1932)]{lamb1932hydrodynamics}
{\sc \au{Lamb, H.}} \yr{1932} {\em Hydrodynamics\/}.  \publ{Cambridge University Press}.

\bibitem[Lhuissier \& Villermaux(2012)]{lhuissier2012bursting}
{\sc \au{Lhuissier, H.} \& \au{Villermaux, E.}} \yr{2012}  \at{Bursting bubble aerosols}.  \jt{Journal of Fluid Mechanics}  \bvol{696},  \pg{5--44}.

\bibitem[Li {\em et~al.\/}(2018)Li, Thoraval, Marston \& Thoroddsen]{li2018early}
{\sc \au{Li, E.Q.}, \au{Thoraval, M.-J.}, \au{Marston, J.~O.} \& \au{Thoroddsen, S.~T.}} \yr{2018}  \at{Early azimuthal instability during drop impact}.  \jt{Journal of Fluid Mechanics}  \bvol{848},  \pg{821--835}.

\bibitem[Ling \& Mahmood(2023)]{ling2023detailed}
{\sc \au{Ling, Y.} \& \au{Mahmood, T.}} \yr{2023}  \at{Detailed numerical investigation of the drop aerobreakup in the bag breakup regime}.  \jt{Journal of Fluid Mechanics}  \bvol{972},  \pg{A28}.

\bibitem[Liu {\em et~al.\/}(2022)Liu, Hernandez-Rueda, Gelderblom \& Versolato]{liu2022speed}
{\sc \au{Liu, B.}, \au{Hernandez-Rueda, J.}, \au{Gelderblom, H.} \& \au{Versolato, O.~O.}} \yr{2022}  \at{Speed of fragments ejected by an expanding liquid tin sheet}.  \jt{Physical Review Fluids}  \bvol{7}~(8),  \pg{083601}.

\bibitem[Liu \& Bothe(2016)]{liu2016numerical}
{\sc \au{Liu, M.} \& \au{Bothe, D.}} \yr{2016}  \at{Numerical study of head-on droplet collisions at high {W}eber numbers}.  \jt{Journal of Fluid Mechanics}  \bvol{789},  \pg{785--805}.

\bibitem[Liu {\em et~al.\/}(2021)Liu, Lo, Li, Liu, Zhao \& Xu]{liu2021role}
{\sc \au{Liu, Q.}, \au{Lo, J. H.~Y.}, \au{Li, Y.}, \au{Liu, Y.}, \au{Zhao, J.} \& \au{Xu, L.}} \yr{2021}  \at{The role of drop shape in impact and splash}.  \jt{Nature Communications}  \bvol{12}~(1),  \pg{3068}.

\bibitem[Lohse(2022)]{lohse2022fundamental}
{\sc \au{Lohse, D.}} \yr{2022}  \at{Fundamental fluid dynamics challenges in inkjet printing}.  \jt{Annual Review of Fluid Mechanics}  \bvol{54},  \pg{349--382}.

\bibitem[Longuet-Higgins(2001)]{longuet2001vertical}
{\sc \au{Longuet-Higgins, M.~S.}} \yr{2001}  \at{Vertical jets from standing waves}.  \jt{Proceedings of the Royal Society of London. Series A: Mathematical, Physical and Engineering Sciences}  \bvol{457}~(2006),  \pg{495--510}.

\bibitem[Marmanis \& Thoroddsen(1996)]{marmanis1996scaling}
{\sc \au{Marmanis, H.} \& \au{Thoroddsen, S.~T.}} \yr{1996}  \at{Scaling of the fingering pattern of an impacting drop}.  \jt{Physics of Fluids}  \bvol{8}~(6),  \pg{1344--1346}.

\bibitem[Mehta {\em et~al.\/}(2017)Mehta, Haj-Ahmad, Rasekh, Arshad, Smith, van~der Merwe, Li, Chang \& Ahmad]{Mehta2017}
{\sc \au{Mehta, P.}, \au{Haj-Ahmad, R.}, \au{Rasekh, M.}, \au{Arshad, M.~S.}, \au{Smith, A.}, \au{van~der Merwe, S.~M.}, \au{Li, X.}, \au{Chang, M.-W.} \& \au{Ahmad, Z.}} \yr{2017}  \at{Pharmaceutical and biomaterial engineering via electrohydrodynamic atomization technologies}.  \jt{Drug Discovery Today}  \bvol{22}~(1),  \pg{157--165}.

\bibitem[Mongruel {\em et~al.\/}(2009)Mongruel, Daru, Feuillebois \& Tabakova]{mongruel2009early}
{\sc \au{Mongruel, A.}, \au{Daru, V.}, \au{Feuillebois, F.} \& \au{Tabakova, S.}} \yr{2009}  \at{Early post-impact time dynamics of viscous drops onto a solid dry surface}.  \jt{Physics of Fluids}  \bvol{21}~(3).

\bibitem[Mostert {\em et~al.\/}(2022)Mostert, Popinet \& Deike]{Mostert2021}
{\sc \au{Mostert, W.}, \au{Popinet, S.} \& \au{Deike, L.}} \yr{2022}  \at{High-resolution direct simulation of deep water breaking waves: transition to turbulence, bubbles and droplets production}.  \jt{Journal of Fluid Mechanics}  \bvol{942},  \pg{A27}.

\bibitem[N{\'e}el {\em et~al.\/}(2020)N{\'e}el, Lhuissier \& Villermaux]{neel2020fines}
{\sc \au{N{\'e}el, B.}, \au{Lhuissier, H.} \& \au{Villermaux, E.}} \yr{2020}  \at{‘fines’ from the collision of liquid rims}.  \jt{Journal of Fluid Mechanics}  \bvol{893},  \pg{A16}.

\bibitem[{\'O}~N{\'a}raigh \& Mairal(2023)]{naraigh2023analysis}
{\sc \au{{\'O}~N{\'a}raigh, L.} \& \au{Mairal, J.}} \yr{2023}  \at{Analysis of the spreading radius in droplet impact: The two-dimensional case}.  \jt{Physics of Fluids}  \bvol{35}~(10).

\bibitem[Obenauf \& Sojka(2021)]{Obenauf2021}
{\sc \au{Obenauf, D.~G.} \& \au{Sojka, P.~E.}} \yr{2021}  \at{Theoretical deformation modeling and drop size prediction in the multimode breakup regime}.  \jt{Physics of Fluids}  \bvol{33}~(9),  \pg{092113}.

\bibitem[Pairetti {\em et~al.\/}(2018)Pairetti, Popinet, Dami{\'a}n, Nigro \& Zaleski]{Pairetti2018}
{\sc \au{Pairetti, C.}, \au{Popinet, S.}, \au{Dami{\'a}n, S.}, \au{Nigro, N.} \& \au{Zaleski, S.}} \yr{2018} Bag mode breakup simulations of a single liquid droplet.  \bt{In {\em 6th European Conference on Computational Mechanics\/}}.

\bibitem[Pal {\em et~al.\/}(2021)Pal, Crialesi-Esposito, Fuster \& Zaleski]{pal2021statistics}
{\sc \au{Pal, S.}, \au{Crialesi-Esposito, M.}, \au{Fuster, D.} \& \au{Zaleski, S.}} \yr{2021}  \at{Statistics of drops generated from ensembles of randomly corrugated ligaments}.  \jt{arXiv preprint arXiv:2106.16192} .

\bibitem[Philippi {\em et~al.\/}(2016)Philippi, Lagr{\'e}e \& Antkowiak]{philippi2016drop}
{\sc \au{Philippi, J.}, \au{Lagr{\'e}e, P.-Y.} \& \au{Antkowiak, A.}} \yr{2016}  \at{Drop impact on a solid surface: short-time self-similarity}.  \jt{Journal of Fluid Mechanics}  \bvol{795},  \pg{96--135}.

\bibitem[Popinet(2009)]{popinet2009accurate}
{\sc \au{Popinet, S.}} \yr{2009}  \at{An accurate adaptive solver for surface-tension-driven interfacial flows}.  \jt{Journal of Computational Physics}  \bvol{228}~(16),  \pg{5838--5866}.

\bibitem[Popinet(2018)]{popinet2018numerical}
{\sc \au{Popinet, S.}} \yr{2018}  \at{Numerical models of surface tension}.  \jt{Annual Review of Fluid Mechanics}  \bvol{50},  \pg{49--75}.

\bibitem[Popinet(2019)]{Popinet2019basilisk}
{\sc \au{Popinet, S.}} \yr{2019} Basilisk flow solver and {PDE} library. Available at: \url{http://basilisk.fr}.

\bibitem[Riboux \& Gordillo(2014)]{riboux2014experiments}
{\sc \au{Riboux, G.} \& \au{Gordillo, J.~M.}} \yr{2014}  \at{Experiments of drops impacting a smooth solid surface: a model of the critical impact speed for drop splashing}.  \jt{Physical Review Letters}  \bvol{113}~(2),  \pg{024507}.

\bibitem[Riboux \& Gordillo(2015)]{riboux2015diameters}
{\sc \au{Riboux, G.} \& \au{Gordillo, J.~M.}} \yr{2015}  \at{The diameters and velocities of the droplets ejected after splashing}.  \jt{Journal of Fluid Mechanics}  \bvol{772},  \pg{630--648}.

\bibitem[Rivi{\`e}re {\em et~al.\/}(2021)Rivi{\`e}re, Mostert, Perrard \& Deike]{riviere2021sub}
{\sc \au{Rivi{\`e}re, A.}, \au{Mostert, W.}, \au{Perrard, S.} \& \au{Deike, L.}} \yr{2021}  \at{Sub-hinze scale bubble production in turbulent bubble break-up}.  \jt{Journal of Fluid Mechanics}  \bvol{917},  \pg{A40}.

\bibitem[Savva \& Bush(2009)]{savva2009viscous}
{\sc \au{Savva, N.} \& \au{Bush, J. W.~M.}} \yr{2009}  \at{Viscous sheet retraction}.  \jt{Journal of Fluid Mechanics}  \bvol{626},  \pg{211--240}.

\bibitem[Tang {\em et~al.\/}(2023)Tang, Adcock \& Mostert]{tang2022bag}
{\sc \au{Tang, K.}, \au{Adcock, T.A.A.} \& \au{Mostert, W.}} \yr{2023}  \at{Bag film breakup of droplets in uniform airflows}.  \jt{Journal of Fluid Mechanics}  \bvol{970},  \pg{A9}.

\bibitem[Tang {\em et~al.\/}(2021)Tang, Mostert, Fuster \& Deike]{tang2021effects}
{\sc \au{Tang, K.}, \au{Mostert, W.}, \au{Fuster, D.} \& \au{Deike, L.}} \yr{2021}  \at{Effects of surface tension on the {R}ichtmyer-{M}eshkov instability in fully compressible and inviscid fluids}.  \jt{Physical Review Fluids}  \bvol{6}~(11),  \pg{113901}.

\bibitem[Taylor(1959)]{taylor1959dynamics}
{\sc \au{Taylor, G.~I.}} \yr{1959}  \at{The dynamics of thin sheets of fluid ii. waves on fluid sheets}.  \jt{Proceedings of the Royal Society of London. Series A. Mathematical and Physical Sciences}  \bvol{253}~(1274),  \pg{296--312}.

\bibitem[Thoraval {\em et~al.\/}(2013)Thoraval, Takehara, Etoh \& Thoroddsen]{thoraval2013drop}
{\sc \au{Thoraval, M.-J.}, \au{Takehara, K.}, \au{Etoh, T.~G.} \& \au{Thoroddsen, S.~T.}} \yr{2013}  \at{Drop impact entrapment of bubble rings}.  \jt{Journal of Fluid Mechanics}  \bvol{724},  \pg{234--258}.

\bibitem[Thoroddsen \& Sakakibara(1998)]{thoroddsen1998evolution}
{\sc \au{Thoroddsen, S.~T.} \& \au{Sakakibara, J.}} \yr{1998}  \at{Evolution of the fingering pattern of an impacting drop}.  \jt{Physics of Fluids}  \bvol{10}~(6),  \pg{1359--1374}.

\bibitem[Thoroddsen {\em et~al.\/}(2012)Thoroddsen, Takehara \& Etoh]{thoroddsen2012micro}
{\sc \au{Thoroddsen, S.~T.}, \au{Takehara, K.} \& \au{Etoh, T.~G.}} \yr{2012}  \at{Micro-splashing by drop impacts}.  \jt{Journal of Fluid Mechanics}  \bvol{706},  \pg{560--570}.

\bibitem[Veron(2015)]{Veron2015}
{\sc \au{Veron, F.}} \yr{2015}  \at{Ocean spray}.  \jt{Annual Review of Fluid Mechanics}  \bvol{47},  \pg{507--538}.

\bibitem[Villermaux(2007)]{Villermaux2007}
{\sc \au{Villermaux, E.}} \yr{2007}  \at{Fragmentation}.  \jt{Annual Review of Fluid Mechanics}  \bvol{39},  \pg{419--446}.

\bibitem[Villermaux \& Bossa(2009)]{Villermaux2009}
{\sc \au{Villermaux, E.} \& \au{Bossa, B.}} \yr{2009}  \at{Single-drop fragmentation determines size distribution of raindrops}.  \jt{Nature Physics}  \bvol{5}~(9),  \pg{697--702}.

\bibitem[Villermaux \& Bossa(2011)]{Villermaux2011}
{\sc \au{Villermaux, E.} \& \au{Bossa, B.}} \yr{2011}  \at{Drop fragmentation on impact}.  \jt{Journal of Fluid Mechanics}  \bvol{668},  \pg{412--435}.

\bibitem[Visser {\em et~al.\/}(2015)Visser, Frommhold, Wildeman, Mettin, Lohse \& Sun]{visser2015dynamics}
{\sc \au{Visser, C.~W.}, \au{Frommhold, P.~E.}, \au{Wildeman, S.}, \au{Mettin, R.}, \au{Lohse, D.} \& \au{Sun, C.}} \yr{2015}  \at{Dynamics of high-speed micro-drop impact: numerical simulations and experiments at frame-to-frame times below 100 ns}.  \jt{Soft Matter}  \bvol{11}~(9),  \pg{1708--1722}.

\bibitem[Vledouts {\em et~al.\/}(2016)Vledouts, Quinard, Vandenberghe \& Villermaux]{vledouts2016explosive}
{\sc \au{Vledouts, A.}, \au{Quinard, J.}, \au{Vandenberghe, N.} \& \au{Villermaux, E.}} \yr{2016}  \at{Explosive fragmentation of liquid shells}.  \jt{Journal of Fluid Mechanics}  \bvol{788},  \pg{246--273}.

\bibitem[Wagner(1932)]{wagner1932uber}
{\sc \au{Wagner, H.}} \yr{1932}  \at{Über stoß- und gleitvorgänge an der oberfläche von flüssigkeiten}.  \jt{Zeitschrift für Angewandte Mathematik und Mechanik}  \bvol{12}~(4),  \pg{193--215}.

\bibitem[Wang {\em et~al.\/}(2023)Wang, Liu, Bayeul-Lain{\'e}, Murphy, Katz \& Coutier-Delgosha]{wang2023analysis}
{\sc \au{Wang, H.}, \au{Liu, S.}, \au{Bayeul-Lain{\'e}, A.-C.}, \au{Murphy, D.}, \au{Katz, J.} \& \au{Coutier-Delgosha, O.}} \yr{2023}  \at{Analysis of high-speed drop impact onto deep liquid pool}.  \jt{Journal of Fluid Mechanics}  \bvol{972},  \pg{A31}.

\bibitem[Wang \& Bourouiba(2017)]{wang2017drop}
{\sc \au{Wang, Y.} \& \au{Bourouiba, L.}} \yr{2017}  \at{Drop impact on small surfaces: thickness and velocity profiles of the expanding sheet in the air}.  \jt{Journal of Fluid Mechanics}  \bvol{814},  \pg{510--534}.

\bibitem[Wang \& Bourouiba(2018)]{wang2018unsteady}
{\sc \au{Wang, Y.} \& \au{Bourouiba, L.}} \yr{2018}  \at{Unsteady sheet fragmentation: droplet sizes and speeds}.  \jt{Journal of Fluid Mechanics}  \bvol{848},  \pg{946--967}.

\bibitem[Wang \& Bourouiba(2021)]{wang2021growth}
{\sc \au{Wang, Y.} \& \au{Bourouiba, L.}} \yr{2021}  \at{Growth and breakup of ligaments in unsteady fragmentation}.  \jt{Journal of Fluid Mechanics}  \bvol{910},  \pg{A39}.

\bibitem[Wang \& Bourouiba(2022)]{wang2022mass}
{\sc \au{Wang, Y.} \& \au{Bourouiba, L.}} \yr{2022}  \at{Mass, momentum and energy partitioning in unsteady fragmentation}.  \jt{Journal of Fluid Mechanics}  \bvol{935},  \pg{A29}.

\bibitem[Wang {\em et~al.\/}(2018)Wang, Dandekar, Bustos, Poulain \& Bourouiba]{wang2018universal}
{\sc \au{Wang, Y.}, \au{Dandekar, R.}, \au{Bustos, N.}, \au{Poulain, S.} \& \au{Bourouiba, L.}} \yr{2018}  \at{Universal rim thickness in unsteady sheet fragmentation}.  \jt{Physical Review Letters}  \bvol{120}~(20),  \pg{204503}.

\bibitem[Wang {\em et~al.\/}(2016)Wang, Yang \& Stern]{wang2016high}
{\sc \au{Wang, Z.}, \au{Yang, J.} \& \au{Stern, F.}} \yr{2016}  \at{High-fidelity simulations of bubble, droplet and spray formation in breaking waves}.  \jt{Journal of Fluid Mechanics}  \bvol{792},  \pg{307--327}.

\bibitem[Wildeman {\em et~al.\/}(2016)Wildeman, Visser, Sun \& Lohse]{wildeman2016spreading}
{\sc \au{Wildeman, S.}, \au{Visser, C.~W.}, \au{Sun, C.} \& \au{Lohse, D.}} \yr{2016}  \at{On the spreading of impacting drops}.  \jt{Journal of Fluid Mechanics}  \bvol{805},  \pg{636--655}.

\bibitem[Worthington(1877)]{worthington1877xxviii}
{\sc \au{Worthington, A.~M.}} \yr{1877}  \at{On the forms assumed by drops of liquids falling vertically on a horizontal plate}.  \jt{Proceedings of the Royal Society of London}  \bvol{25}~(171-178),  \pg{261--272}.

\bibitem[Yarin(2006)]{yarin2006drop}
{\sc \au{Yarin, A.~L.}} \yr{2006}  \at{Drop impact dynamics: Splashing, spreading, receding, bouncing…}.  \jt{Annual Review of Fluid Mechanics}  \bvol{38},  \pg{159--192}.

\bibitem[Yarin \& Weiss(1995)]{yarin1995impact}
{\sc \au{Yarin, A.~L.} \& \au{Weiss, D.~A.}} \yr{1995}  \at{Impact of drops on solid surfaces: self-similar capillary waves, and splashing as a new type of kinematic discontinuity}.  \jt{Journal of Fluid Mechanics}  \bvol{283},  \pg{141--173}.

\bibitem[Zhang {\em et~al.\/}(2010)Zhang, Brunet, Eggers \& Deegan]{zhang2010wavelength}
{\sc \au{Zhang, L.~V.}, \au{Brunet, P.}, \au{Eggers, J.} \& \au{Deegan, R.~D.}} \yr{2010}  \at{Wavelength selection in the crown splash}.  \jt{Physics of Fluids}  \bvol{22}~(12).

\end{thebibliography}

\end{document}